\documentclass{jfm}
\usepackage{graphicx}
\usepackage{amsmath}
\usepackage{epstopdf,epsfig}
\usepackage{float}
\usepackage{amssymb}
\usepackage{color,contour}
\usepackage{overpic}
\usepackage[dvipsnames]{xcolor}
\usepackage{tabularx}
\usepackage{booktabs}
\usepackage[colorlinks,citecolor = blue, linkcolor=red,hyperindex,CJKbookmarks]{hyperref}

\title[From Rayleigh-B\'enard convection to porous-media convection]{From Rayleigh-B\'enard convection to porous-media convection: how porosity affects heat transfer and flow structure}

\author[S. Liu and others]%
{Shuang Liu$^{1,2}$
, Linfeng Jiang$^1$
, Kai Leong Chong$^3$
, Xiaojue Zhu$^{4}$
, Zhen-Hua Wan$^{5}$
, Roberto Verzicco$^{6,7,3}$
, Richard J. A. M. Stevens$^3$
, Detlef Lohse$^{3,8}$
, Chao Sun$^{1,2}$ \thanks{Email address for correspondence: chaosun@tsinghua.edu.cn}}

\affiliation{ 
	$^1$Center for Combustion Energy, Key Laboratory for Thermal Science and Power Engineering of Ministry of Education, International Joint Laboratory on Low Carbon Clean Energy Innovation, Department of Energy and Power Engineering, Tsinghua University, Beijing 100084, China \\[\affilskip]	
 	$^2$ Department of Engineering Mechanics, School of Aerospace Engineering, Tsinghua University, Beijing 100084, China \\ [\affilskip]
 	$^3$Physics of Fluids Group and Max Planck Center Twente, University of Twente, PO Box 217, 7500AE Enschede, The Netherlands \\ [\affilskip]
 	$^4$Center of Mathematical Sciences and Applications, and School of Engineering and Applied Sciences,  Harvard University, Cambridge, MA 02138, USA \\ [\affilskip]
 	$^5$Department of Modern Mechanics, University of Science and Technology of China, Hefei, Anhui 230027, China \\ [\affilskip]
 	$^6$Dipartimento di Ingegneria Industriale, University of Rome `Tor Vergata', Via del Politecnico 1, 00133 Roma, Italy  \\ [\affilskip]
 	$^7$Gran Sasso Science Institute - Viale F. Crispi, 7, 67100 L’Aquila, Italy \\[\affilskip]
 	$^8$Max Planck Institute for Dynamics and Self-Organization, 37077 G{\"o}ttingen, Germany
 }

\pubyear{2010}
\volume{650}
\pagerange{119--126}
\date{?; revised ?; accepted ?. - To be entered by editorial office}
\begin{document}

\maketitle

%\vspace{-2.2 mm}
\begin{abstract}
	We perform a numerical study of the heat transfer and flow structure of Rayleigh-B\'enard (RB) convection in {\color{black}(in most cases regular)} porous media, which are comprised of circular, solid obstacles {\color{black}located} on a square lattice. This study is focused on the role of porosity $\phi$ in the flow properties {\color{black}during} the transition process from the traditional RB convection with $\phi=1$ (so no obstacles included) to Darcy-type porous-media convection with $\phi$ approaching 0. Simulations are carried out in a cell with unity aspect ratio, for the Rayleigh number $Ra$ from $10^5$ to $10^{10}$ and varying porosities $\phi$, at a fixed Prandtl number $Pr=4.3$, and we restrict ourselves to the two dimensional case. For fixed $Ra$, the Nusselt number $Nu$ {\color{black}is found to} vary non-monotonously as a function of $\phi$; namely, with decreasing $\phi$, it first increases, before it decreases for $\phi$ approaching 0. The non-monotonous behaviour of $Nu(\phi)$ originates from two competing effects of the porous structure on the heat transfer. On the one hand, the flow coherence is enhanced in the porous media, which is beneficial for the heat transfer. On the other hand, the convection is slowed down by the enhanced resistance due to the porous structure, leading to heat transfer reduction. For fixed $\phi$, depending on $Ra$, two different heat transfer regimes are identified, with different effective power-law behaviours of $Nu$ vs $Ra$, namely, a steep one for low $Ra$ when viscosity dominates, and the standard classical one for large $Ra$. The scaling crossover occurs when the thermal boundary layer thickness and the pore scale are comparable. The influences of the porous structure on the temperature and velocity fluctuations, convective heat flux, and energy dissipation rates are analysed, further demonstrating the competing effects of the porous structure to enhance or reduce the heat transfer.
\end{abstract}
%\vspace{-2.2 mm}


\section{Introduction}
Thermal convection is an omnipresent phenomenon in nature and technology.
One of the paradigms for thermal convection studies is Rayleigh-B{\'e}nard (RB) convection, i.e., convection in a container heated from below and cooled from above, and it has been studied extensively over the last few decades \citep{ahlers2009heat,lohse2010small,chilla2012new,xia2013current}.
Also the related problem of convection in a fluid-saturated porous medium has received increasing attention owing to its importance in a wide range of natural and industrial processes, such as geothermal energy recovery and geological sequestration of carbon dioxide \citep{cinar2009experimental,hassanzadeh2007scaling,orr2009onshore,huppert2014fluid,riaz2014carbon,cinar2014carbon,emami2015two,de2016influence,soltanian2016critical,amooie2018solutal}. {\color{black}It is indeed of both fundamental and practical interest to study RB convection in porous media, and considerable progress has been achieved over the years based on the combinations of experimental, numerical, and theoretical studies} \citep{lapwood1948convection,wooding1957steady,joseph1982nonlinear,otero2004high,nield2006convection,araujo2006distribution,landman2007heat,hewitt2012ultimate,hewitt2014high,wen2015structure,keene2015thermal,ataei2019flow,chakkingal2019numerical}.

{\color{black}For pure RB convection, in particular, the heat transfer and flow structure have been studied extensively \citep{ahlers2009heat}.} There also has been rapid progress in the modelling of RB convection with additional effects, such as multiphase RB convection \citep{lakkaraju2013heat,wang2019self}, convection with rough walls \citep{shishkina2011modelling,wagner2015heat,zhu2017roughness,jiang2018controlling,zhu2019nu}, {\color{black}tilted convection \citep{shishkina2016thermal,zwirner2018confined,wang2018multiple,jiang2019robustness},} and partitioned RB convection \citep{bao2015enhanced}.
{\color{black}In the numerical study of confined inclined RB convection in low-Prandtl-number fluids, \cite{zwirner2018confined} identified significant heat transfer enhancement, which is closely related to the organization of the plumes in inclined convection, namely, the formation of system-sized plume columns impinging on the opposite boundary layers.}
Also porous-media convection can be understood as geometrically modified RB convection, with correspondingly modified heat transfer and flow structure.
In this study we want to investigate {\color{black}how these flow properties are affected by the porous structure}.

{\color{black}When a porous medium is present in a convection cell, the convection is blocked and the flow is stabilized. {\color{black}For given strength of the driving buoyancy force, the strength of the stabilizing effect can be quantified by the porosity.} The stabilizing effect {\color{black}is stronger} for greater blockage, i.e., for smaller porosity $\phi$. However, the interplay between stabilizing and driving forces can result in surprising effects.
Examples include confined RB convection \citep{chong2015condensation,chong2018effect}, rotating RB convection \citep{zhong2009prandtl,stevens2009transitions}, and double diffusive convection \citep{yang2015salinity,yang2016scaling}. \cite{chong2017confined} compared these three cases and found that in all three cases, the appropriate strength of the stabilizing force leads to significant heat transfer enhancement due to increased flow coherence, in particular, revealing a universal mechanism of the turbulent transport enhancement in the presence of stabilizing forces. Obviously, when the stabilizing force becomes even stronger, the flow motion is eventually suppressed, leading to heat transfer reduction. Consequently, the heat transport varies non-monotonously with the strength of the stabilizing force.
Also from this comparative perspective between the different systems, it is of interest to investigate how the stabilizing effect of a porous structure affects the heat transfer and flow structure of RB convection.}

{\color{black}For modelling porous-media convection, various studies have been conducted focusing on the Darcy-type convection.} Related numerical simulations are generally performed based on coarse-grained macroscopic models, such as Darcy's law and its extensions \citep{otero2004high,hewitt2012ultimate,hewitt2014high,wen2015structure}. In these macroscopic models, it is assumed that a macroscopic index, the permeability, relates the average fluid velocity through the pores to the pressure drop.

{\color{black}Recently, the numerical study of Darcy convection has been extended to very high Rayleigh numbers and the linear classical scaling for the Nusselt number with respect to the so-called Darcy Rayleigh number $Ra^*=RaDa$ has been observed \citep{hewitt2012ultimate,hewitt2014high}. (All these dimensionless numbers are exactly defined later in the paper.)}
{\color{black}Studies of non-Darcy porous-media convection commonly address the extension of the Darcy regime, such as the inclusion of the inertial effect \citep{nield2006convection}.}
{\color{black}\cite{nithiarasu1997natural} developed a coarse-grained, generalized model of non-Darcy convection, which yields the single-phase model in the limit of unity porosity. Based on this model, they found that in the Darcy regime $Nu$ is determined by the Darcy Rayleigh number $Ra^*$ and is independent of the individual values of the Rayleigh number $Ra$ and Darcy number $Da$. In contrast, in the non-Darcy flow regime, $Nu$ is significantly affected by both the Rayleigh number and the Darcy number, and by the porosity.}

{\color{black}Although both the traditional RB convection without a porous structure and the Darcy-type porous-media convection have been studied extensively, fewer studies have been done to reveal the physics of the transition process between these two extreme cases. One of the few studies focuses on the effect of the buoyancy strength on the flow properties of RB convection in specified porous media \citep{keene2015thermal}.} That study involved experimental measurements of heat transfer properties of RB convection in a cubic enclosure containing a packed bed of spheres, and it was found that the heat transfer properties at high Rayleigh numbers approach the behaviour of a homogeneous fluid layer without spheres. %The important role of a thin thermal boundary layer on heat transfer is demonstrated.
\cite{ataei2019flow} performed an experimental study of RB convection in an enclosure, filled with solid packing of relatively large spheres. The influences of the packing type, size, and conductivity of the spheres on the flow and heat transfer were investigated for varying Rayleigh numbers, and two heat transfer regimes were observed. One is a reduced heat transfer regime at lower Rayleigh numbers, and the other one is the asymptotic regime at high Rayleigh numbers, where the Nusselt number lines up with the results of the traditional RB convection without spheres.

In related work, \cite{chakkingal2019numerical} carried out numerical simulations of RB convection in a cubic cell packed with relatively large solid spheres and identified three different flow regimes depending on the solid-to-fluid thermal conductivity and Rayleigh number.
{\color{black} At low Rayleigh numbers, the convective heat transfer is effectively suppressed, whereas the {\color{black}overall} heat transfer can be enhanced for high-conductivity packings due to the significant contribution of the conductive heat transfer. At intermediate Rayleigh numbers, the convective flow is highly suppressed for the high-conductivity packing case, due to the strongly stabilizing effect of the packing on the stratified temperature distribution. The total heat transfer is then lower than that in the classical RB convection case. At higher Rayleigh numbers, convection is the dominant mechanism of heat transfer and then the convective heat transfer is close to that of classical RB convection.}

{\color{black}Despite all of these studies, the nature of the transition between the traditional RB convection and Darcy-type porous-media convection is still not well understood, and further studies are required to investigate the effects of the porosity on the heat transfer and flow structure. Detailed examinations of microscale flow field are needed to gain a comprehensive understanding of this transition process and its underlying physical mechanism.}

\begin{figure}
	\centering
	\includegraphics[width=0.55\linewidth]{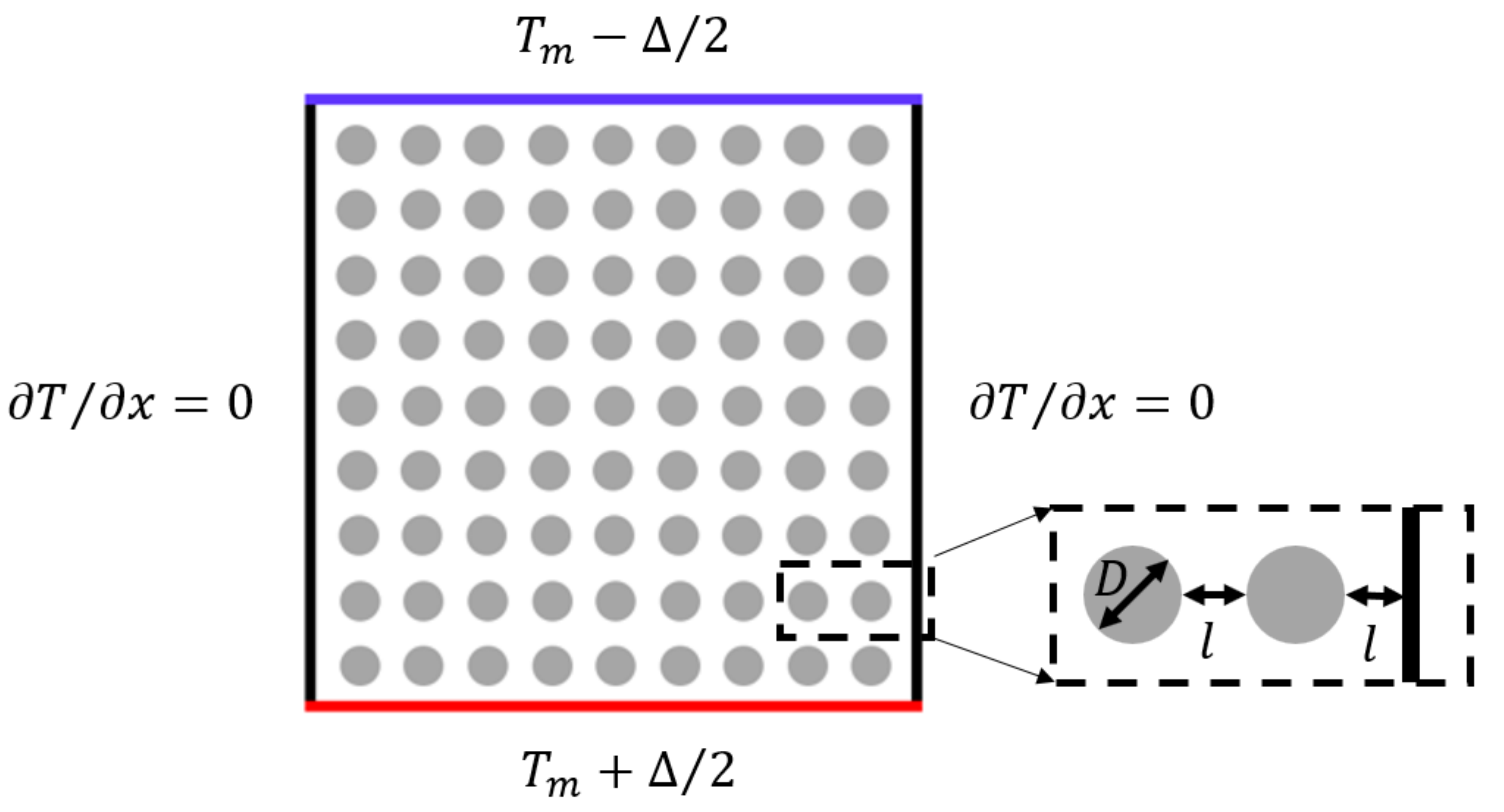}
	\caption{\label{schematic} Schematic configuration of the RB convection in regular porous media.}
\end{figure}

{\color{black} In this study we perform a numerical investigation of a representative porous medium model, namely two-dimensional (2D) RB convection in regular porous media, trying to gain an understanding of the transition process between the two distinct convection regimes. A schematic of the flow configuration is shown in figure \ref{schematic}. Circular, solid obstacles are spaced uniformly on a square lattice.
In such a pore-scale model, the detailed flow in the pores is resolved and the interaction between the porous medium and various flow structures of convection is faithfully captured, which are essential to connect the macroscopic properties with the microscale mechanisms.}

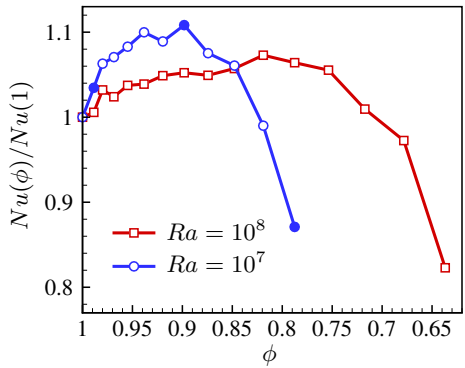
\begin{figure}
	\centering
	\hspace{-4 mm}
	\includegraphics[width=0.45\linewidth]{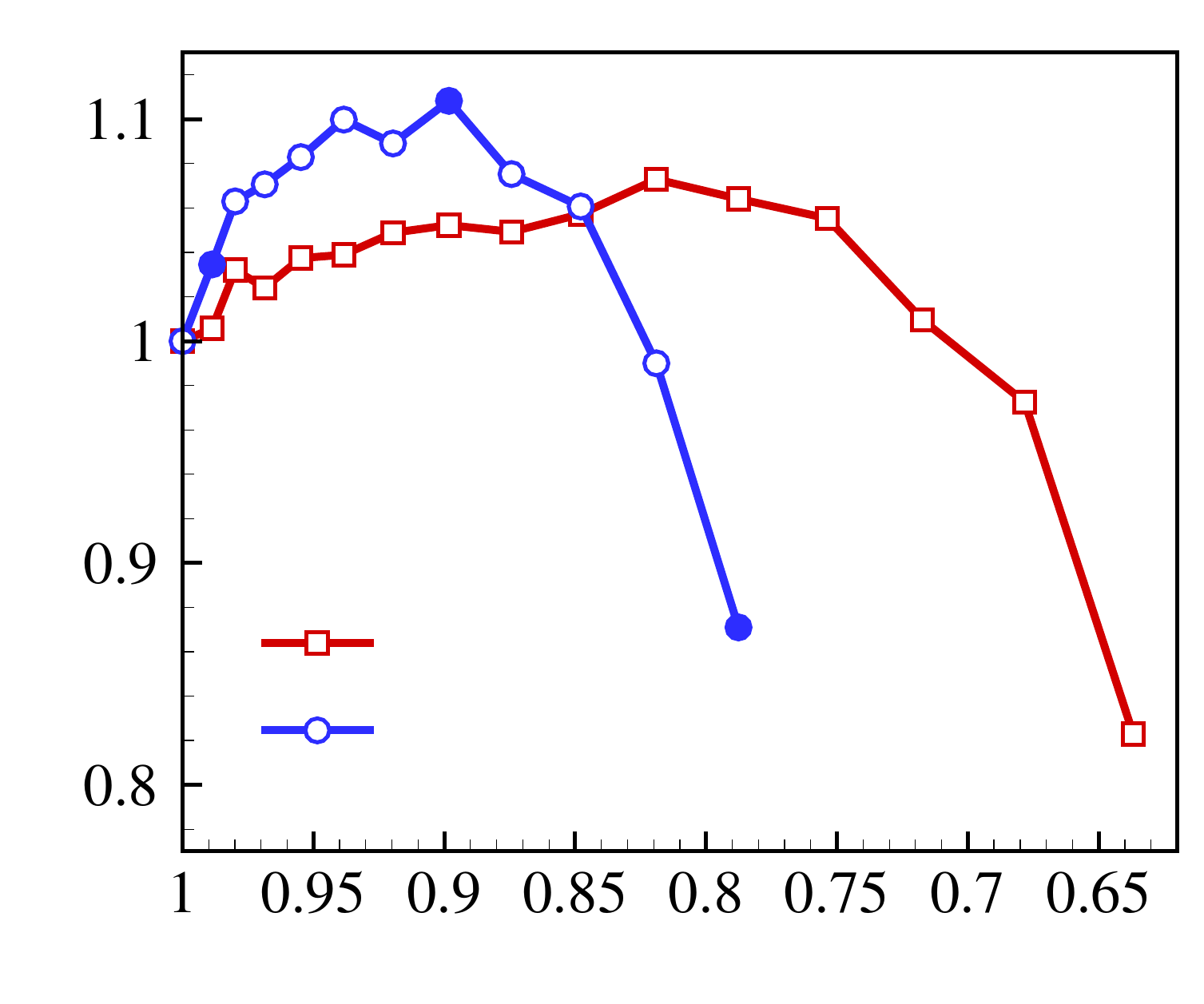}
	\put(-174,53){\rotatebox{90}{$Nu(\phi)/Nu(1)$}}
	\put(-79,4){$\phi$}
	\put(-115,49){$Ra=10^8$}
	\put(-115,37){$Ra=10^7$}
	\caption{\label{optimal_spacing} Normalized Nusselt number $Nu(\phi)/Nu(1)$ at $Ra=10^7$ and $10^8$. Snapshots of the instantaneous temperature field corresponding to the filled symbols will be shown in figure \ref{optimal_spacing_1}.}
\end{figure}

The heat transfer properties and flow structures of the porous-media convection are investigated for varying Rayleigh numbers $Ra$ and porosities $\phi$, at a fixed Prandtl number $Pr=4.3$. 
{\color{black}Figure \ref{optimal_spacing} shows the effect of the porosity $\phi$ on the heat transfer (in its dimensionless form, the Nusselt number $Nu$) for $Ra=10^7$ and $10^8$. It is observed that $Nu(\phi)$ varies non-monotonously as $\phi$ is decreased from 1 (so no obstacles included). First, when $\phi$ is slightly decreased, the heat transfer is enhanced, and then when $\phi$ is sufficiently small, it is reduced as compared to the $\phi=1$ case. Correspondingly, there is an optimum porosity for the heat transfer, which depends on $Ra$.}
{\color{black}The non-monotonic behaviour of $Nu(\phi)$ is reminiscent of the influence of other stabilizing force on turbulent transport through coherent structure manipulation \citep{chong2017confined}, which we discussed above.}

{\color{black}The objective of this study is to further understand this non-monotonic {\color{black}behaviour} of $Nu(\phi)$ and to clarify the connection of this system to other stabilizing-destabilizing flows}. 
{\color{black}In particular, we will reveal how the porous structure induces the two competing effects on the heat transfer and how they depend on the Rayleigh number and on the porosity $\phi$, and will connect the observed global transport properties to the local energy dissipation rates.}

Although the real physical system is three-dimensional (3D), we employ the 2D configuration, as we want to reveal the dominant effects of porous media over a wide range of parameters, for which 3D simulations are still prohibitive. Moreover, as the 2D numerical simulations are computationally less demanding, a sufficient flow resolution can be guaranteed.
{\color{black}In a detailed examination of the differences and similarities between two- and three-dimensional RB convection, \cite{Poel2013Comparison} found that for a large range of $Ra$ and large $Pr>1$, the $Nu$ vs $Ra$ scaling {\color{black}behaviours} in two- and three-dimensional convection are comparable up to constant prefactors, which justifies our 2D simulation at $Pr=4.3$ chosen by us (corresponding to water) {\color{black}for the traditional RB convection without obstacles. The phenomena observed in the 2D problem with obstacles may also be relevant to the 3D case. The investigation of 3D cases will be the focus of a subsequent study.}

The remainder of this paper is organized as follows. The numerical model is described in detail in \S \ref{sec:numerical}, including the governing equations {\color{black}of the pore-scale simulations} and the numerical methods. We present the main results in \S \ref{sec:sec_heat_transfer}, \S \ref{sec:sec_flow_structure}, and \S \ref{sec:sec_energy_dissipation}, focusing on the heat transfer properties, the flow structures, and the energy dissipation rates, respectively. In \S \ref{sec:summary} we will summarize our findings and give an outlook on further work.

\section{Numerical model}\label{sec:numerical}

\subsection{Governing equations}
We consider a Boussinesq fluid in a 2D RB cell with unity aspect ratio, {\color{black}containing an obstacle array located on a square lattice}, as depicted in figure \ref{schematic}.
{\color{black}The fluid is heated from below and cooled from above with a temperature difference $\Delta$ between two horizontal plates a distance $L$ apart in the vertical direction.}
The diameter of the circular, solid obstacles is $D$. The {\color{black}minimum value of the} obstacle-obstacle and obstacle-wall separations is $l$. 
The porous structures are characterized by the porosity $\phi$, which measures the volume fraction of the fluid phase. A value of $\phi=1$ corresponds to the traditional RB convection without the obstacle array.
{\color{black}For the convection cell with unity aspect ratio, we have $l=(L-N\times D)/(N+1)$, where $N$ is the number of the obstacles along the horizontal and vertical directions. The corresponding porosity is $\phi=1-N^2\pi D^2/(4L^2)$. For fixed $D$, only specific values of $\phi$ (or $l$) can be achieved, since $N$ is an integer.}
{\color{black}We note that, besides the case with regular obstacle arrangement on a square lattice, the effect of obstacle arrangement will also be considered in this study (see figure \ref{obstacle_arrangement}).}
 The non-dimensional governing equations describing the flow dynamics in the pores read
\begin{eqnarray}
\begin{split}
&\frac{\partial \vec{u}}{\partial t}+\vec{u} \cdot \nabla \vec{u}=-\nabla p+\sqrt{\frac{P r}{R a}} \nabla^{2} \vec{u}+T \vec{e}_{z} + \vec{f}, \\
&~~~~~~~\frac{\partial T}{\partial t}+\nabla \cdot (\vec{u} T)=\frac{1}{\sqrt{Pr R a}} \nabla^{2} T, \\
&~~~~~~~~~~~~~~~~~~~~~\nabla \cdot \vec{u}=0,
\end{split}
\label{governing_equations}
\end{eqnarray}
where $\vec{u}$ is the velocity vector with components $(u,v)$ along the horizontal and vertical directions $(x,z)$, $p$ is the pressure, $T$ is the temperature and $\vec{e}_{z}$ is the unit vector in the vertical direction. The immersed boundary force term $\vec{f}$ is added to account for the presence of the obstacle array.
The governing equations were non-dimensionalized using $L$ for length, $\Delta$ for temperature, the free-fall velocity $U=\sqrt{g \beta \Delta L}$ for velocity and $L/U$ for time, where $g$ is the gravitational acceleration and $\beta$ the isobaric expansion coefficient.

The two dimensionless parameters in the governing equations (\ref{governing_equations}) are the Rayleigh number, {\color{black}$R a=g \beta \Delta L^{3} / (\nu \kappa)$}, and the Prandtl number, $P r=\nu / \kappa$, where $\nu$ is the kinematic viscosity and $\kappa$ the thermal diffusivity.
The heat transfer property is measured by the Nusselt number, $Nu= \sqrt{Ra Pr} \langle v T\rangle_{x,t}-\langle\partial_z T \rangle_{x,t}$, where $\langle \cdot \rangle_{x,t}$ denotes taking averages over any horizontal plane and time. In practice, the average is taken over the top and bottom plates.
No-slip and no-penetration boundary conditions are imposed at all solid surfaces, including the fluid-obstacle interfaces. The horizontal top and bottom plates are isothermal and the sidewalls are thermally insulated.

Note that the heat transfer between the fluid and solid phases is considered. For simplicity, we assume the same thermal properties for the two phases, including $\rho c_p$, thermal conductivity $k$, and thermal diffusivity $\kappa=k/(\rho c_p)$, where $\rho$ is the density and $c_p$ the specific heat capacity. 
The temperature equation for the fluid and solid phases reads
\begin{equation}
\frac{\partial T}{\partial t}+\nabla \cdot (\vec{u}_{cp} T)=\frac{1}{\sqrt{Pr R a}} \nabla^{2} T,
\label{temperature_eqn}
\end{equation}
where $\vec{u}_{cp}$ {\color{black}(to be defined below)} is the velocity for the fluid and solid phases depending on the position in the domain  \citep{verzicco2002sidewall,verzicco2004effects,stevens2014sidewall,ardekani2018heat,ardekani2018numerical,sardina2018buoyancy}.

\subsection{Numerical methods}
In this subsection the essential elements of the numerical methods are presented.
The governing equations are discretized using a second-order finite-difference method with a pressure-correction scheme.
A fractional-step third-order Runge-Kutta scheme is employed for the time stepping of the explicit terms and a Crank-Nicholson scheme for the implicit terms.
A uniform, staggered, Cartesian grid is used in this study. The pressure Poisson equation is solved efficiently using a FFT-based solver. 
For more details of the numerical schemes of the governing equations, we refer the reader to \cite{van2015pencil}.

To account for the obstacle array, a direct-forcing immersed boundary method (IBM) in the Euler-Lagrange framework is adopted \citep{uhlmann2005immersed,breugem2012second}.
{\color{black}For the resolution of the obstacles, $N_L$ markers are distributed uniformly along the boundary of each obstacle, with the Lagrangian grid size being about 0.7 times of the Eulerian grid size.}
In each Runge-Kutta substep, a first prediction velocity is obtained by advancing the momentum equations in time without considering the IBM force $\vec{f}$. The first prediction velocity is then interpolated from the Eulerian grid to the Lagrangian grid using the moving-least-squares approach \citep{vanella2009short,de2016moving,spandan2017parallel,spandan2018a}.
The IBM force $\vec{f}$ required on each Lagrangian marker for satisfying no-slip and no-penetration conditions is computed, which is then spread back to the Eulerian grid using the moving-least-squares approach. The force $\vec{f}$ is used to update the velocity, followed by a standard pressure correction step.
{\color{black}We note that the pressure field required to enforce the incompressibility condition has vanishing gradient at the immersed boundary. Thus, the pressure correction step does not change practically the numerical accuracy of the IBM scheme, as observed by \cite{fadlun2000combined} and \cite{kempe2012An}.}
{\color{black}As a validation of the implementation of the IBM, we considered a steady, axisymmetric shear flow inside a circular domain driven by the rotation of the circular boundary at a fixed rotation rate without body force. The numerical simulation showed that the azimuthal velocity increases linearly with the distance away from the rotation axis and the vorticity is uniform, in quantitative agreement with the theoretical results.}

The heat transfer between the fluid and solid phases is realized by solving the temperature equation (\ref{temperature_eqn}) in both phases \citep{ardekani2018heat,ardekani2018numerical,sardina2018buoyancy}.
A phase indicator $\xi$ is introduced to represent the solid volume fraction and to distinguish the fluid and solid phases on each grid point of two components of velocity. 
The value of $\xi$ is determined from a level-set function $\zeta$ given by the signed distance of four corner nodes to the obstacle surface using the formula \citep{ardekani2018heat,kempe2012An}
\begin{equation}
\xi=\frac{\sum_{n=1}^{4}-\zeta_{n} \mathcal{H}\left(-\zeta_{n}\right)}{\sum_{n=1}^{4}\left|\zeta_{n}\right|},
\end{equation}
where $\mathcal{H}$ is the Heaviside step function.
Then the velocity of the combined phase is defined at each point in the computational domain as
\begin{equation}
\vec{u}_{c p}=(1-\xi) \vec{u}_{f}+\xi \vec{u}_{p},
\end{equation}
where $\vec{u}_{f}$ and $\vec{u}_{p}$ denote velocities of the fluid and solid phases, respectively. For fixed obstacles considered in this study we obviously have $\vec{u}_{p}=0$.
{\color{black}We note that the coupling of the IBM and heat transfer between different phases has been considered in other convection problems, such as the turbulent convection confined by walls with finite conductivities} \citep{verzicco2002sidewall,verzicco2004effects,stevens2014sidewall}.

\begin{table}
	\centering
	\caption{\label{tab:resolution} {\color{black}Numerical details of the typical grid resolutions. The columns from left to right denote the Rayleigh number $Ra$, the number of  obstacles $N_o$, the porosity $\phi$, the grid resolution $N_x\times N_z$, the Nusselt number $Nu$, the number of grid nodes $N_{BL}$ in the thermal boundary layers, and the number of grid nodes $N_D$ per obstacle diameter. The thickness of thermal boundary layer is estimated as $\delta_{th}=L/(2Nu)$. We note that the table only presents some typical cases. {\color{black}Totally, 67 different cases were simulated, and the results are collected in figure \ref{Nu_Ra}.}}}
	\renewcommand\arraystretch{1.2}
	\setlength{\tabcolsep}{3mm}
	\begin{tabular}{ccccccc}
		\hline
		$Ra$      &	$N_o$ & $\phi$ & $N_x\times N_z$  & $Nu$   & $N_{BL}$ & $N_{D}$ \\ \hline
		$10^6$    & 144 &  0.82  & 540$\times$540   & 2.77   & 97       & 22      \\
		$10^7$    & 144 &  0.82  & 540$\times$540   & 12.89  & 21       & 22      \\
		$10^8$    & 144 &  0.82  & 540$\times$540   & 28.16  & 10       & 24      \\
		$10^9$    & 144 &  0.82  & 1080$\times$1080 & 53.40  & 10       & 43      \\
		$10^{10}$ & 144 &  0.82  & 3072$\times$3072 & 106.1  & 14       & 123     \\
	\end{tabular}
\end{table}

In this study a uniform Eulerian grid is used and the resolution is chosen to fully resolve the boundary layers and the smallest scales in the bulk \citep{stevens2010radial,shishkina2010boundary,zhang2017statistics}. %With regard to the resolution of obstacles, we note that the flow strength is reduced due to the drag of obstacle array, and the pore Reynolds numbers are not too large, particularly for small $Ra$.
{\color{black}The numerical details of some typical grid resolutions are given in table \ref{tab:resolution}.
For the small-$Ra$ cases, the grid is restricted by the resolution of the obstacles. For these cases a grid of 540$\times$540 is used, and the number of grid nodes per obstacle diameter is 22. For larger $Ra$, the mesh size chosen is adequate to achieve a full resolution of the thermal boundary layer, with the thermal boundary layer resolved by at least 10 grid points.
For the highest $Ra$ ($Ra=10^{10}$), a grid of 3072$\times$3072 is used. This ensures that the turbulent flow in the bulk and boundary layers is fully resolved.
For the traditional RB convection at $Ra=10^{10}$, the typical Kolmogorov length scale $\eta$ is estimated by the global criterion $\eta=LPr^{1/2}/[Ra(Nu-1)]^{1/4}$, and the Batchelor scale $\eta_B$ is estimated by $\eta_B=\eta Pr^{-1/2}$ \citep{shishkina2010boundary}. We find that the grid spacing $\Delta_g$ satisfies $\Delta_g\lesssim 0.16 \eta$ and $\Delta_g\lesssim 0.33\eta_B$.
The thermal boundary layers are resolved with 14 grid points, which agrees with the recommendations of \cite{shishkina2010boundary}.}
{\color{black}Long-time averages are conducted for the calculation of $Nu$}.
The difference of $Nu$ obtained by time averages over the first and second halves of the simulations (both taken after initial transients) was smaller than 1\%.
Besides, the relative difference between the top- and bottom-wall $Nu$ was smaller than 1\%.

Simulations were performed for varying Rayleigh numbers (from $10^{5}$ to $10^{10}$) and porosities (from $0.75$ to $1$). The Prandtl number and obstacle diameter were fixed at $Pr=4.3$ and $D=0.04$, respectively.

\section{Heat transfer}\label{sec:sec_heat_transfer}
In this section the influence of the obstacle array on the heat transfer properties is examined. The presence of the obstacle array has a significant influence on the heat transfer properties. The normalized Nusselt number $Nu(\phi)$ for two fixed $Ra$ was already shown in figure \ref{optimal_spacing}. In figure \ref{Nu_Ra}, $Nu(Ra)$ for five fixed porosities $\phi$ is shown.
In the traditional RB convection with $\phi=1$, $Nu$ increases with $Ra$ following an effective power law  $Nu\sim Ra^{0.30}$, consistent with the results in the literature for 2D RB convection \citep{Poel2013Comparison,zhu2017roughness,zhang2017statistics}. 
For fixed $\phi<1$, it is found that the variation of $Nu$ with $Ra$ exhibits two scaling regimes for the parameter range studied. {\color{black}Compared with the results of traditional RB convection with $\phi=1$, the heat transfer is reduced for small $Ra$} even if $Nu$ increases with $Ra$ with a steep effective power law $Nu\sim Ra^{0.65}$; on the other hand, for large enough $Ra$, {\color{black}$Nu$ becomes larger than the corresponding value without obstacles and increases with $Ra$ with an effective power law  $Nu\sim Ra^{0.30}$,} similar to that of traditional RB convection. The increase of $Nu$ in the large-$Ra$ regime is more visible in the compensated plot of figure \ref{Nu_Ra}($b$) and in figure \ref{optimal_spacing}.
{\color{black}Note that the regime of heat transfer enhancement with decreasing $\phi$ displayed in figure \ref{optimal_spacing} corresponds to the large-$Ra$ regime shown in figure \ref{Nu_Ra}.}
{\color{black}Figure \ref{Nu_Ra}$(a,b)$ shows that} the trends of variation of $Nu$ with $Ra$ for different values of $\phi$ are similar, and the critical $Ra$ for scaling crossover increases as $\phi$ is decreased.
{\color{black}Considering that the values of $\phi$ and $Nu$ determine the characteristic length scales of the regular porous medium and flow structures, respectively, the scaling crossover indicates a competition of the length scale of porosity and that of the flow structures.} %$\phi$ is determined by the obstacle separation $l$, and the thicknesses of thermal boundary layers and plumes are decreasing functions of $Ra$.
When we rescale the thermal boundary layer thickness $\delta_{th}=L/(2Nu)$ with the obstacle separation $l$, it is found that the results for different $\phi$ approximately collapse, and the scaling crossover occurs when $\delta_{th}/l\approx 1$, as shown in figure \ref{Nu_Ra}$(c)$.
The data collapse indicates that the heat transfer properties are determined by the length scales of the flow structures and the pore scale of the regular porous medium.

{\color{black}
It is interesting to examine the relevance of our results to pure Darcy-type convection. For this, it is instructive to show how $Nu$ varies with the so-called Darcy Rayleigh number $Ra^*=RaDa$, {\color{black}where $Da=K/L^2$ is the Darcy number. Here $K$ is the permeability, which depends on the geometry of the porous media and represents the resistance for the fluid to flow through the porous media.}
We {\color{black}estimate} $Da$ with the porosity $\phi$ and the obstacle diameter $D$ {\color{black}based on the Kozeny's equation} \citep{nithiarasu1997natural,nield2006convection}:
\begin{equation}
Da=\frac{\phi^{3} D^{2}}{150(1-\phi)^{2}L^2}.
\end{equation}
The results are shown in figure \ref{Nu_Ra}$(d)$.
We find that in the small-$Ra^*$ regime, $Nu(Ra^*)$ shows a trend to collapse as $\phi$ is decreased, which is a signature of the transition to Darcy-type flow and consistent with the observation based on the generalized non-Darcy model \citep{nithiarasu1997natural}.}
{\color{black}We note that, in porous media there may exist multiple flow states with different heat transfer efficiencies for the same parameters. The solution may not be able to sample all the flow configurations due to the suppression of fluctuations in the presence of an obstacle array. The numerically realized steady state may be different for different initial conditions. Despite this, we expect that the differences of the statistics of multiple flow states are relatively small, particularly for the large-$Ra$ cases, and the global trend of variation of $Nu$ with $Ra$ and $\phi$ will not be qualitatively affected by the existence of multiple flow states.
}

\begin{figure}
	\centering
	\includegraphics[width=0.45\linewidth]{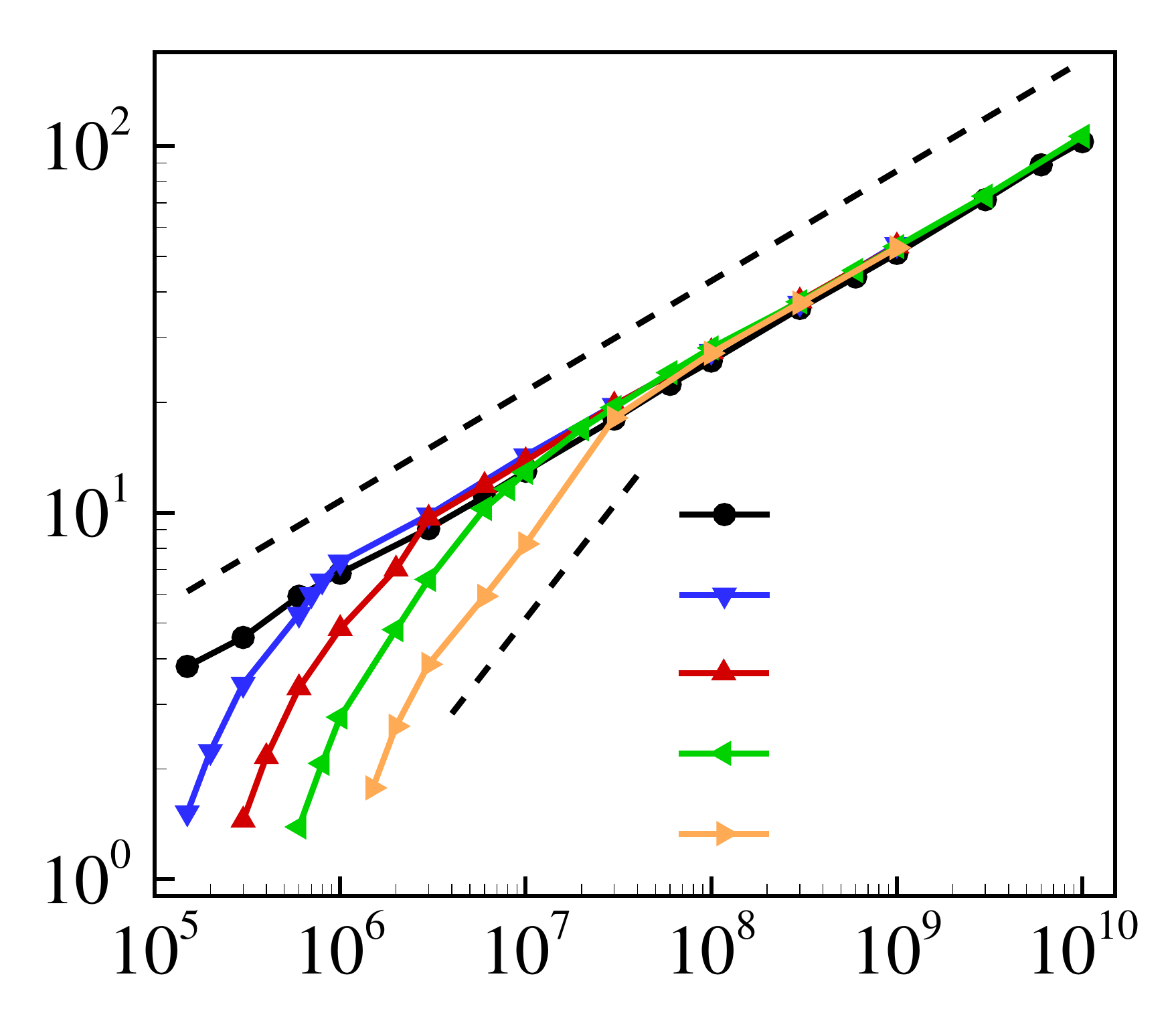}
	\put(-178,133){$(a)$}
	\put(-178,74){\rotatebox{90}{$Nu$}}
	\put(-89,-5){$Ra$}
	\put(-55,81){ $\phi$,~~$Da$}
	\put(-57,70){~~1}
	\put(-57,58){0.92, 1.3e-3}
	\put(-57,46.5){0.87, 4.5e-4}
	\put(-57,35){0.82, 1.8e-4}
	\put(-57,23){0.75, 7.5e-5}
	\put(-102,39){$Ra^{0.65}$}
	\put(-110,98){$Ra^{0.30}$}
	\hspace{3 mm}
	\includegraphics[width=0.45\linewidth]{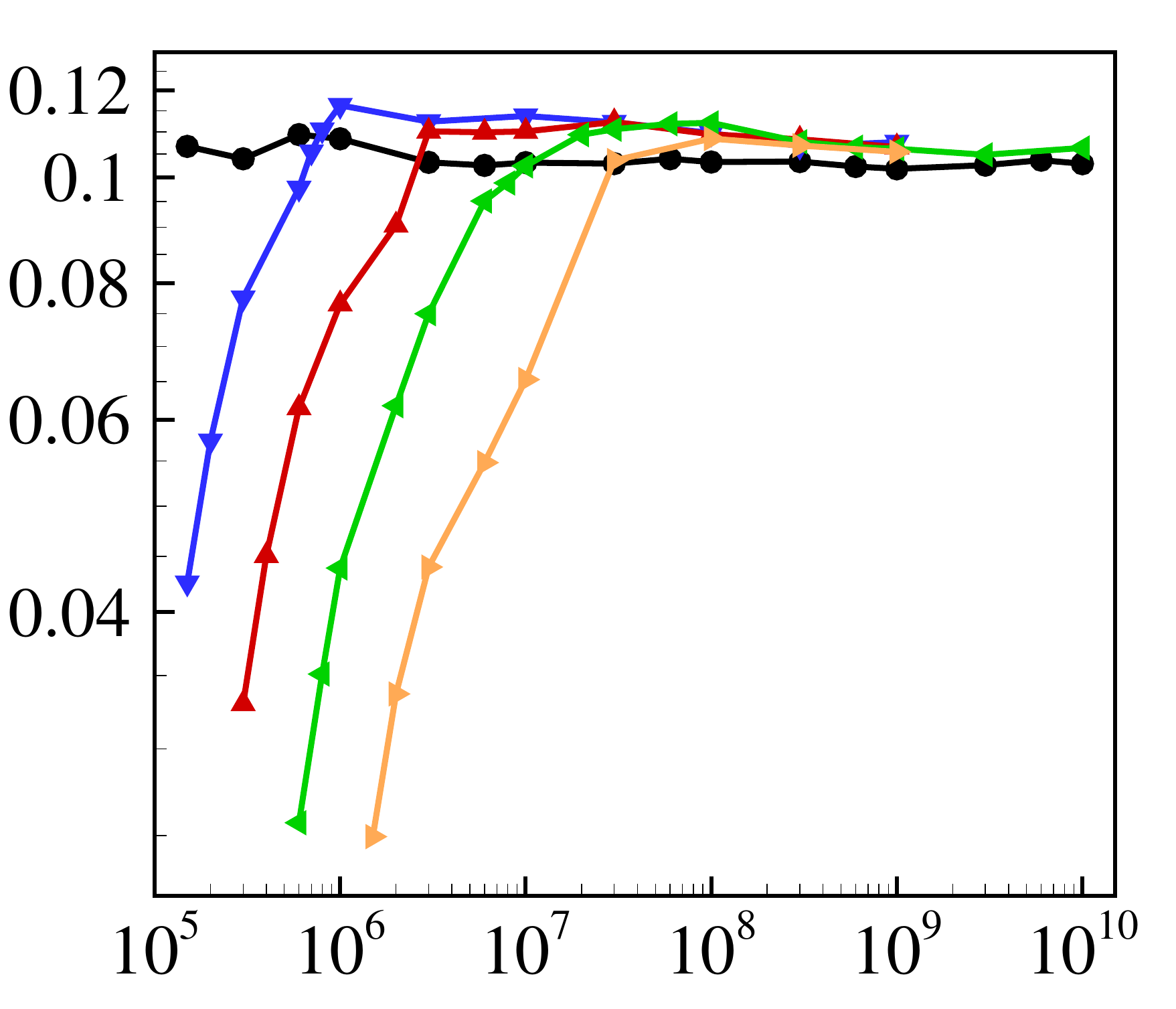}
	\put(-185,133){$(b)$}
	\put(-183,53){\rotatebox{90}{$NuRa^{-0.30}$}}
	\put(-89,-5){$Ra$}
	\\
	\vspace{1 mm}
	\includegraphics[width=0.45\linewidth]{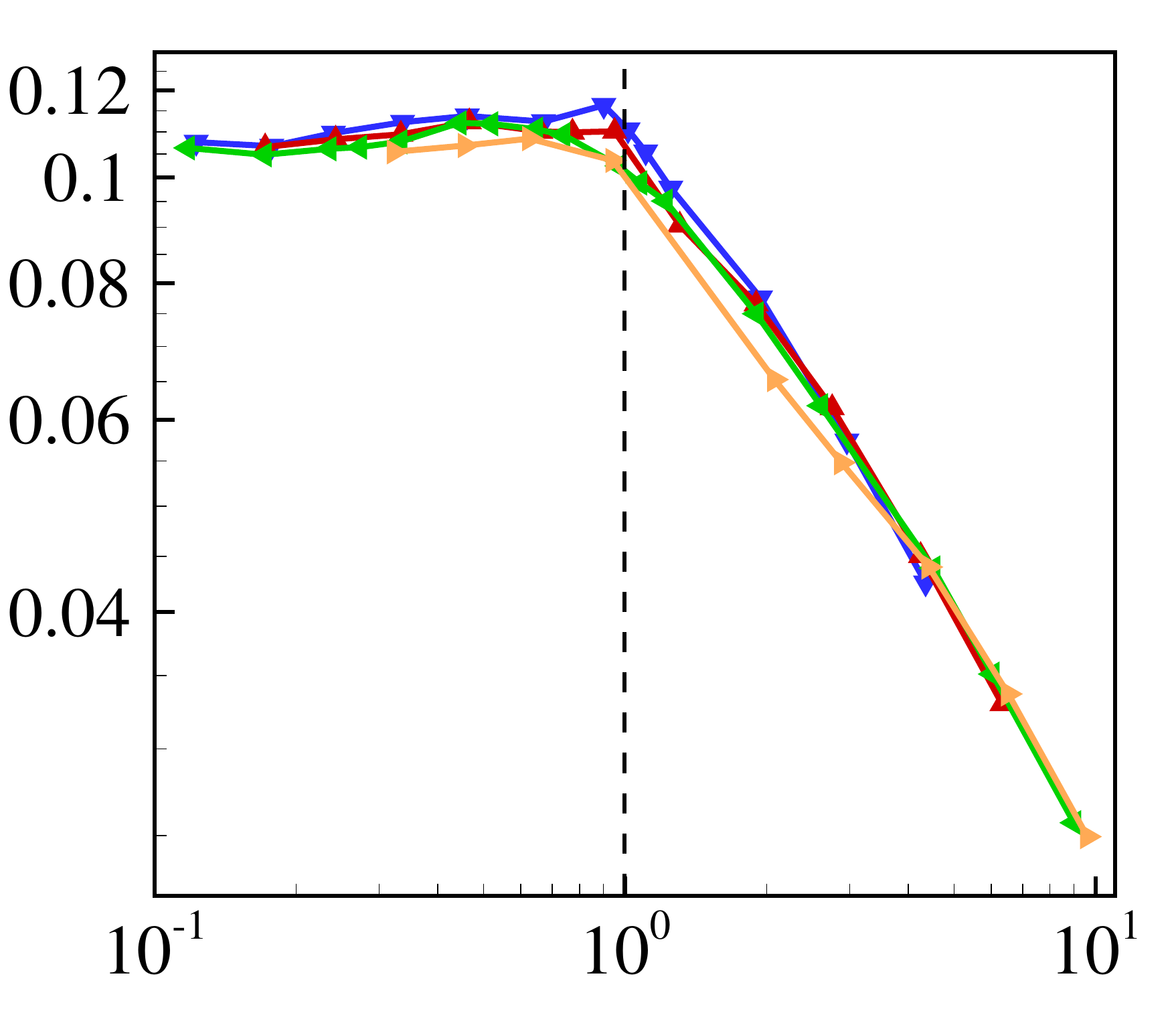}
	\put(-185,133){$(c)$}
	\put(-183,53){\rotatebox{90}{$NuRa^{-0.30}$}}
	\put(-89,-5){$\delta_{th}/l$}
	\hspace{3 mm}
	\includegraphics[width=0.45\linewidth]{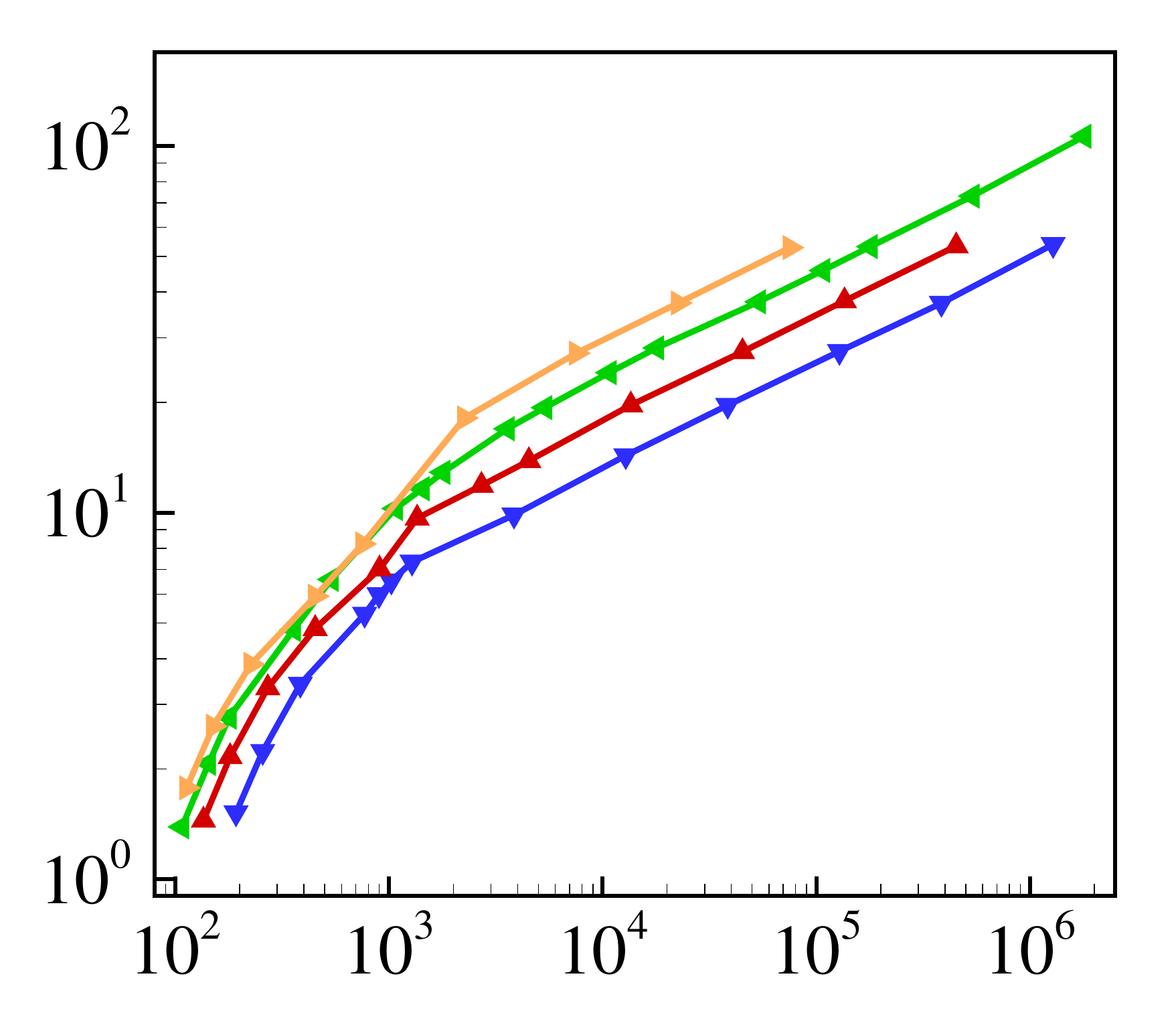}
	\put(-185,133){$(d)$}
	\put(-178,74){\rotatebox{90}{$Nu$}}
	\put(-89,-5){$Ra^*$}	
	\caption{\label{Nu_Ra}$(a)$ Variations of $Nu$ with $Ra$ for different $\phi$. $(b)$ The compensated plot of $(a)$. $(c)$ Results for different $\phi$ collapse by rescaling the thermal boundary layer thickness $\delta_{th} = L/(2 N u)$ with the pore scale $l$. {\color{black} $(d)$ Variation of $Nu$ with the Darcy Rayleigh number $Ra^*=Ra Da$.}
	}
\end{figure}

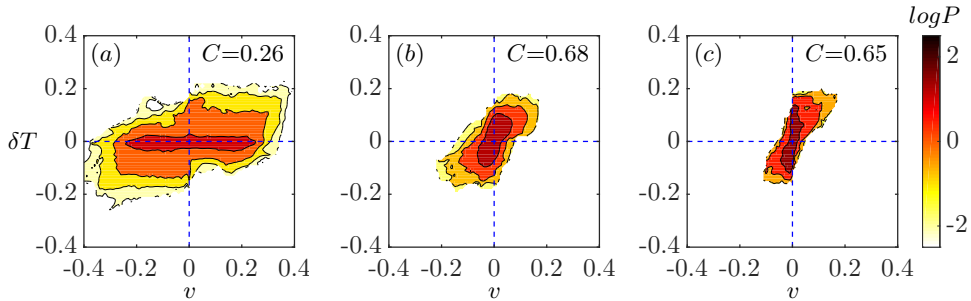
\begin{figure}
	\centering
	\vspace{2 mm}
	\hspace{-2 mm}
	\includegraphics[width=0.35\linewidth]{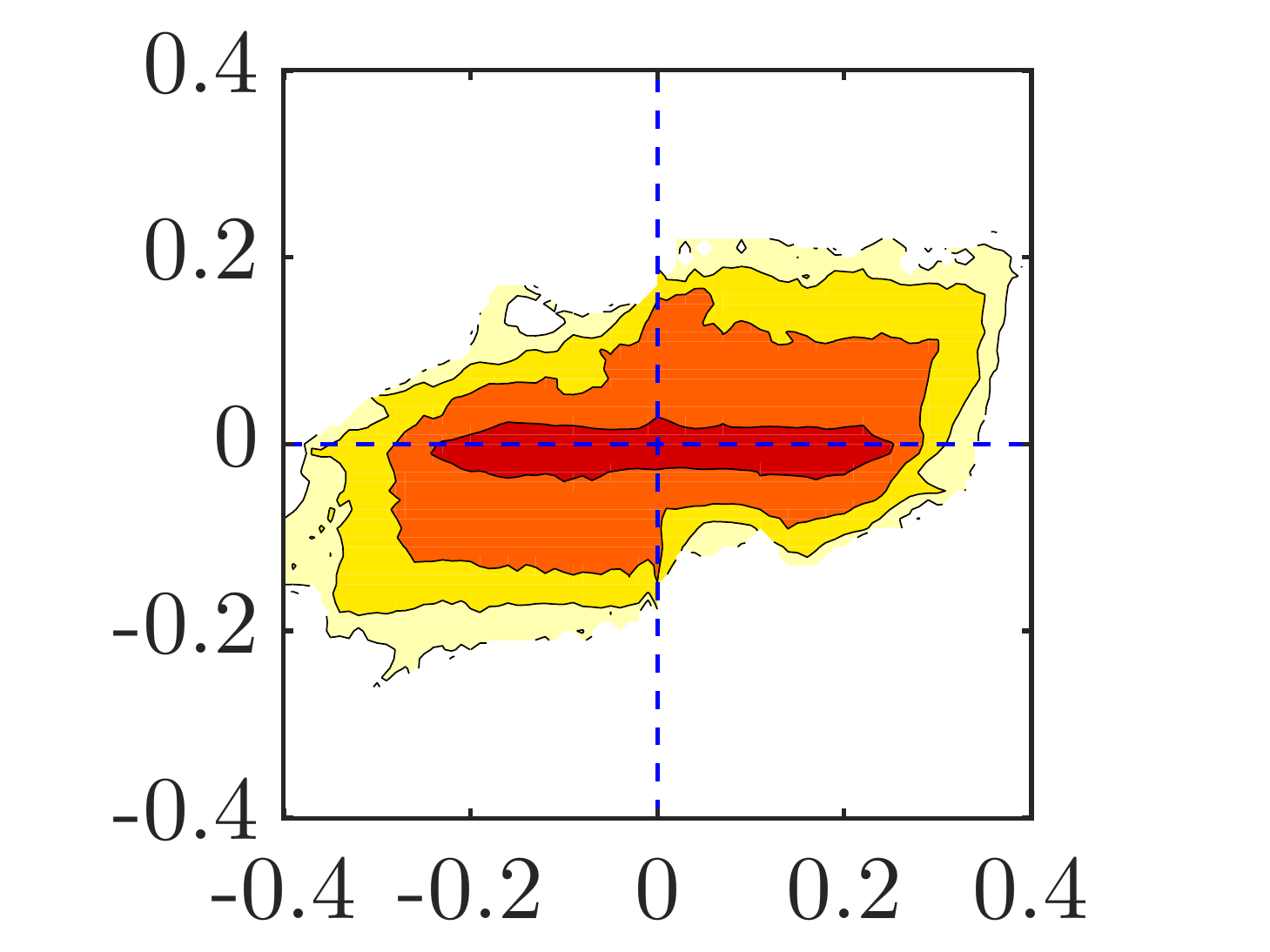}
	\put(-102,84){$(a)$}
	\put(-133,51){$\delta T$}
	\put(-60,84){{\small $C$=$0.26$}}
	\put(-67,-5){$v$}
	\hspace{-8 mm}
	\includegraphics[width=0.35\linewidth]{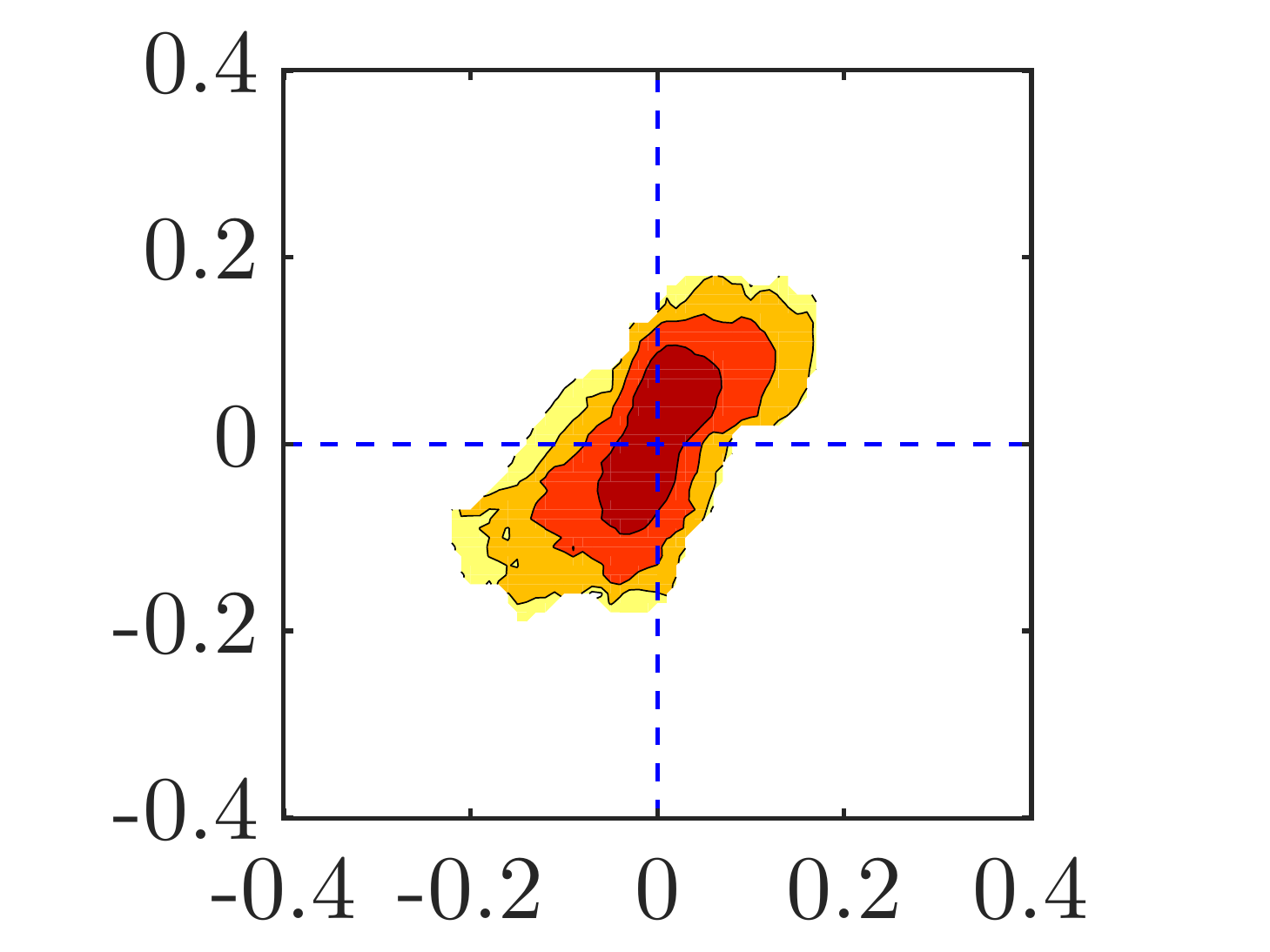}
	\put(-102,84){$(b)$}
	\put(-67,-5){$v$}
	\put(-60,84){{\small $C$=$0.68$}}
	\hspace{-5 mm}
	\includegraphics[width=0.35\linewidth]{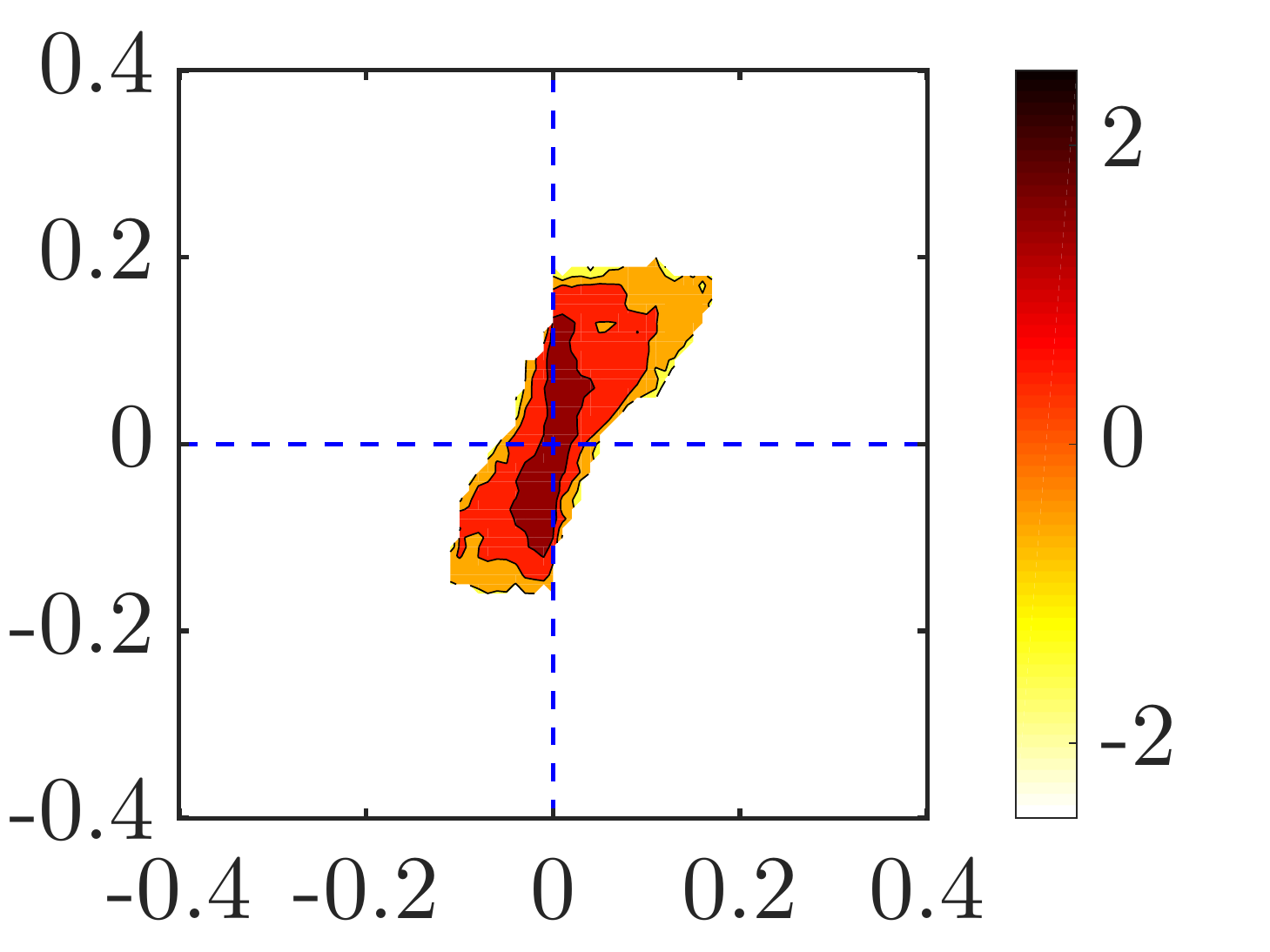}
	\put(-113,84){$(c)$}
	\put(-78,-5){$v$}
	\put(-71,84){{\small$C$=$0.65$}}
	\put(-32,98){{\footnotesize $logP$}}
	\vspace{2 mm}
	\caption{\label{horizsect_flux_jpdf} The joint PDFs of the vertical velocity $v$ and the temperature fluctuation $\delta T=T-T_m$ on the horizontal mid-plane $z=0.5$ for different $\phi$ at $Ra=10^8$: $(a)$ $\phi=1$, $(b)$ $\phi=0.92$, $(c)$ $\phi=0.82$. Corresponding values of the cross-correlation $C$ between $v$ and $\delta T$ are labeled at the top right corner of each panel.}
\end{figure}

We now focus on the statistics of the velocity and temperature fields on the horizontal mid-plane $z=0.5$, {\color{black}to quantify the modifications of heat transfer properties as $\phi$ is decreased, particularly in the regime where $Nu(\phi)$ varies non-monotonously.}
Figure \ref{horizsect_flux_jpdf} plots the joint probability density function (PDF) $P(v,\delta T)$ of vertical velocity $v$ and temperature fluctuation $\delta T$ for different $\phi$ at $Ra=10^8$, where $\delta T=T-T_m$ and $T_m=0.5$ is the arithmetic mean temperature.
{\color{black}In the absence of the obstacle array (i.e., for $\phi=1$), 
the velocity can attain considerably larger values than that in porous media at the same $Ra$.}
The temperature in the bulk is well mixed with $\delta T\approx 0$. Thus, $P(v,\delta T)$ is concentrated {\color{black}in a slender region} {\color{black}located} on the horizontal axis, as shown in figure \ref{horizsect_flux_jpdf}($a$).
From $P(v,\delta T)$ the correlation between $v$ and $\delta T$ can be identified. Due to buoyancy, flow elements with positive (negative) $\delta T$ are more likely to move upward (downward).
Note that there is a non-negligible probability for the occurrence of counter-gradient convective heat transfer, which is manifested in the finite values of $P(v,\delta T)$ in the second and fourth quadrants, consistent with the observation of counter-gradient convective heat transfer in \cite{sugiyama2010flow} and \cite{huang2013counter}.
The properties of temperature fluctuations and convective heat transfer can also be quantified by the normalized PDFs of $\delta T$ and $v\cdot\delta T$ on the horizontal mid-plane $z=0.5$, which are given in figure \ref{horisect_pdf}.
The strong mixing of temperature in the bulk is manifested by the peak value of $P(\delta T)$ at $\delta T=0$ in figure \ref{horisect_pdf}($a$), and the correlation between $v$ and $\delta T$ is reflected by {\color{black}the asymmetry of the left- and right-hand branches of $P(v\cdot \delta T)$} in figure \ref{horisect_pdf}($b$). The counter-gradient convective heat transfer is quantified by the left-hand branch of $P(v\cdot \delta T)$.

As $\phi$ is decreased, it is found that the fluctuation properties are significantly influenced, as shown in figures \ref{horizsect_flux_jpdf}($b,c$).
The velocity distribution is narrowed down, showing that the convection strength is decreased for the flow through regular porous media.
Compared with the results of traditional RB convection with $\phi=1$, the temperature mixing is less efficient and large temperature fluctuations are more probable to appear.
The decreased mixing efficiency is also manifested by the flattening of the peak of $P(\delta T)$ around $\delta T=0$ in figure \ref{horisect_pdf}($a$).
From $P(v,\delta T)$ it is found that the counter-gradient convective heat flux is suppressed, which is also manifested by the rapid decrease of the left-hand branch of $P(v\cdot \delta T)$ as $v\cdot \delta T$ is decreased from 0 in figure \ref{horisect_pdf}($b$).
Note that as $\phi$ is decreased, the distribution of $P(v,\delta T)$ is more compact and concentrated along a line with positive slope, which suggests that the correlation between $v$ and $\delta T$ is enhanced in regular porous media.
We here quantify the correlation between $v$ and $\delta T$ using the cross-correlation $C\equiv\langle g_v g_T \rangle$, where $g_\chi\equiv[\chi-\langle \chi \rangle]/\sigma_\chi$, $\sigma_\chi\equiv\sqrt{\langle \chi^2\rangle-\langle \chi \rangle^2}$, and $\langle\cdot\rangle$ denotes the average over the horizontal mid-plane and time.
Values of $C=1$ and $-1$ correspond to perfect correlations and anti-correlations, respectively, and $C=0$ corresponds to no correlation.
The values of $C$ for the three cases of figure \ref{horizsect_flux_jpdf} are shown in the top right-hand corner of each panel, which definitely demonstrate the enhancement of the correlation between $v$ and $\delta T$ with decreasing $\phi$.
For both the traditional RB convection and convection in regular porous media, the cross-correlation between temperature fluctuation $\delta T$ and horizontal velocity $u$ on the horizontal mid-plane is small with $|C|<0.1$.

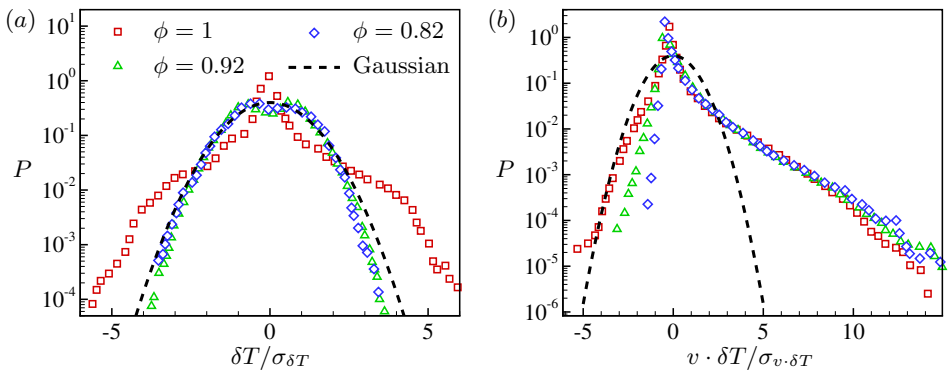
\begin{figure}
	\centering
	\includegraphics[width=0.45\linewidth]{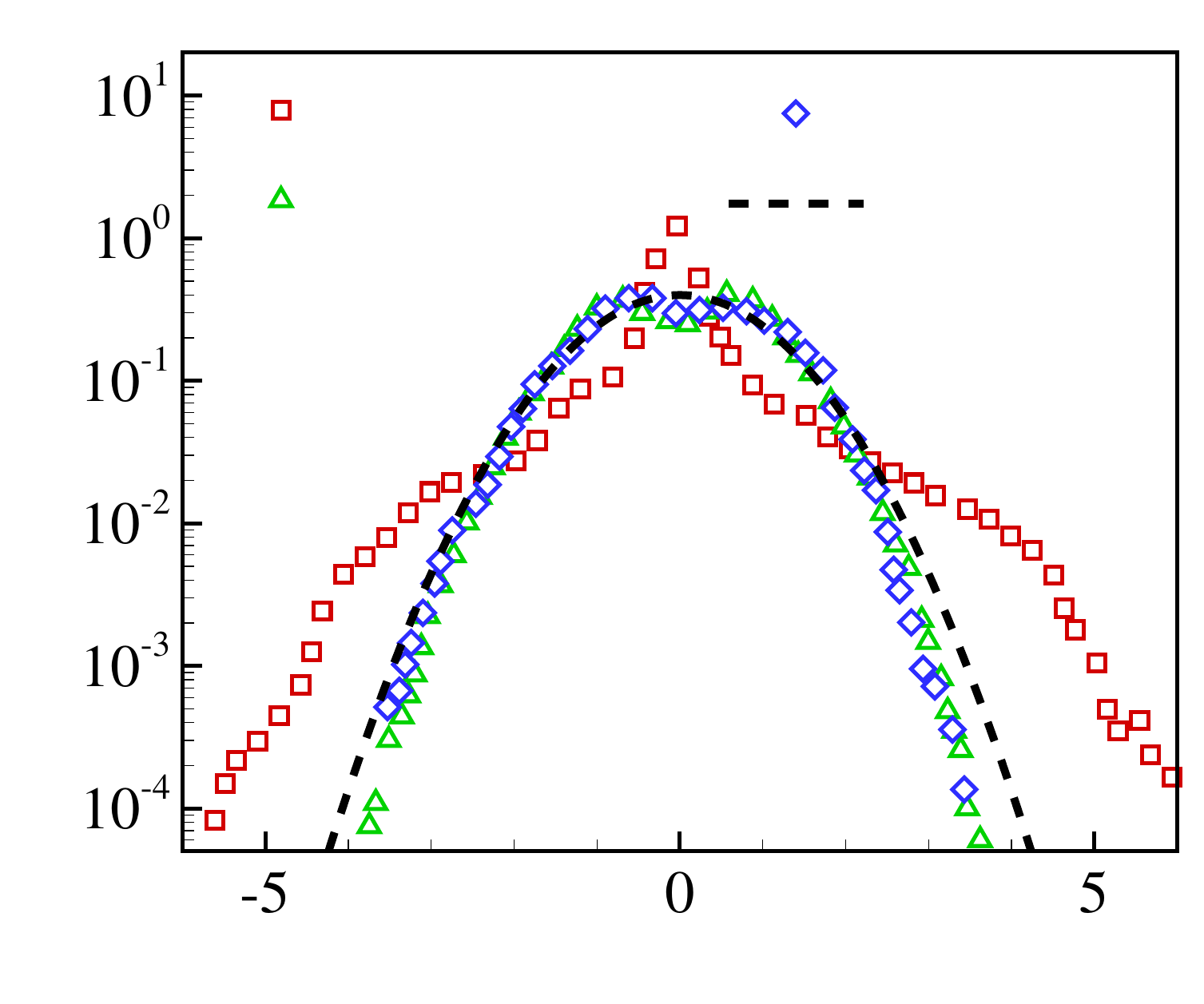}
	\put(-175,130){$(a)$}
	\put(-172,74){$P$}
	\put(-90,4){{\color{black}$\delta T/\sigma_{\delta T}$}}
	\put(-119,126){$\phi=1$}
	\put(-119,113){$\phi=0.92$}
	\put(-44,126){$\phi=0.82$}
	\put(-44,113){Gaussian}
	\hspace{2 mm}
	\includegraphics[width=0.45\linewidth]{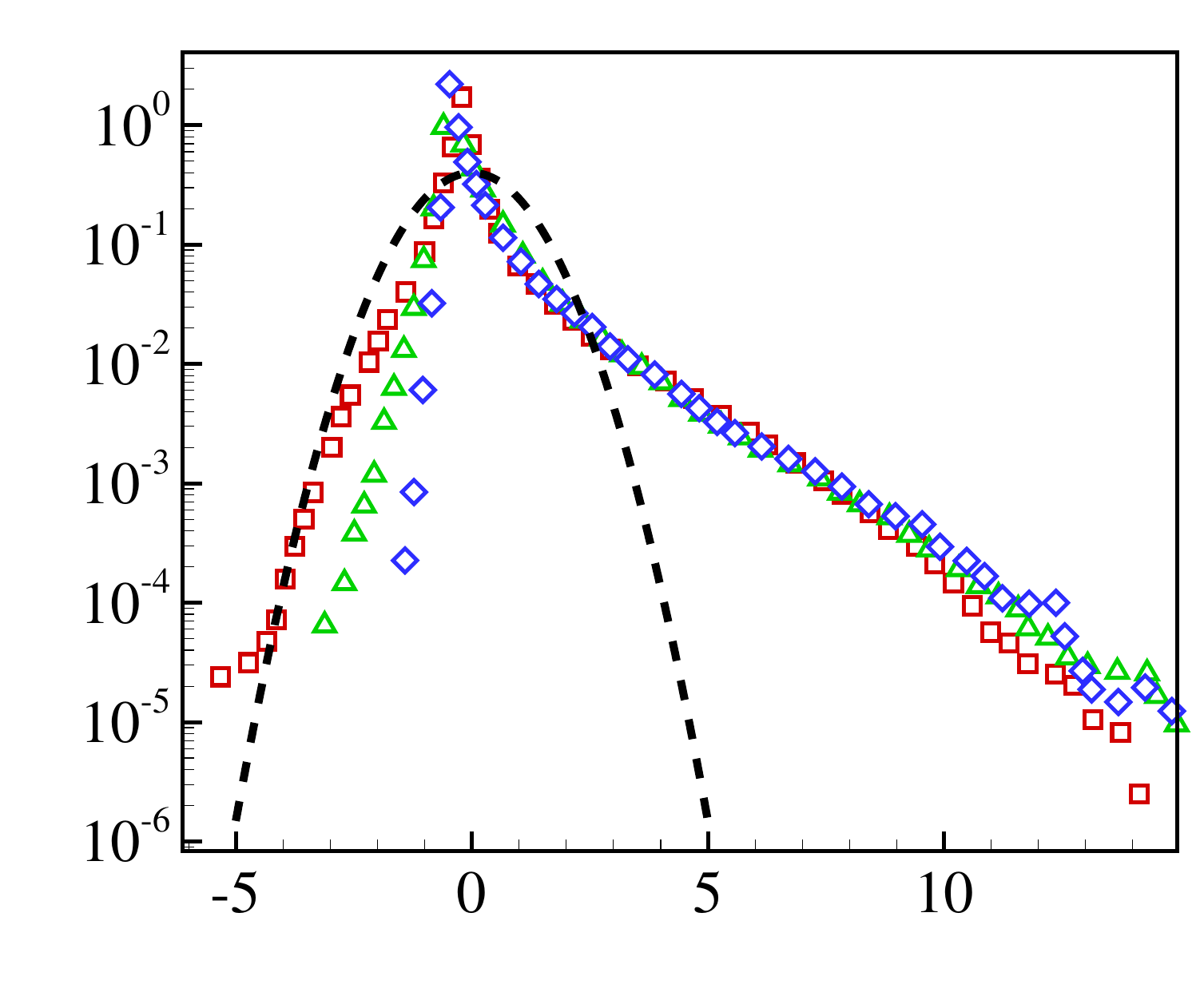}
	\put(-175,130){$(b)$}
	\put(-172,74){$P$}
	\put(-100,4){{\color{black}$v\cdot\delta T/\sigma_{v\cdot \delta T}$}}
	\caption{\label{horisect_pdf} The normalized PDFs of ($a$) {\color{black}$\delta T/\sigma_{\delta T}$ and $(b)$ $v\cdot\delta T/\sigma_{v\cdot\delta T}$} on the horizontal mid-plane $z=0.5$ for different $\phi$ at $Ra=10^8${\color{black}, where $\sigma_{\delta T}$ and $\sigma_{v\cdot\delta T}$ are the standard deviations of $\delta T$ and $v\cdot \delta T$, respectively}. The dashed lines indicate the Gaussian distribution.}
\end{figure}

The {\color{black}variation} of $P(v,\delta T)$ {\color{black}with $\phi$ shown in figure \ref{horizsect_flux_jpdf}} leads us to identify two competing effects of the obstacle array on the heat transfer. On the one hand, the flow becomes more coherent with enhanced correlation between temperature fluctuation and vertical velocity. {\color{black}This} is beneficial for the heat transfer. On the other hand, the obstacle array slows down the convection due to enhanced resistance, which is unfavorable for the heat transfer.
{\color{black}The competing effects of the obstacle array on the heat transfer result in the non-monotonous variation of $Nu$ with $\phi$ as was already shown in figure \ref{optimal_spacing}.}
{\color{black}As $\phi$ is slightly decreased from 1, the enhancement of the flow coherence increases the heat transfer efficiency.
While when $\phi$ is small enough, the convection is strongly suppressed, leading to the heat transfer reduction.}
{\color{black}The heat transfer enhancement by increasing flow coherence is a widespread phenomenon and has been observed in various flow configurations, such as confined RB convection \citep{huang2013confinement,chong2018effect} and partitioned RB convection \citep{bao2015enhanced}.}
{\color{black}\cite{chong2017confined} revealed a universal, non-monotonous behaviour of turbulent transport in systems with competing stabilizing and destabilizing forces.}
{\color{black}The non-monotonous variation of $Nu$ with the increase of the stabilizing effect of porous structure is consistent with the observation of \cite{chong2017confined}.}

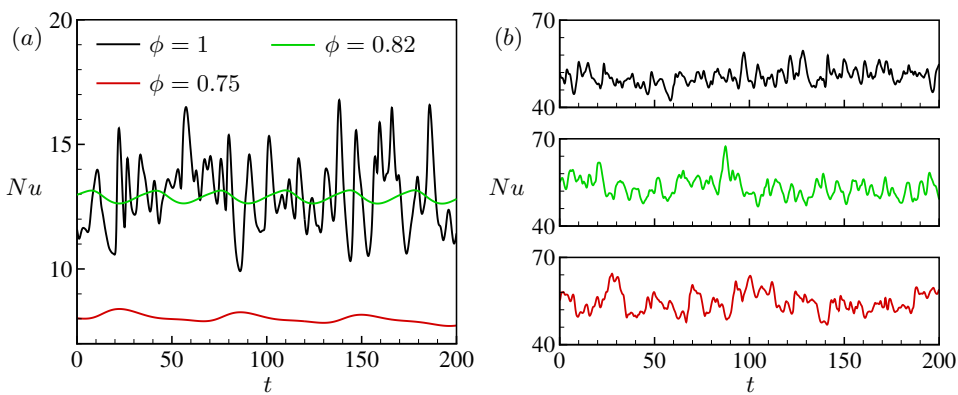
\begin{figure}
	\centering
	\hspace{2 mm}
	\includegraphics[width=0.45\linewidth]{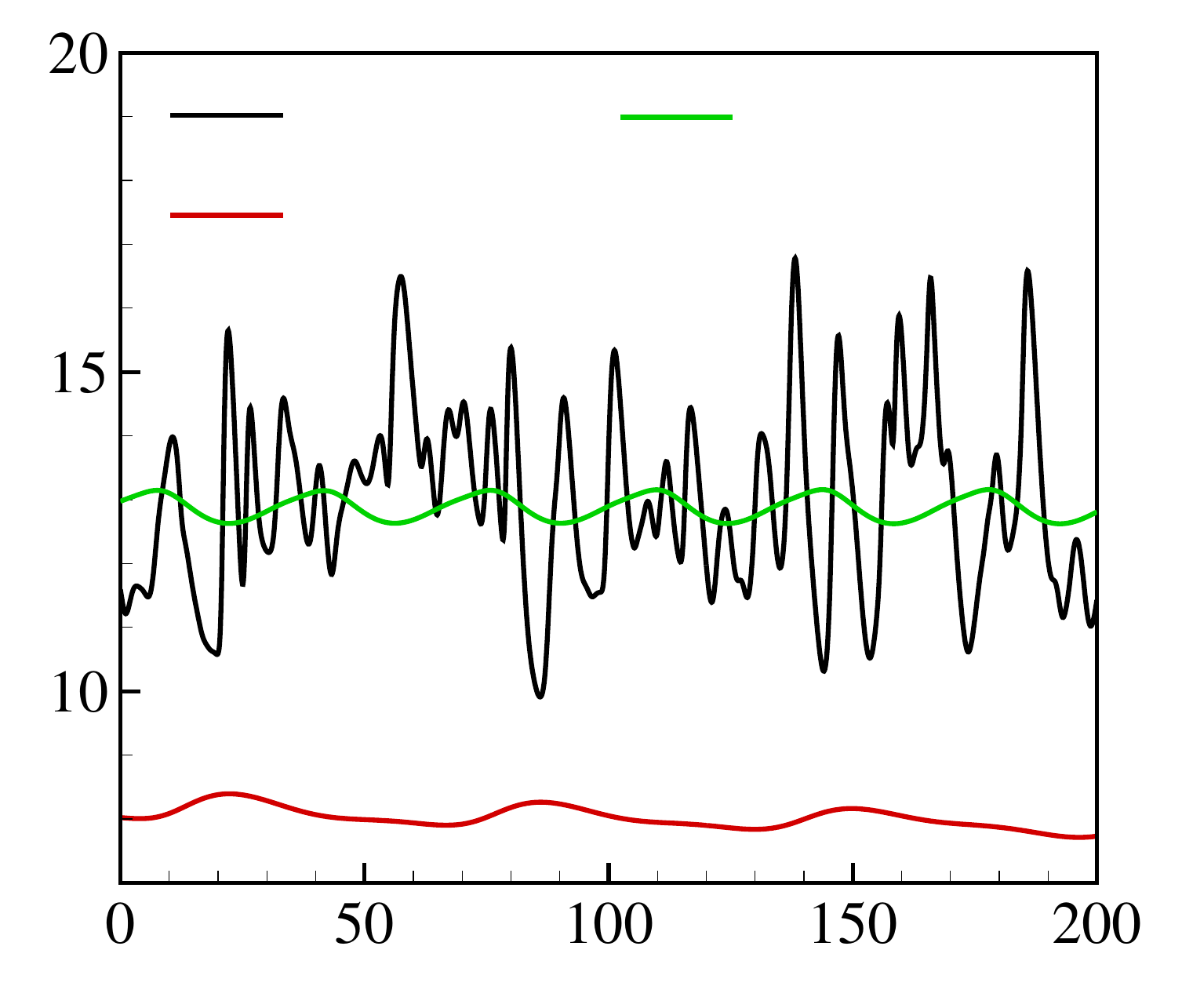}
	\put(-180,130){$(a)$}
	\put(-182,74){$Nu$}
	\put(-85,0){$t$}
	\put(-128,127.5){$\phi=1$}
	\put(-128,113){$\phi=0.75$}
	\put(-62,127.5){$\phi=0.82$}
	\hspace{2 mm}
	\includegraphics[width=0.45\linewidth]{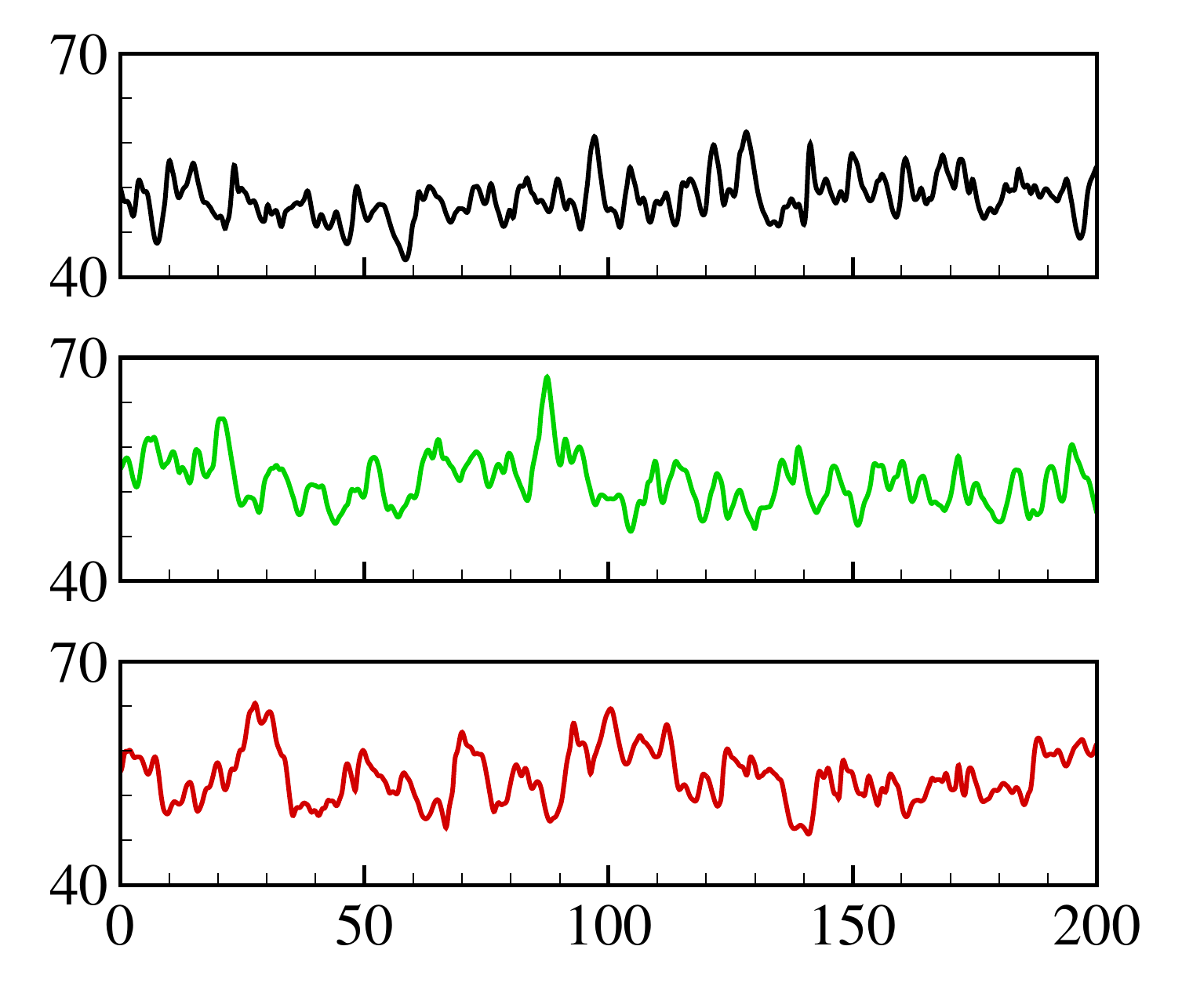}
	\put(-180,130){$(b)$}
	\put(-182,74){$Nu$}
	\put(-85,0){$t$}
	\caption{\label{nu_series}Nusselt number $Nu(t)$ in a time interval of 200 dimensionless units for different $\phi$ at $(a)$ $Ra=10^7$ and $(b)$ $Ra=10^9$.}
\end{figure}

Figure \ref{nu_series} shows sampled time records of $Nu$ for different $\phi$ at $Ra=10^7$ and $10^9$, showing the influence of the obstacle array on the temporal characteristics of fluctuations.
It is found that for relatively small $Ra$, the high-frequency fluctuations are significantly suppressed as $\phi$ is decreased, while for large enough $Ra$, the suppression of high-frequency fluctuations is less visible for the values of $\phi$ studied.
The influence of the obstacle array on the fluctuations is dependent on the relative magnitudes of the spatial coherence length and the pore scale.
At small $Ra$, the length scales of the flow structures are large compared to the pore scale, and the resistance of the obstacle array to the convection is strong. Thus flow fluctuations are significantly suppressed, and there is no developed turbulence in the pores.
As $Ra$ is increased, the convection is more energetic and the length scales of the flow structures are decreased. Thus the obstacle array has smaller effects on the fluctuations, and the flow in the pores becomes chaotic or even turbulent at large enough $Ra$.

\section{Flow structure}\label{sec:sec_flow_structure}
In this section we focus on the flow structures of RB convection in regular porous media. 
Figure \ref{flow_structures} displays typical snapshots of the instantaneous temperature $T$, the velocity magnitude $|\vec{u}|$, and the convective heat flux $v\cdot \delta T$ in the vertical direction. In the traditional RB convection with $\phi=1$, the flow consists of a well-organized large-scale circulation (LSC) with two well-established counter-rotating corner rolls \citep{sugiyama2010flow}, as shown in figures \ref{flow_structures}$(a-c)$. 
Due to the formation of the LSC, the velocity in the bulk is non-uniform.
Following the LSC, plumes detaching from the thermal boundary layers mainly go up and down near the sidewalls. The temperature at the cell center is well mixed with $T\approx T_m$. From the distribution of $v\cdot\delta T$ {\color{black}it is confirmed} that counter-gradient convective heat flux appears locally around the LSC and corner rolls, which is due to the bulk dynamics and the competition between the corner-flow rolls and the LSC \citep{sugiyama2010flow,huang2013counter}.
The inclusion of an array of obstacles has a significant influence on the flow structures, as displayed in figures \ref{flow_structures}($d-i$).
Convection strength is reduced and the LSC is suppressed due to the impedance of the obstacle array. Temperature mixing at the cell center is less efficient. {\color{black}Thermal plumes detaching from the thermal boundary layers can penetrate deep in the bulk, forming convection channels, namely, the regions with strong flows that are formed between the obstacles.}
When $Ra$ is relatively large, the characteristic length scales of the flow structures are smaller than the pore scale, and the flow is chaotic and dominated by fragmented plumes, as shown in figure \ref{flow_structures}($d$); while when $Ra$ is relatively small the flow is dominated by large-scale plumes, as shown in figure \ref{flow_structures}($g$). %The local velocities at the pore throats are larger due to mass conservation. 
Comparing figures \ref{flow_structures}($f,i$) and \ref{flow_structures}($c$), it is found that the counter-gradient convective heat flux is reduced in regular porous media, which is attributed to the suppression of the LSC and coherence of plume dynamics due to the impedance of the obstacle array.

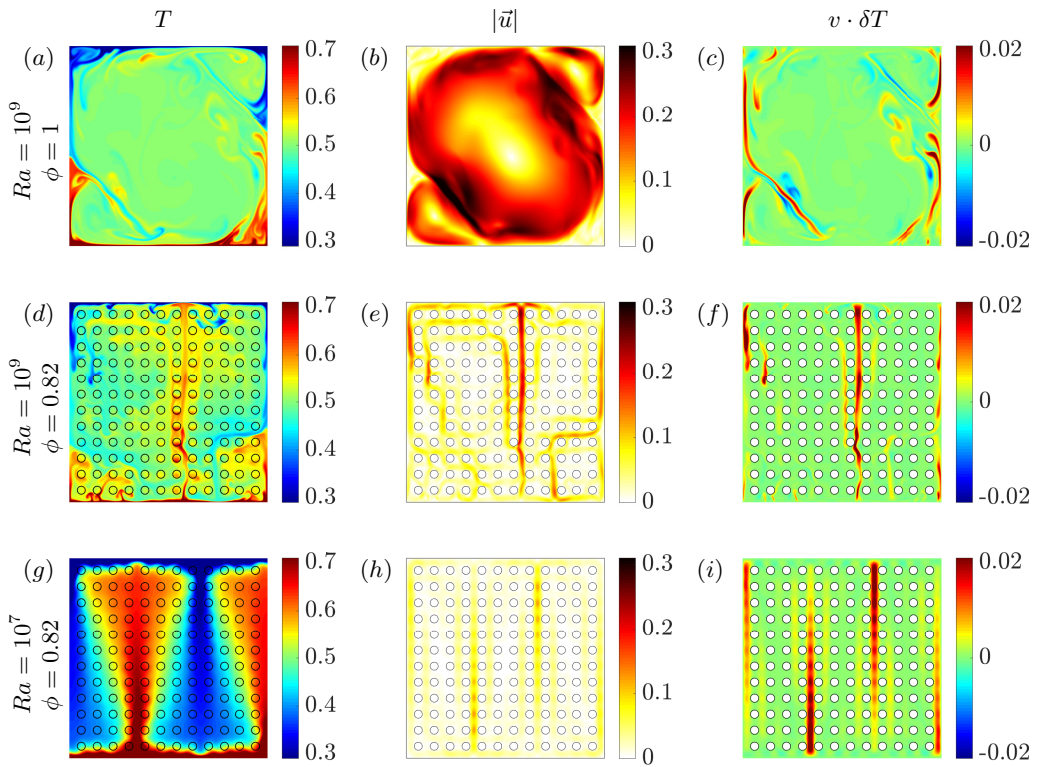
\begin{figure}
	\centering
	\vspace{3 mm}
	\includegraphics[width=0.33\linewidth]{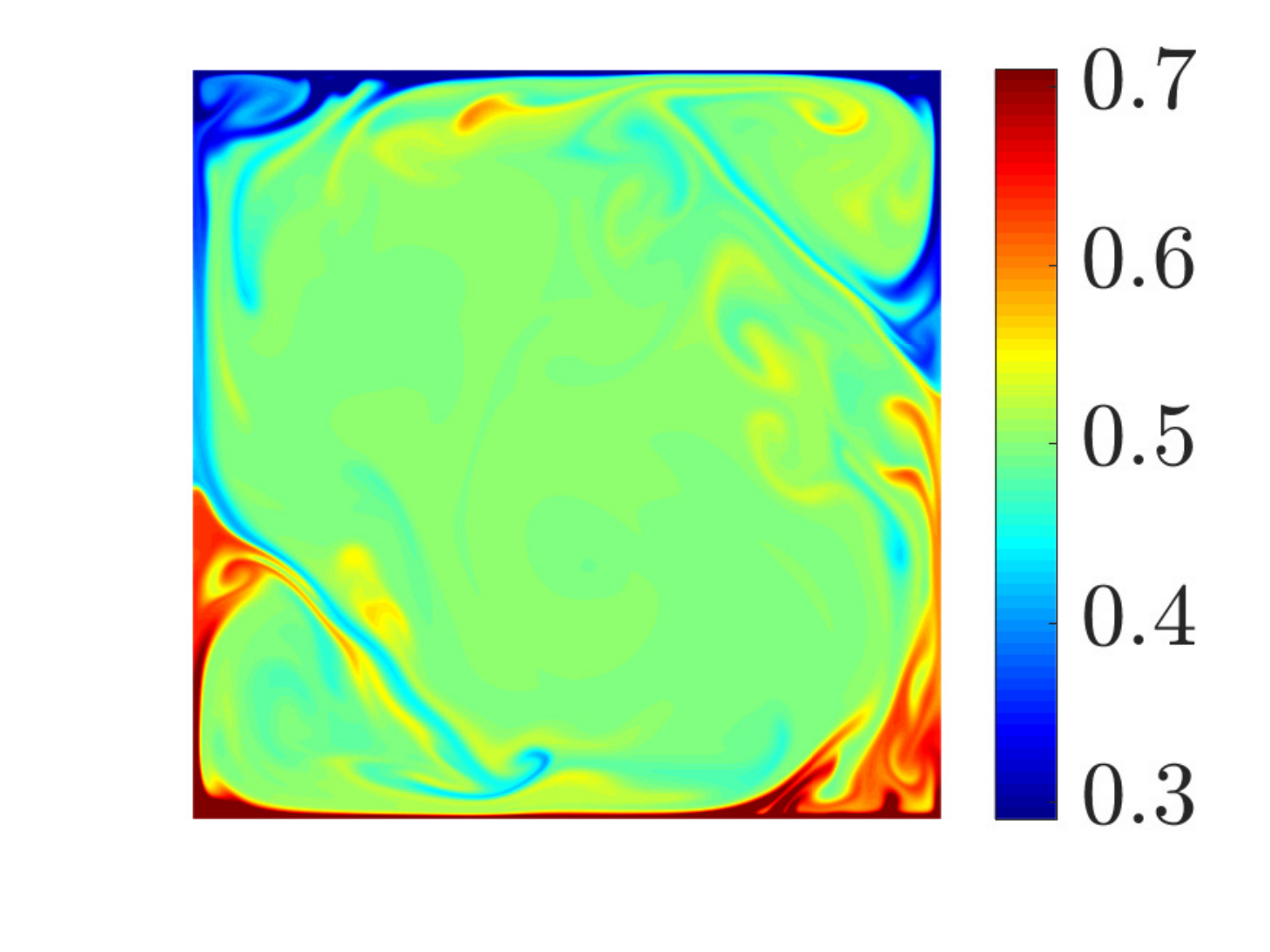}
	\put(-125,81){$(a)$}
	\put(-75.5,95){$T$}
	\put(-130,30){\rotatebox{90}{{\small $Ra=10^9$}}}
	\put(-118,37){\rotatebox{90}{{\small $\phi=1$}}}
	\includegraphics[width=0.33\linewidth]{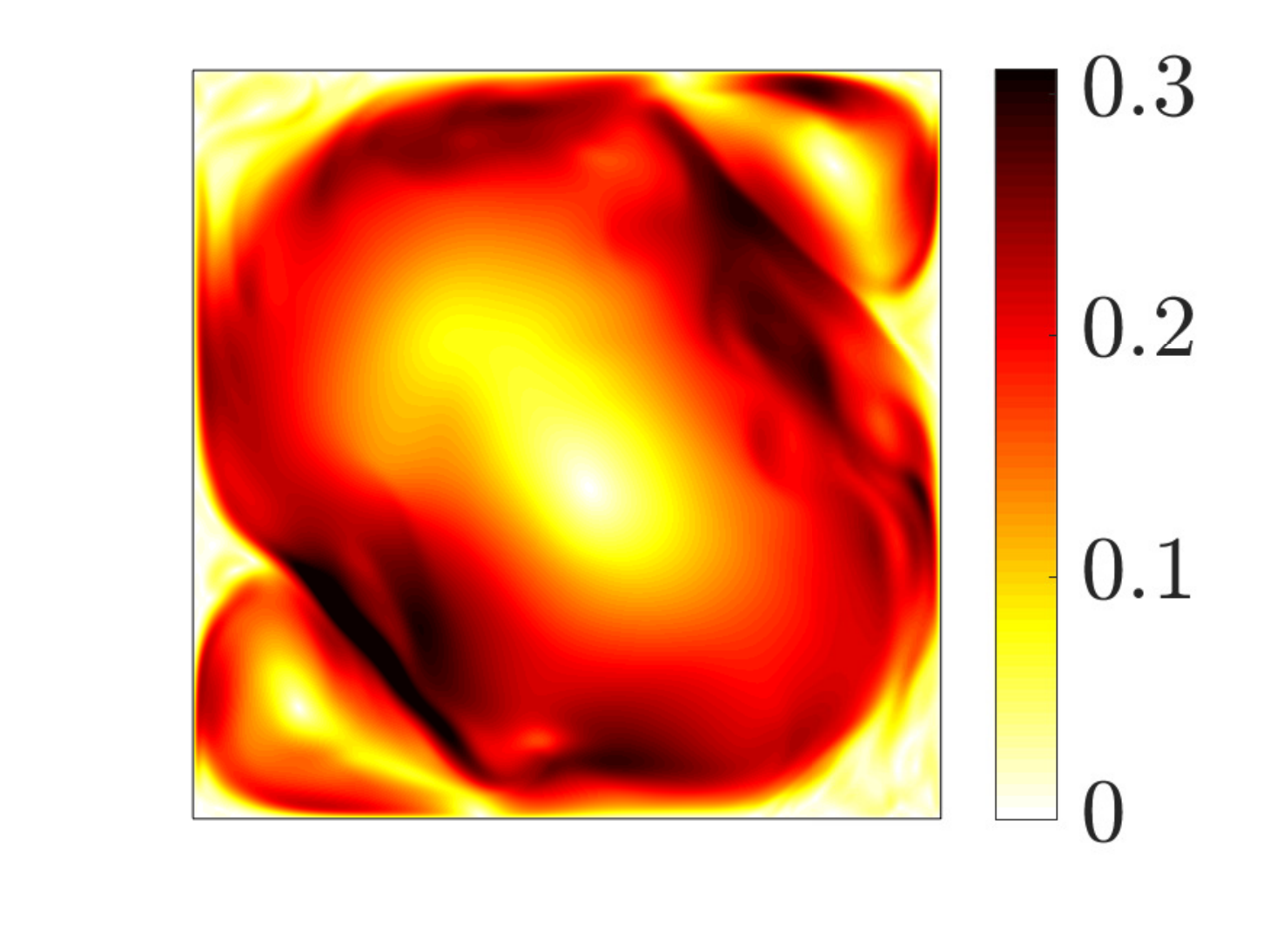}
	\put(-125,81){$(b)$}
	\put(-75.5,95){$|\vec{u}|$}
	\includegraphics[width=0.33\linewidth]{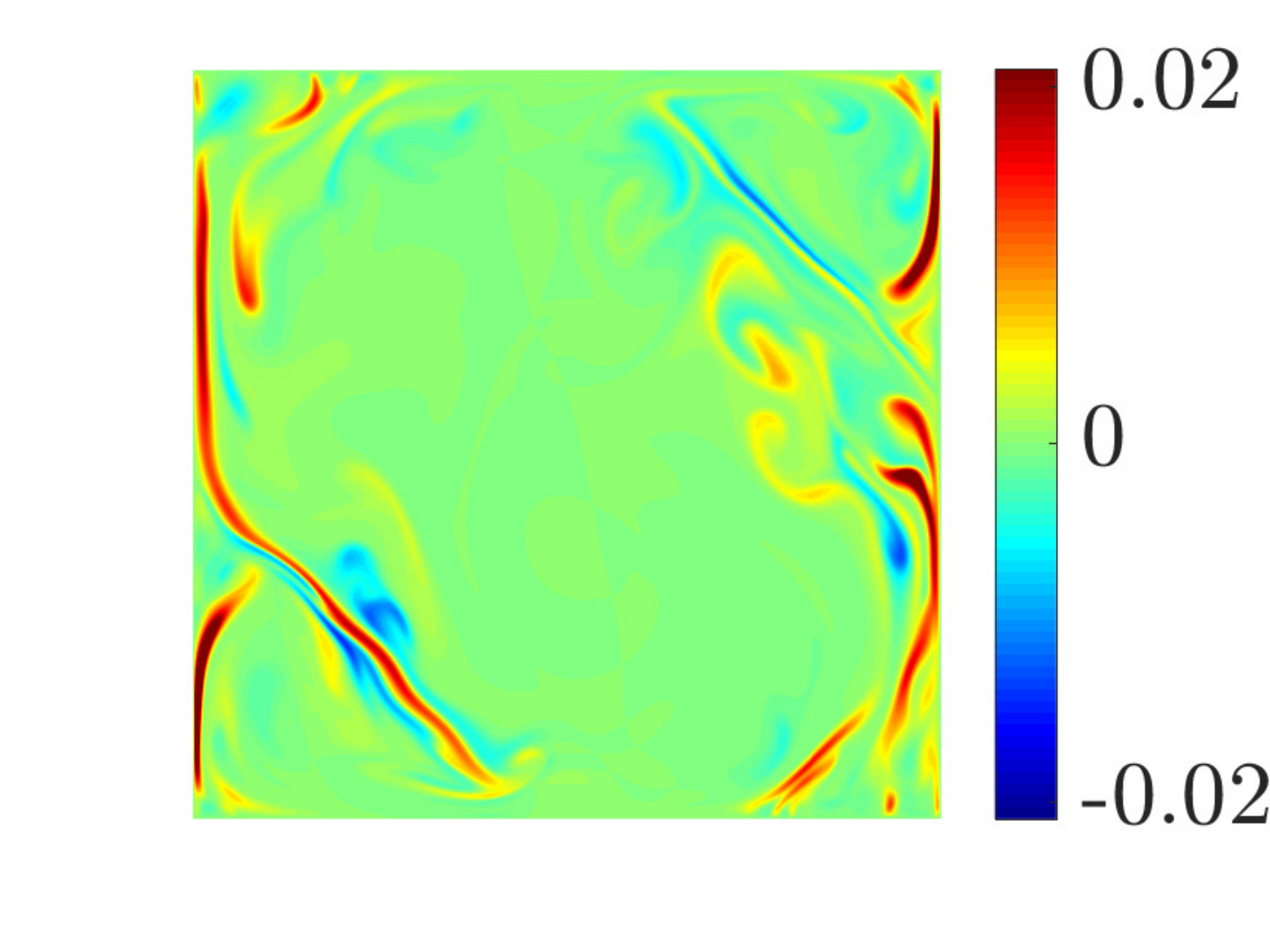}
	\put(-125,81){$(c)$}
	\put(-75.5,95){$v\cdot \delta T$}
	\\
	\includegraphics[width=0.33\linewidth]{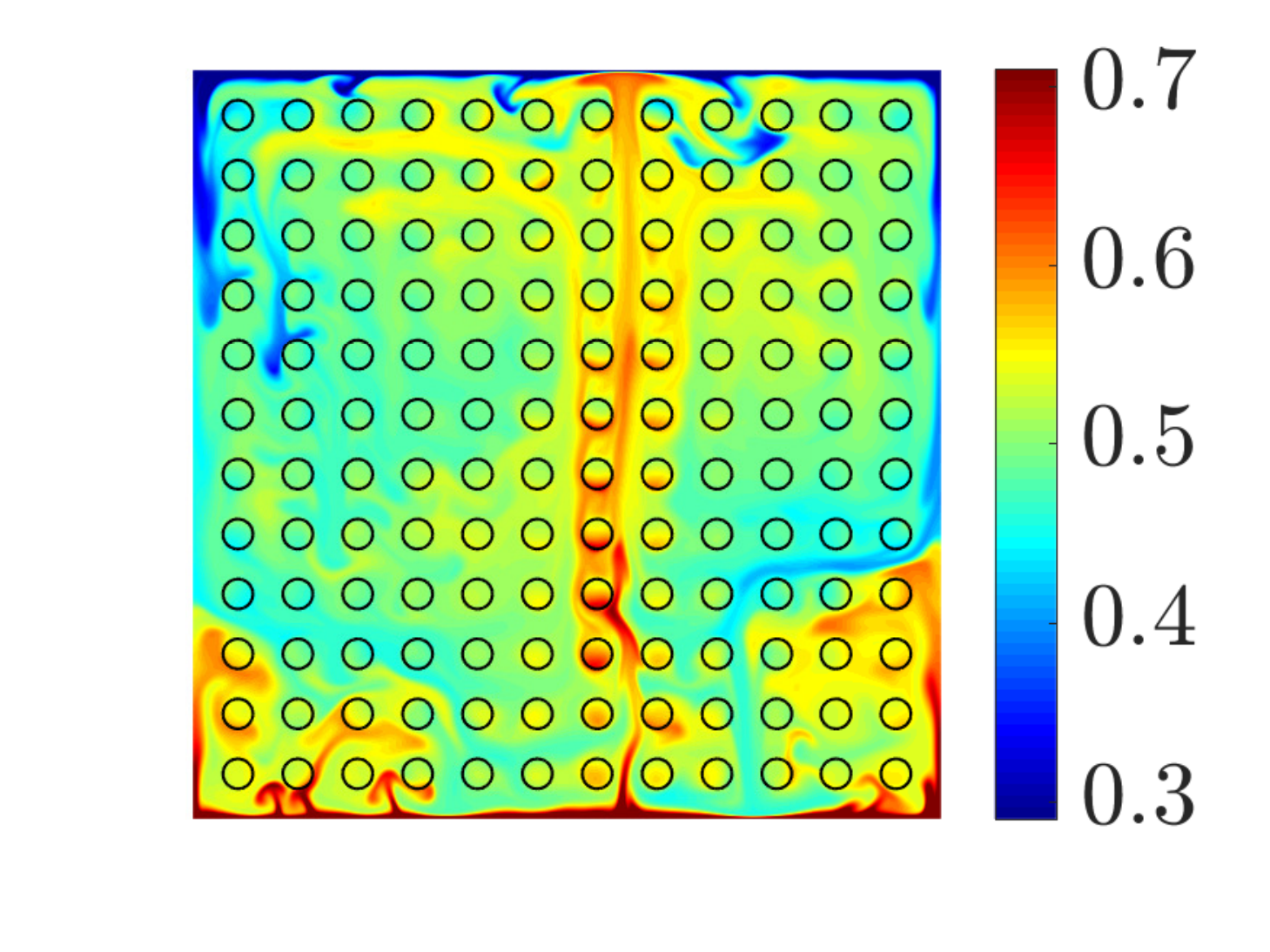}
	\put(-125,81){$(d)$}
	\put(-130,30){\rotatebox{90}{{\small $Ra=10^9$}}}
	\put(-118,31){\rotatebox{90}{{\small $\phi=0.82$}}}
	\includegraphics[width=0.33\linewidth]{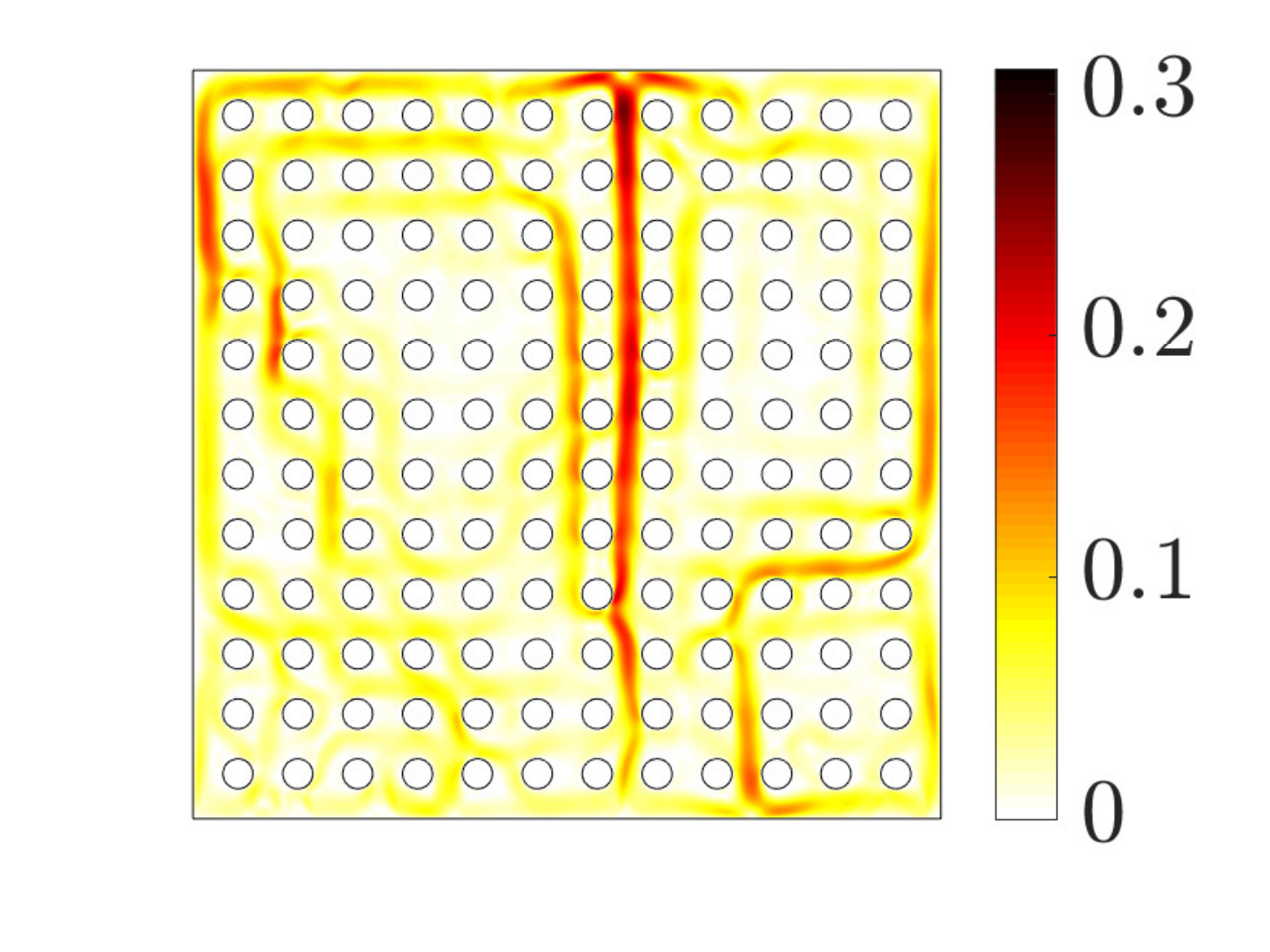}
	\put(-125,81){$(e)$}
	\includegraphics[width=0.33\linewidth]{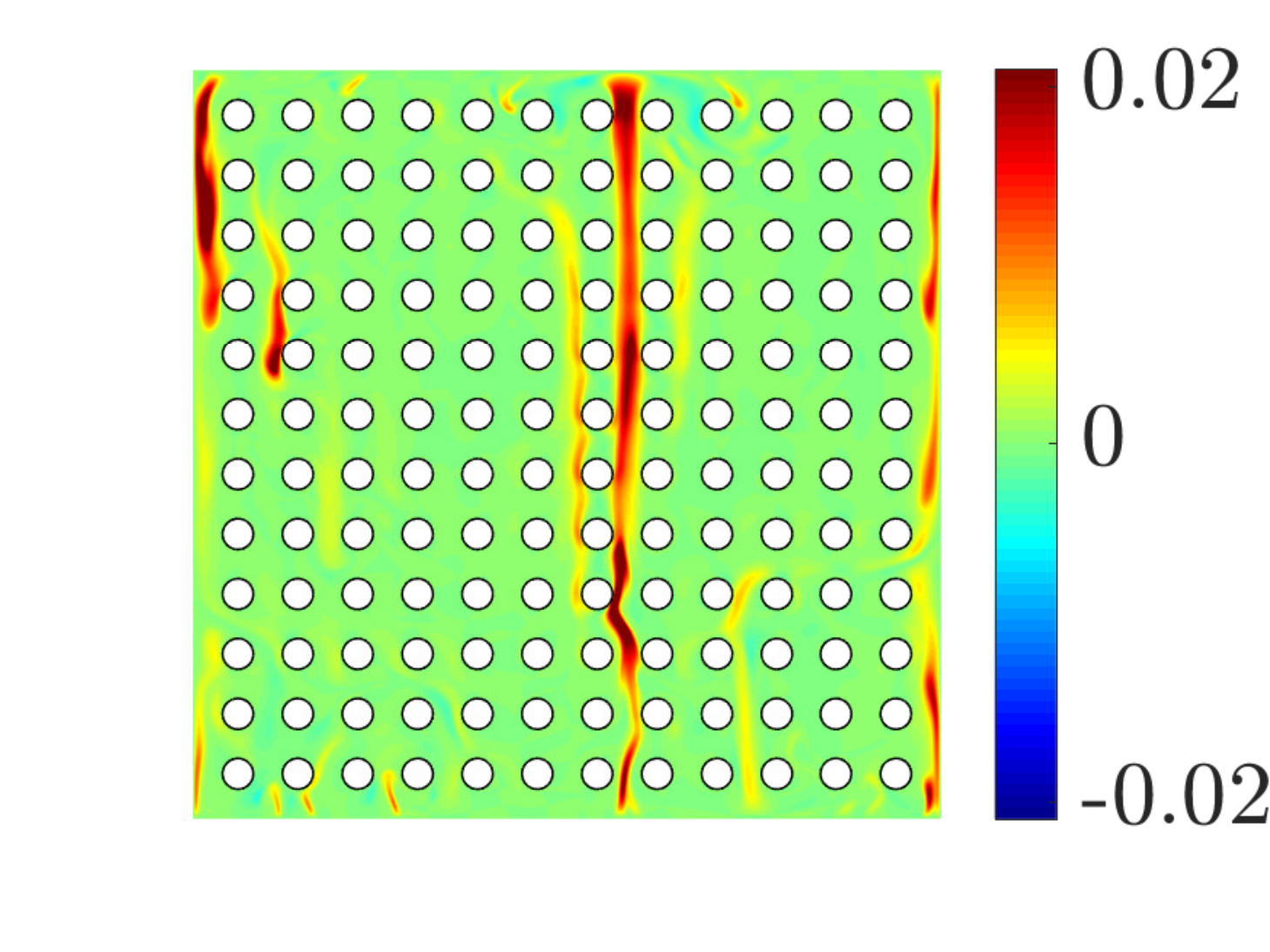}
	\put(-125,81){$(f)$}
	\\
	\includegraphics[width=0.33\linewidth]{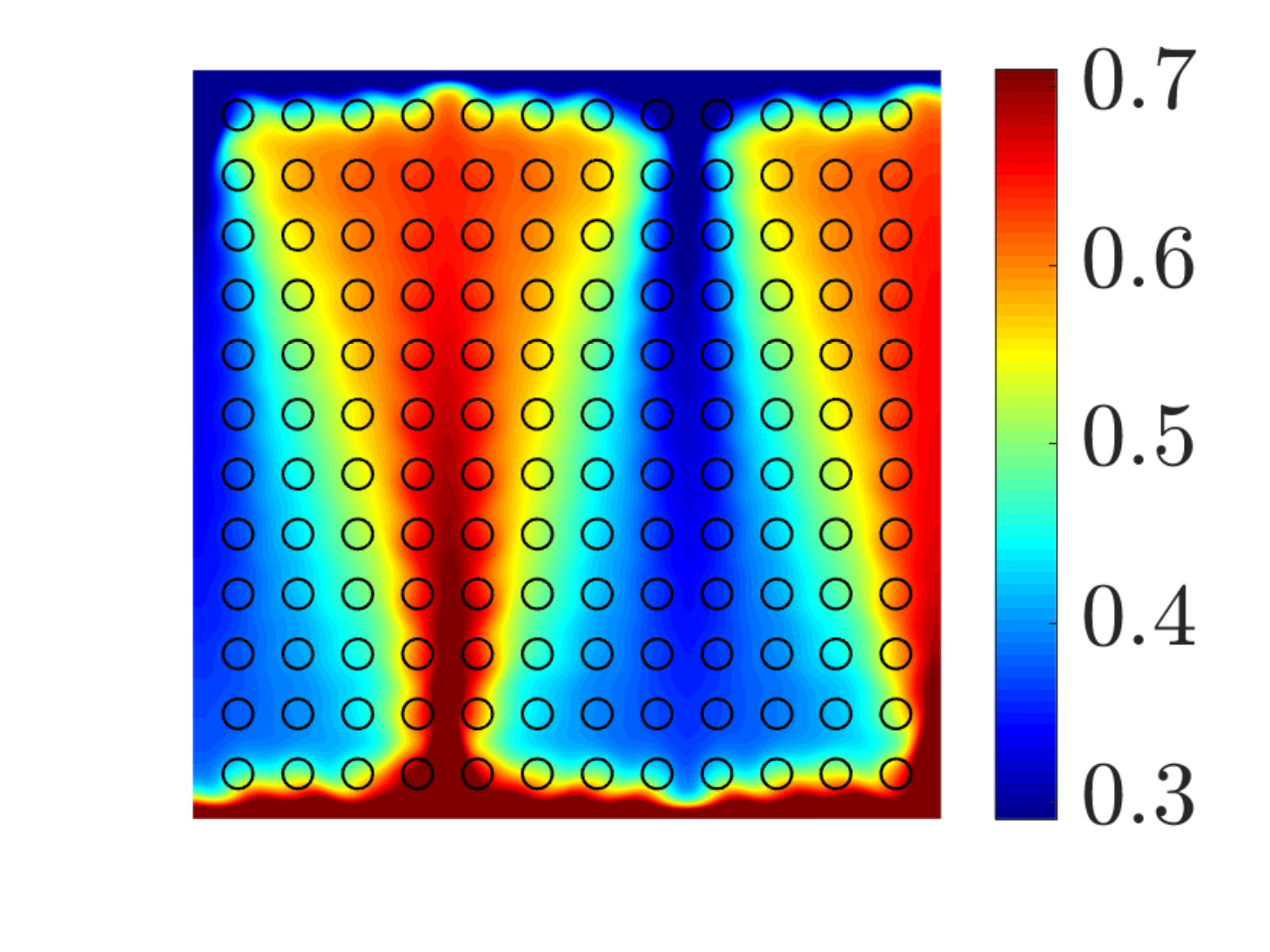}
	\put(-125,81){$(g)$}
	\put(-130,30){\rotatebox{90}{{\small $Ra=10^7$}}}
	\put(-118,31){\rotatebox{90}{{\small $\phi=0.82$}}}
	\includegraphics[width=0.33\linewidth]{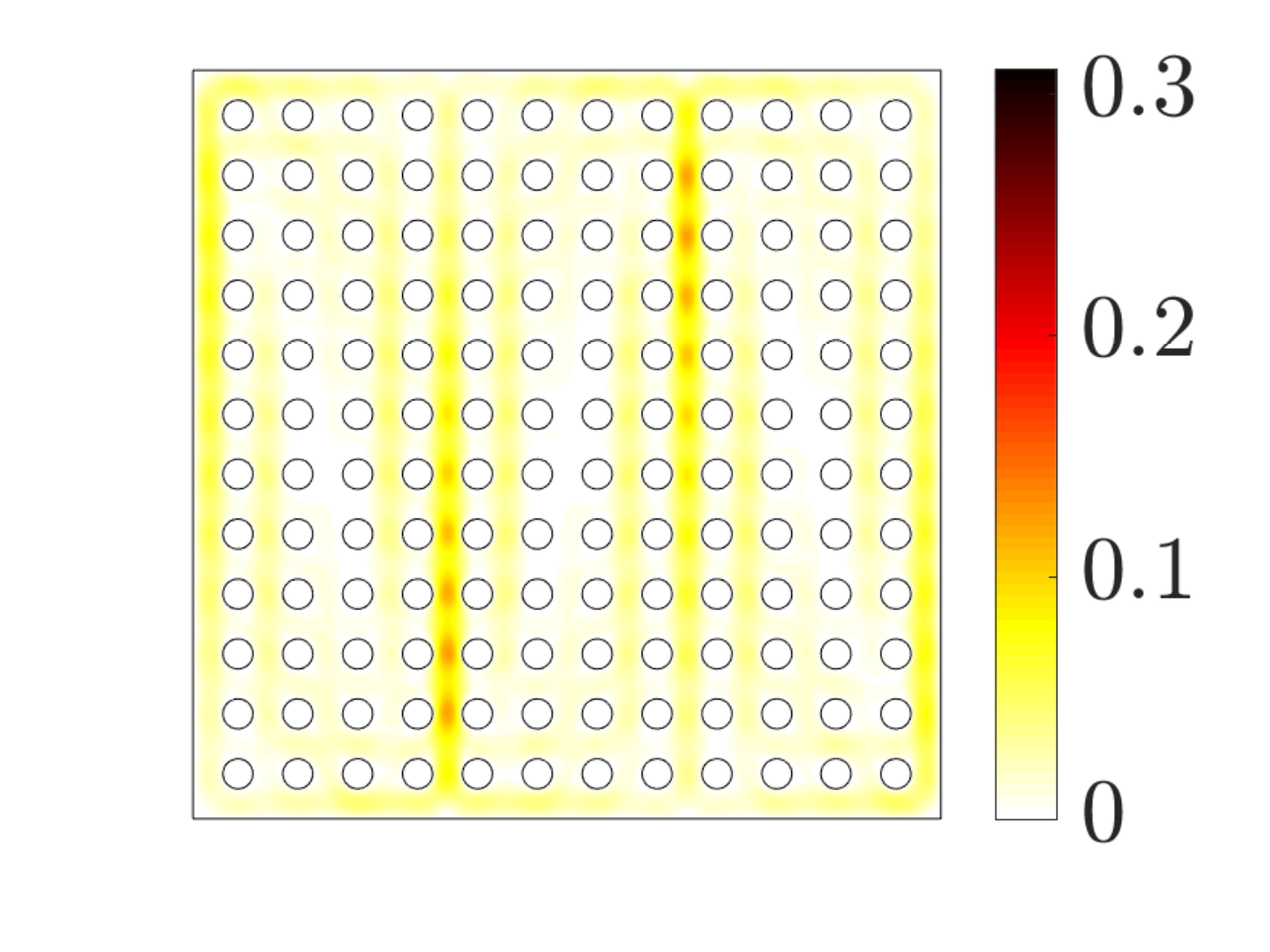}
	\put(-125,81){$(h)$}
	\includegraphics[width=0.33\linewidth]{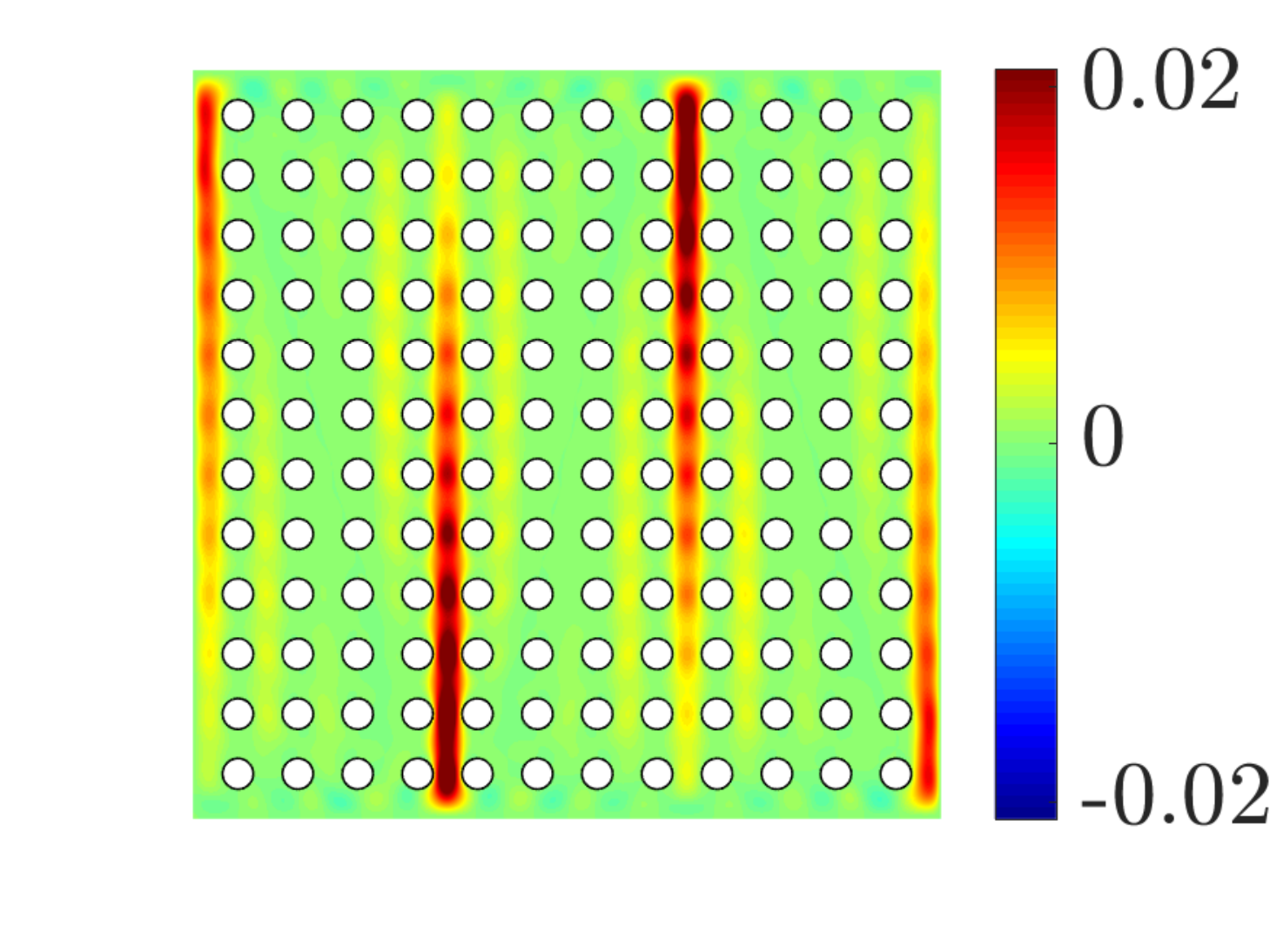}
	\put(-125,81){$(i)$}
	\caption{\label{flow_structures}Typical snapshots of the instantaneous ($a,d,g$) temperature $T$, ($b,e,h$) the velocity magnitude $|\vec{u}|$, and ($c,f,i$) the local convective heat flux $v\cdot \delta T$ in the vertical direction  at $(a,b,c)$ $(Ra,\phi)=(10^9,1)$, ($d,e,f$) $(Ra,\phi)=(10^9,0.82)$, and $(g,h,i)$ $(Ra,\phi)=(10^7,0.82)$. Circles in $(d-i)$ indicate the obstacle array. Note that according to the temperature equation (\ref{temperature_eqn}), the temperature in the obstacles is well defined.}
\end{figure}

Snapshots of temperature fields at $Ra=10^7$ and different porosities are shown in {\color{black}figure} \ref{optimal_spacing_1}, demonstrating the enhancement of flow coherence with decreasing $\phi$. For relatively large $\phi$, plumes wander randomly in the {\color{black}bulk, as} shown in figure \ref{optimal_spacing_1}($a$). As $\phi$ is further decreased, plumes cannot move freely due to the impedance of the obstacle array and prefer moving along convection channels, as shown in figures \ref{optimal_spacing_1}($b,c$).

\begin{figure}
	\centering
	\vspace{3 mm}
	\hspace{-5 mm}
	\includegraphics[width=0.33\linewidth]{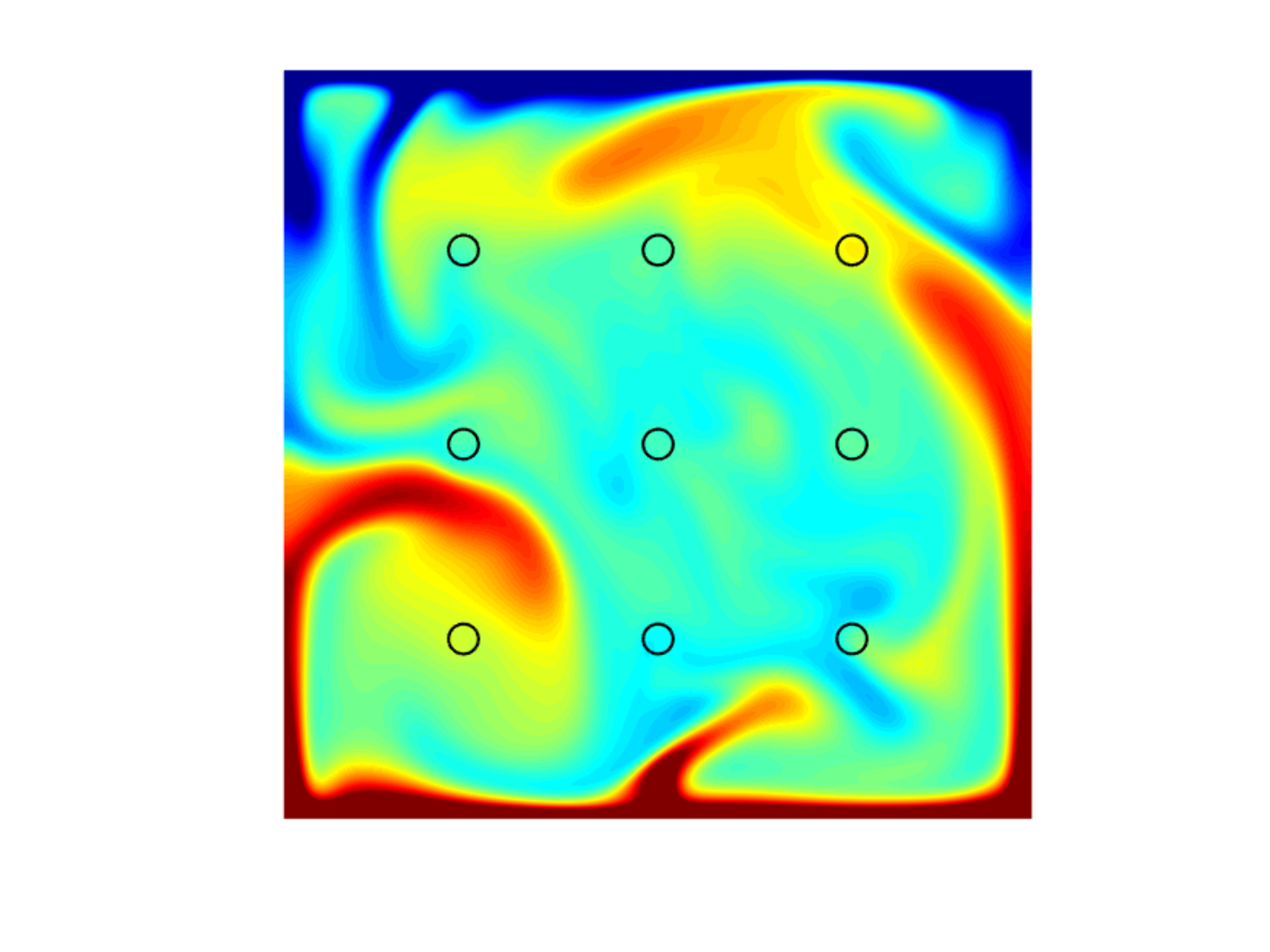}
	\put(-116,81){$(a)$}
	\put(-77,95){$\phi=0.99$}
	\hspace{-1 mm}
	\includegraphics[width=0.33\linewidth]{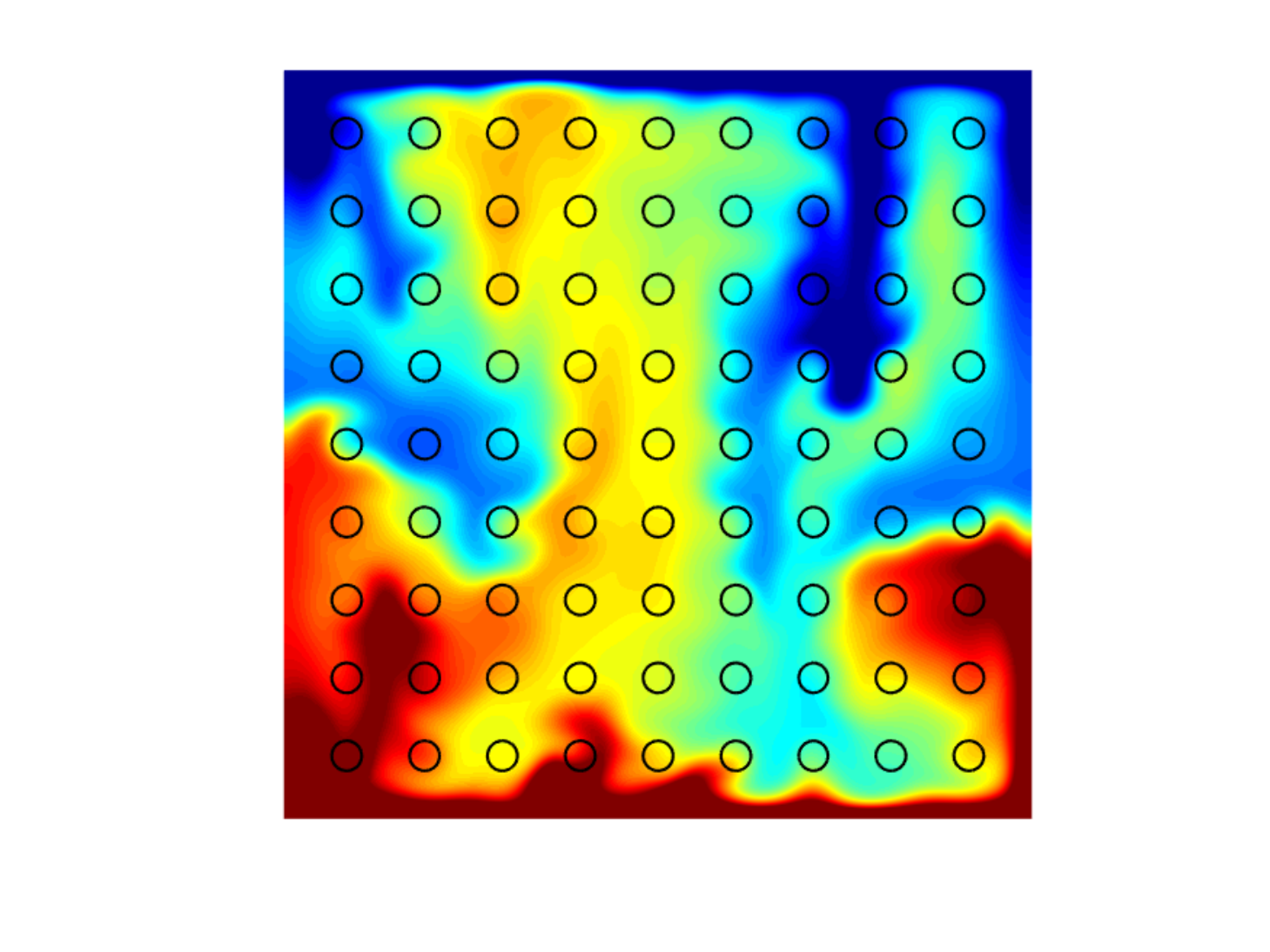}
	\put(-116,81){$(b)$}
	\put(-78,95){$\phi=0.90$}
	\hspace{2 mm}
	\includegraphics[width=0.33\linewidth]{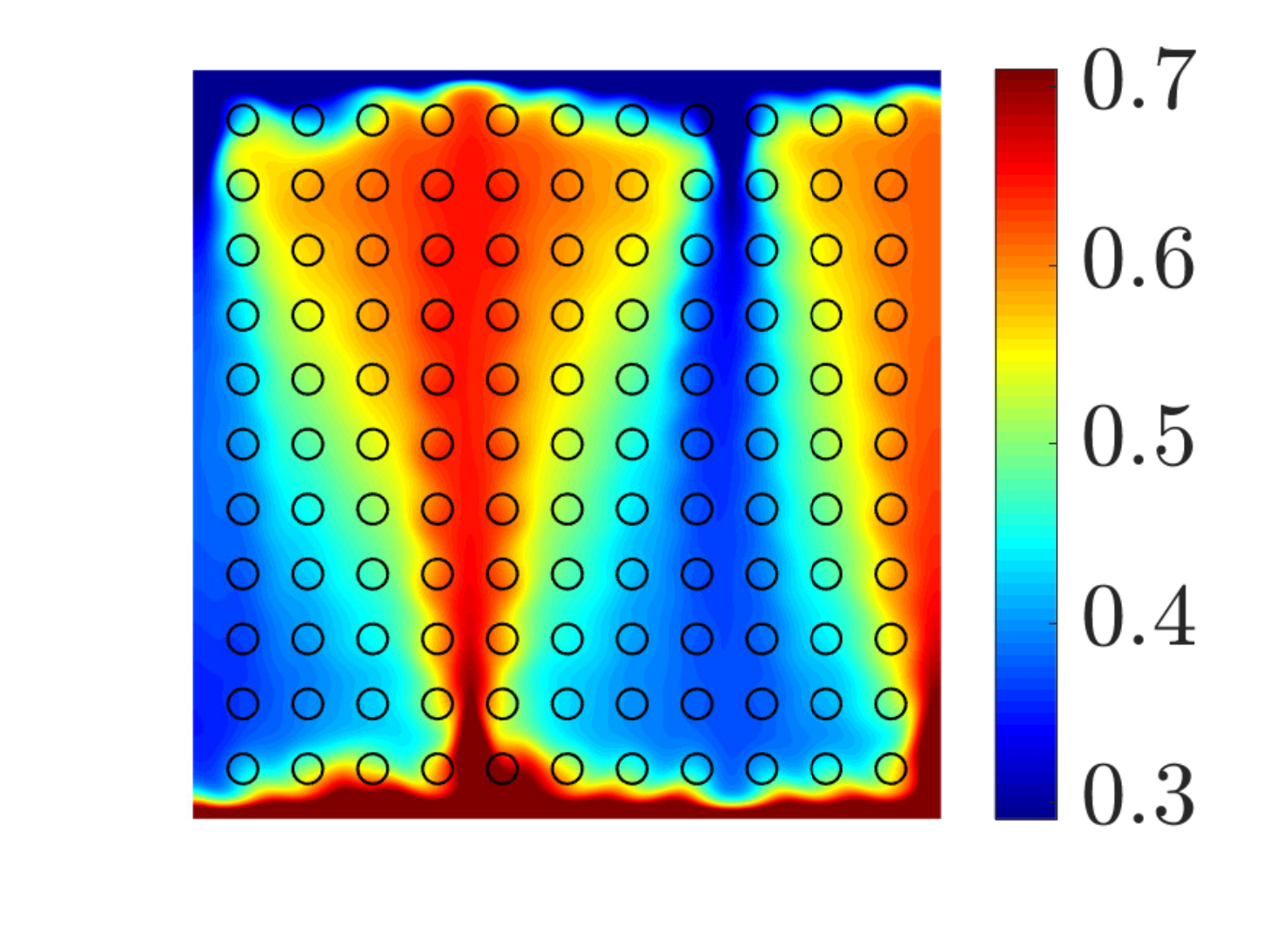}
	\put(-125,81){$(c)$}
	\put(-87,95){$\phi=0.85$}
	\caption{\label{optimal_spacing_1} Snapshots of the instantaneous temperature field at $Ra=10^7$ and different $\phi$, corresponding to the filled symbols in figure \ref{optimal_spacing}: $(a)$ $\phi=0.99$, $(b)$ $\phi=0.90$, and $(c)$ $\phi=0.85$.}
\end{figure}

\begin{figure}
	\centering
	\includegraphics[width=0.45\linewidth]{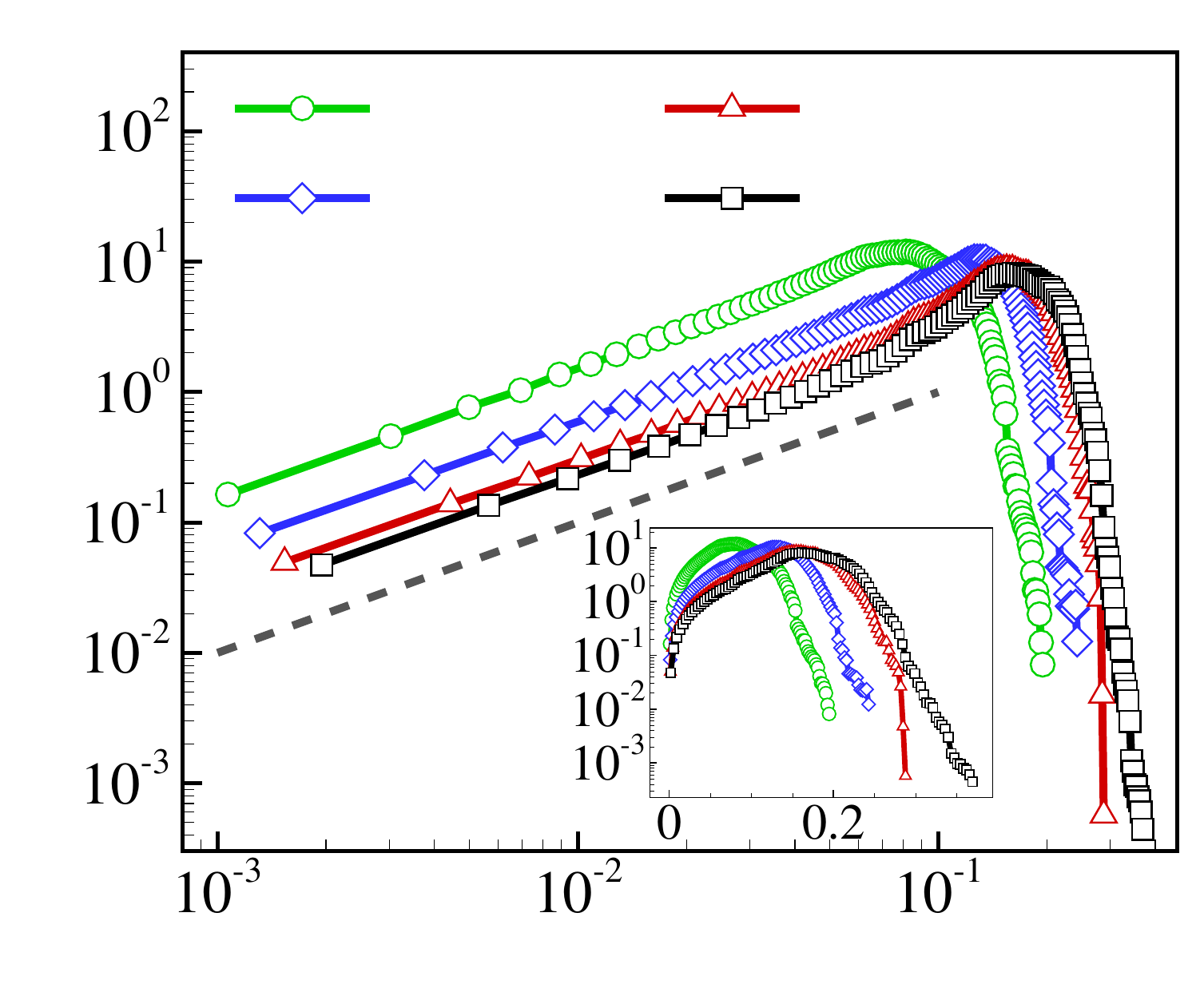}
	\put(-175,130){$(a)$}
	\put(-172,74){$P$}
	\put(-80,5){$|\vec{u}|$}
	\put(-121,126){ $r=0.15$}
	\put(-121,113){ $r=0.25$}
	\put(-59,126){ $r=0.35$}
	\put(-59,113){ $r=0.40$}
	\put(-130,45){ $|\vec{u}|$}
	\hspace{2 mm}
	\includegraphics[width=0.45\linewidth]{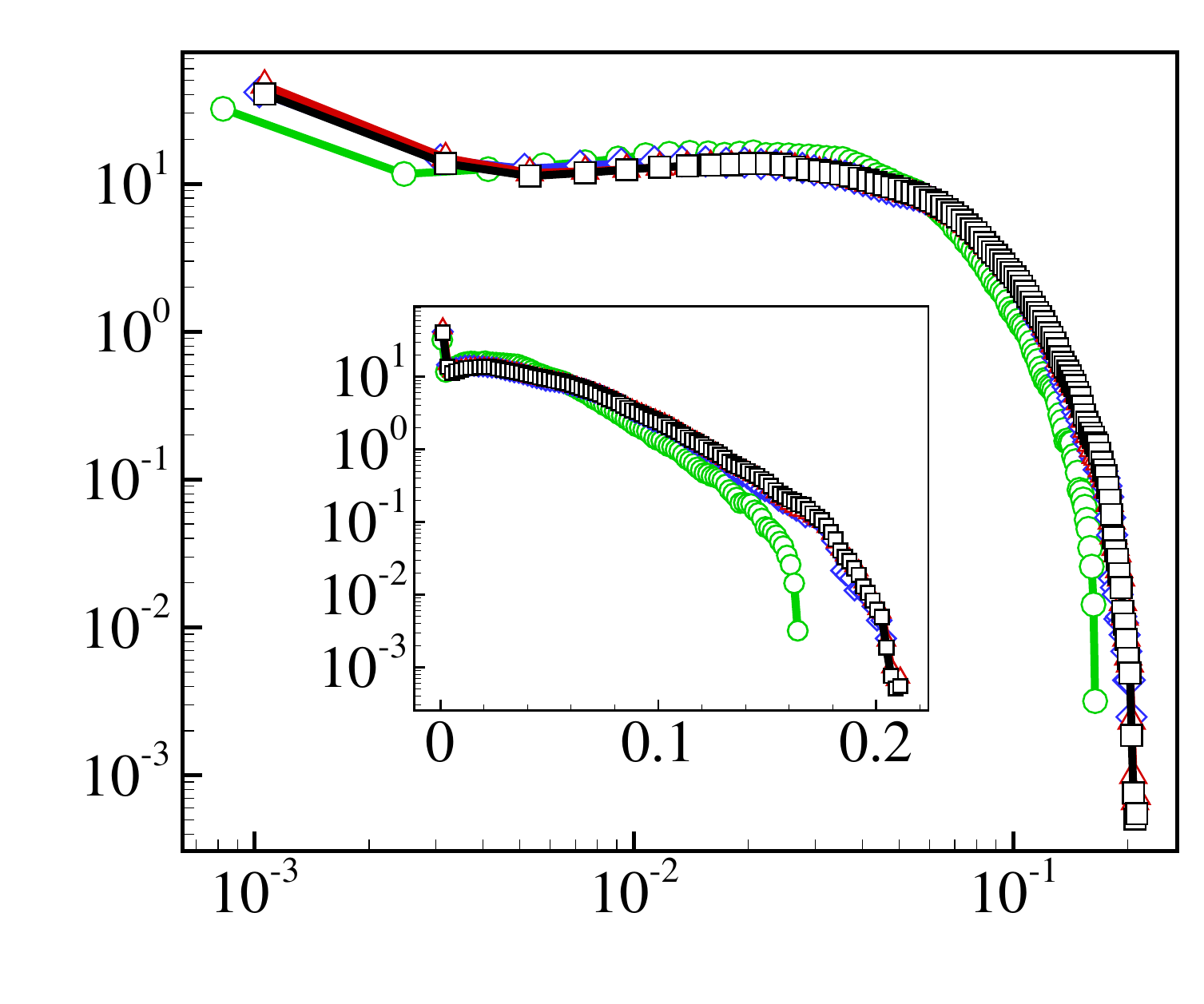}
	\put(-175,130){$(b)$}
	\put(-172,74){$P$}
	\put(-80,5){$|\vec{u}|$}
	\\
	\includegraphics[width=0.45\linewidth]{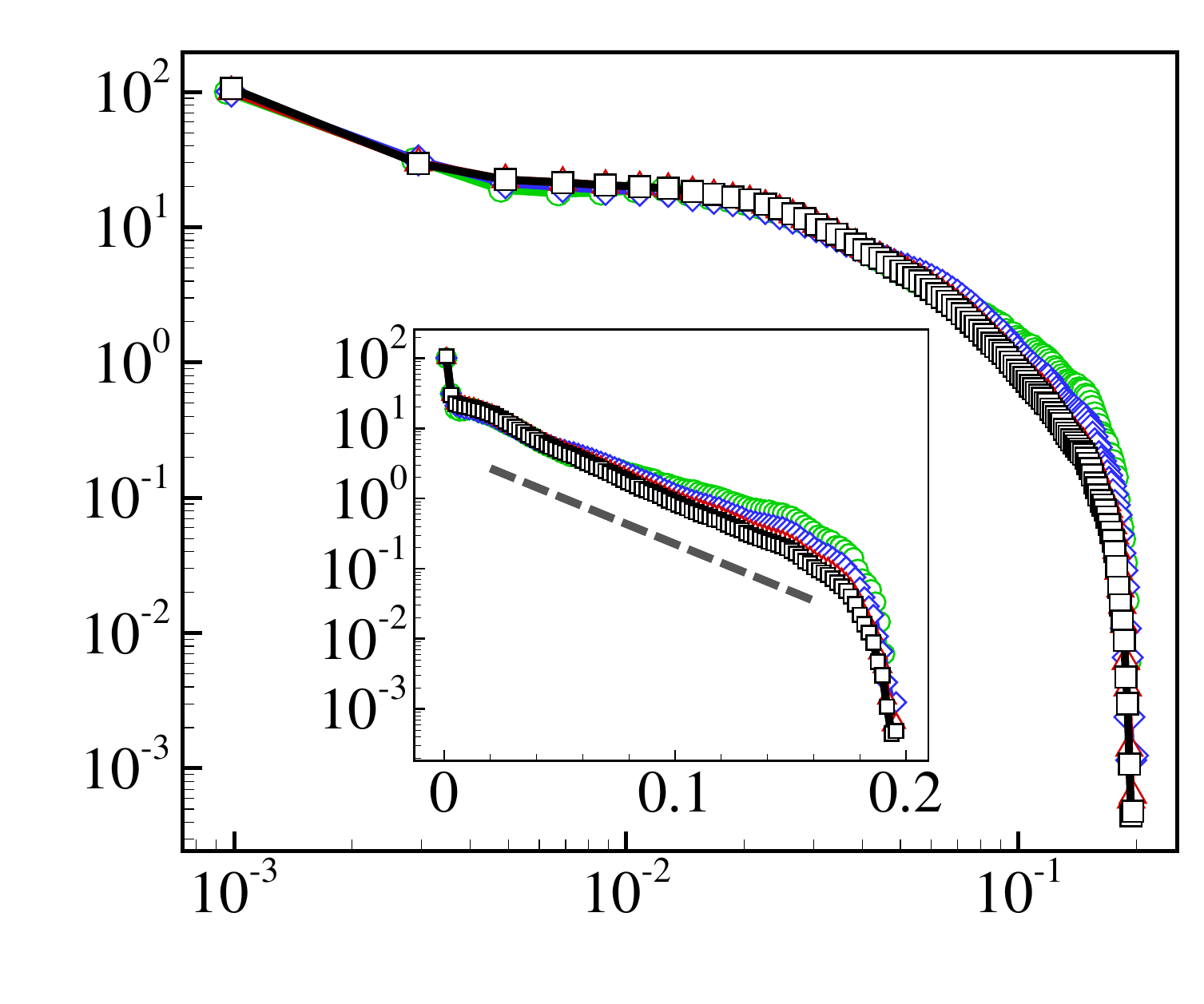}
	\put(-175,130){$(c)$}
	\put(-172,74){$P$}
	\put(-80,5){$|\vec{u}|$}
	\put(-96,50){$e^{-31|\vec{u}|}$}
	\caption{\label{velo_dist} Log-log plots of the PDFs of the velocity magnitude $|\vec{u}|$ inside a circular domain of radius $r$ at the cell center for different $\phi$ at $Ra=10^8$. $(a)$ $\phi=1$, $(b)$ $\phi=0.92$, $(c)$ $\phi=0.82$. The insets are the same results shown in semi-log plots. {\color{black}In $(a)$ and the inset of $(c)$ guiding lines of constant slope are included, describing the respective scaling behaviours of $P(|\vec{u}|)$. As $|\vec{u}|$ increases, $P(|\vec{u}|)$ increases linearly for small $|\vec{u}|$ in $(a)$, and it decreases exponentially in the intermediate range of $|\vec{u}|$ in $(c)$.} }
\end{figure}

\begin{figure}
	\centering
	\vspace{2 mm}
	\hspace{-2 mm}
	\includegraphics[width=0.35\linewidth]{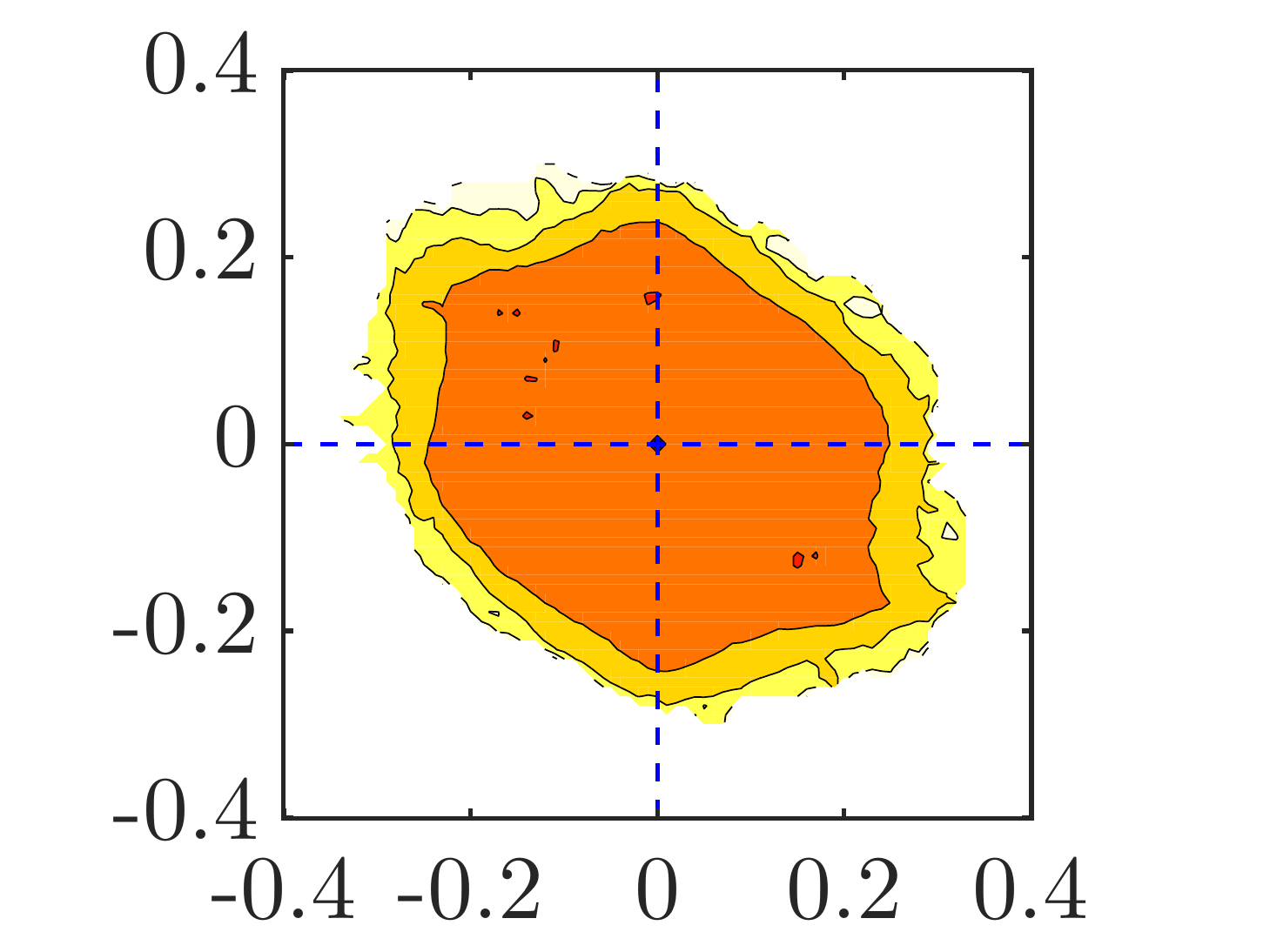}
	\put(-102,84){$(a)$}
	\put(-130,52){$v$}
	\put(-67,-5){$u$}
	\hspace{-8 mm}
	\includegraphics[width=0.35\linewidth]{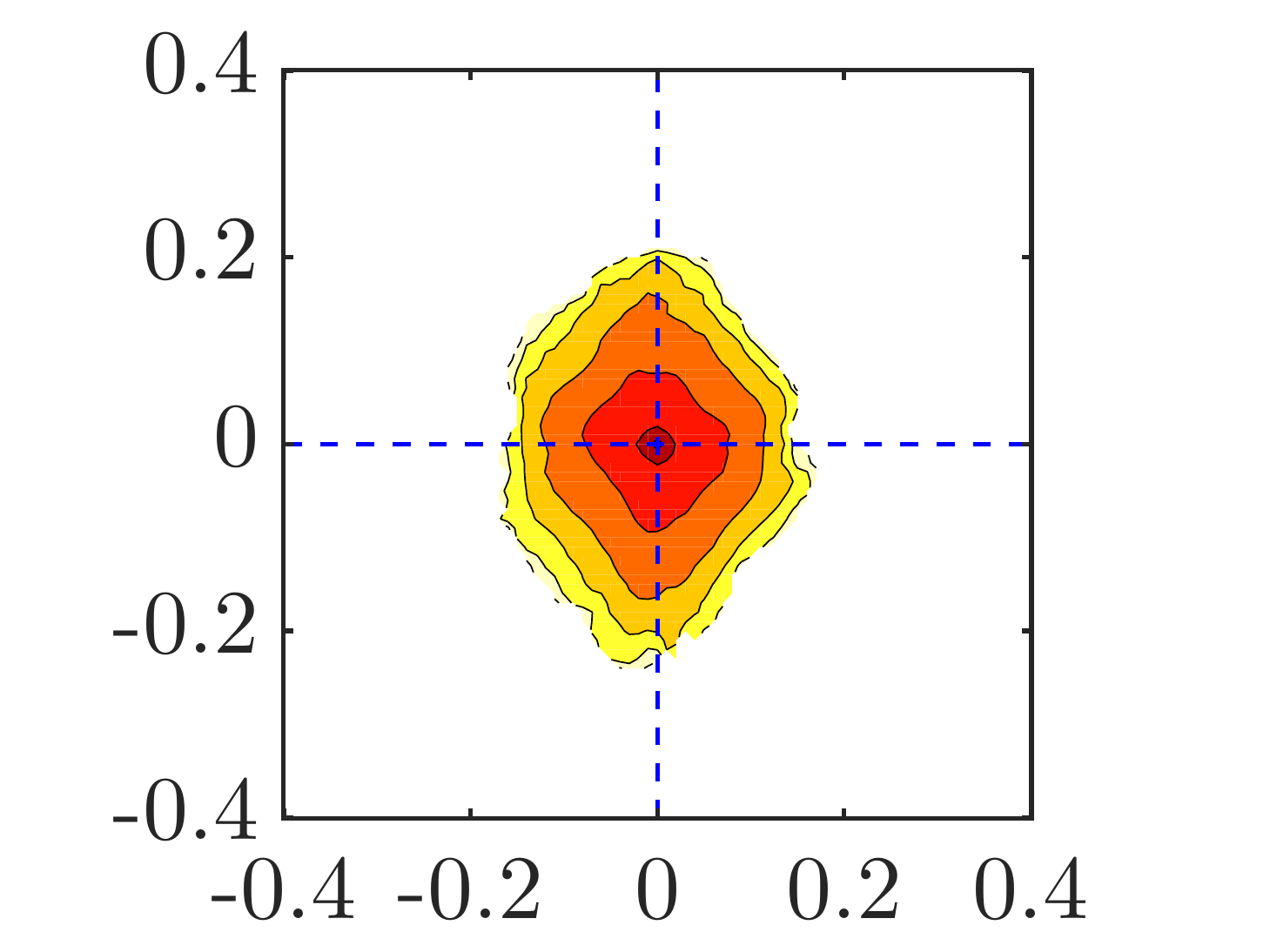}
	\put(-102,84){$(b)$}
	\put(-67,-5){$u$}
	\hspace{-5 mm}
	\includegraphics[width=0.35\linewidth]{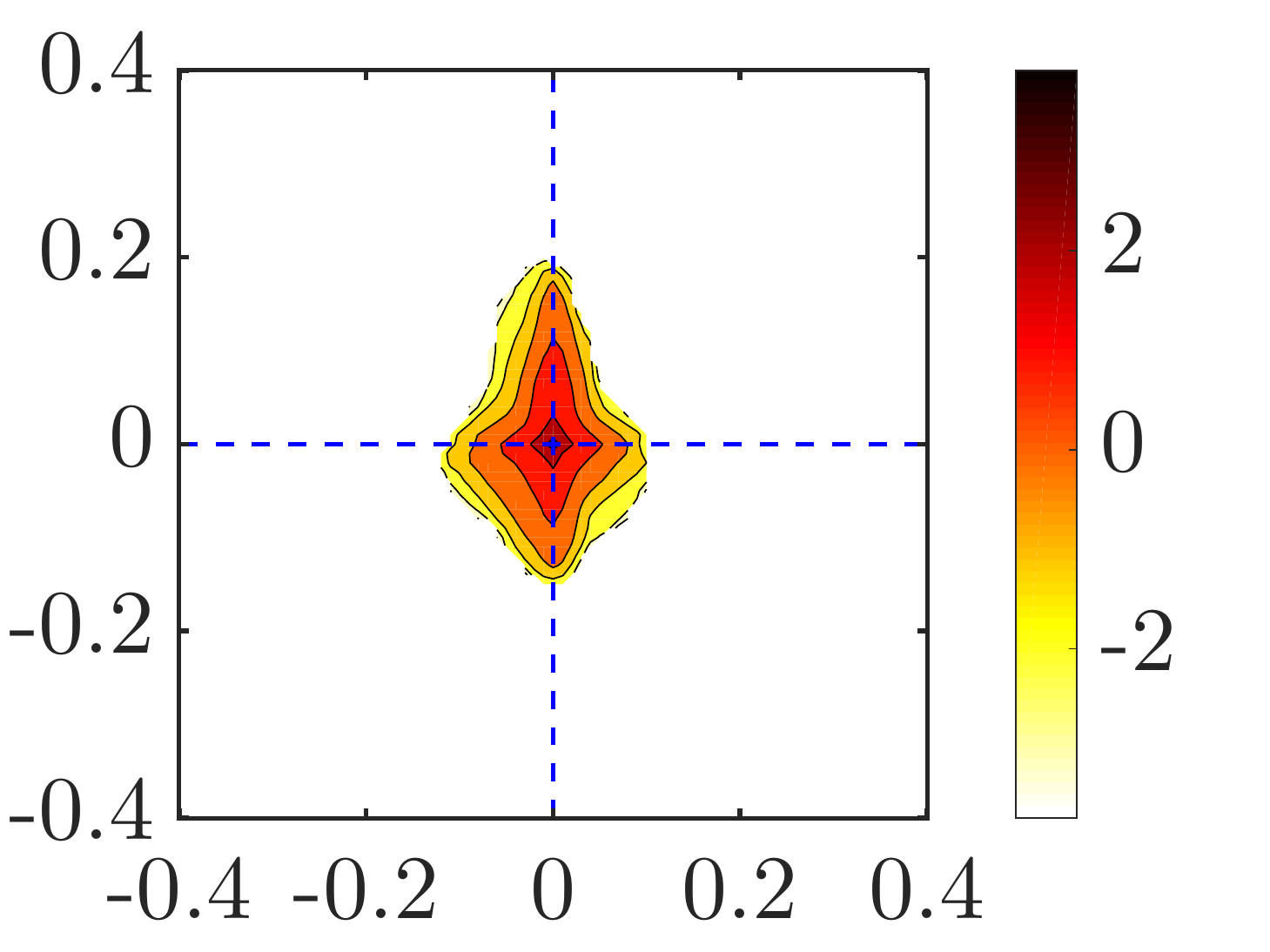}
	\put(-113,84){$(c)$}
	\put(-78,-5){$u$}
	\put(-32,98){{\footnotesize $logP$}}
	\vspace{2 mm}
	\caption{\label{velo_dist_jpdf} The joint PDFs of two velocity components $(u,v)$ inside a circular domain of radius $r=0.4$ at the cell center for different $\phi$ at $Ra=10^8$: $(a)$ $\phi=1$, $(b)$ $\phi=0.92$, $(c)$ $\phi=0.82$.}
\end{figure}

To further investigate the influence of the obstacle array on the convection, we study the velocity distribution in the bulk region. Figure \ref{velo_dist} shows the PDFs of the velocity magnitude $|\vec{u}|$ inside a circular domain of radius $r$ at the cell center for different $\phi$ at $Ra=10^8$.
In the traditional RB convection with $\phi=1$, the velocity magnitude is spatially non-uniform due to the formation of the LSC. The mean velocity vanishes at the cell center and increases linearly with $r$ {\color{black}at leading order}. 
{\color{black}Due to this non-uniformity,} $P(|\vec{u}|)$ shifts along the $|\vec{u}|$ axis as $r$ is increased, and follows a linear relation for small $|\vec{u}|$, $P(|\vec{u}|)\sim|\vec{u}|$, as shown in figure \ref{velo_dist}$(a)$.
The obstacle array has a significant influence on the velocity distribution, as shown in figures \ref{velo_dist}$(b,c)$.
The maximum convection velocity is decreased due to the drag of the obstacle array, and the results for different values of $r$ collapse, indicating that the velocity distribution is spatially uniform in the bulk. 
The shape of $P(|\vec{u}|)$ also changes qualitatively and deviates from the power-law distribution as $\phi$ is decreased. In regular porous media the probability density is larger for $|\vec{u}|$ to take smaller values, and when $\phi$ is small enough, $P(|\vec{u}|)$ satisfies an exponential distribution, as shown in figure \ref{velo_dist}$(c)$.
{\color{black}The change of velocity distribution demonstrates the transition of flow organization as $\phi$ is decreased, from the formation of the LSC in the traditional RB convection to the wandering motion of plumes penetrating in the pores in the porous-media convection.}

Figure \ref{velo_dist_jpdf} shows the joint PDFs of two velocity components ($u,v$) inside a circular domain of radius $r=0.4$ at the cell center for different $\phi$ at $Ra=10^8$. It is found that both the size and shape of $P(u,v)$ change as $\phi$ is decreased from 1.
The shrinking of $P(u,v)$ indicates the reduction of convection strength.
The change of contour levels manifests the increased probability for $|\vec{u}|$ to take smaller values.
When $\phi$ is small enough, the contour lines of $P(u,v)$ protrude along the horizontal and vertical coordinate axes, indicating the formation of convection channels.

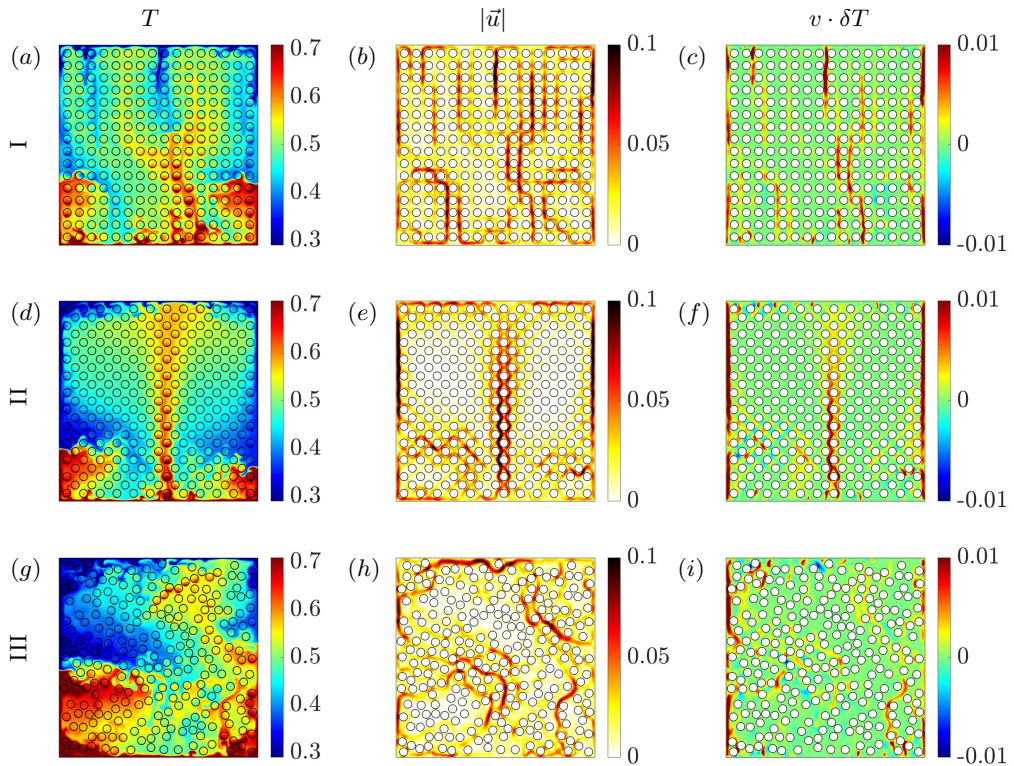
\begin{figure}
	\centering
	\includegraphics[width=0.33\linewidth]{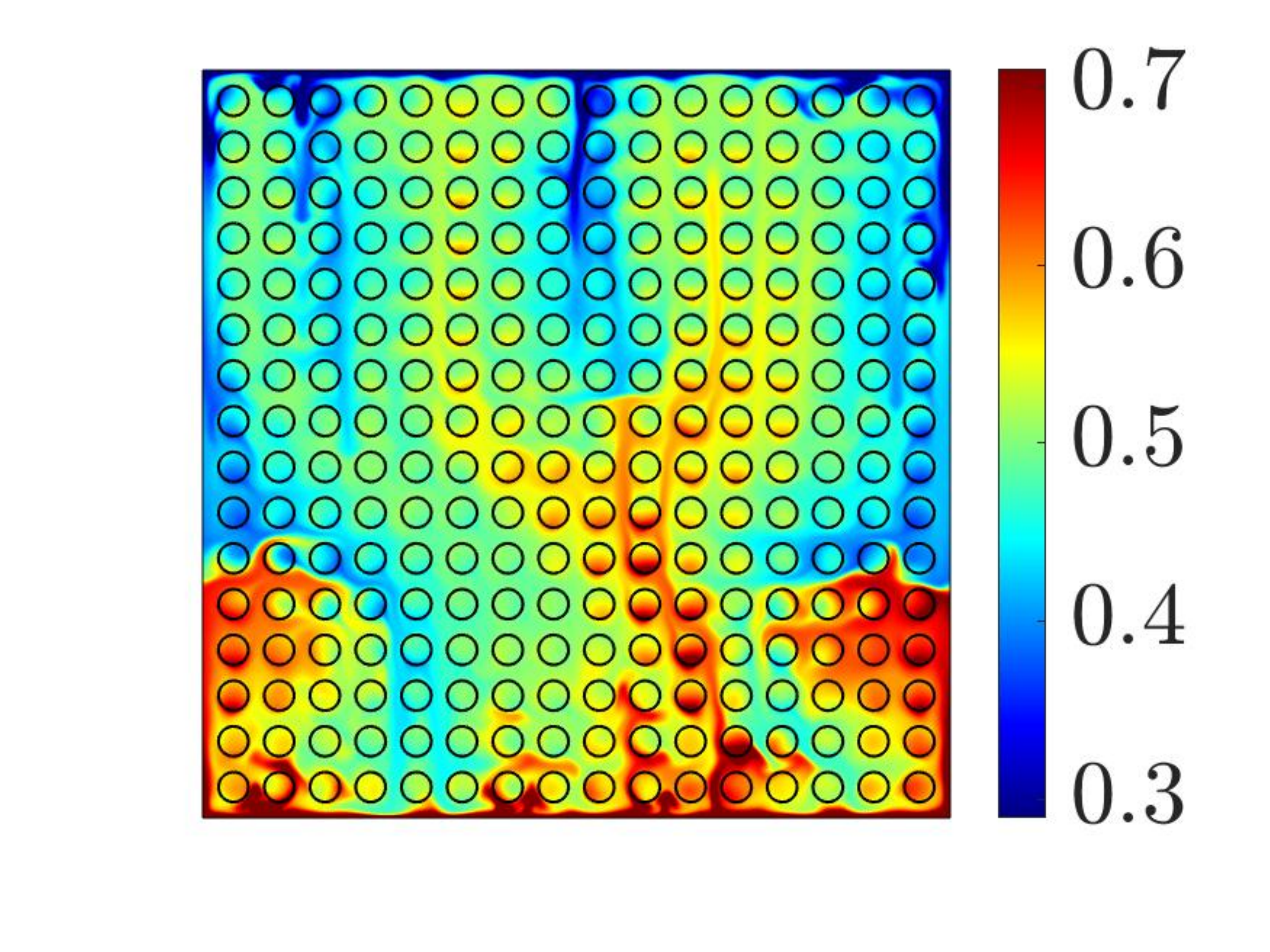}
	\put(-125,81){$(a)$}
	\put(-75.5,95){$T$}
	\put(-125,49){\rotatebox{90}{I}}
	\hspace{-1 mm}
	\includegraphics[width=0.33\linewidth]{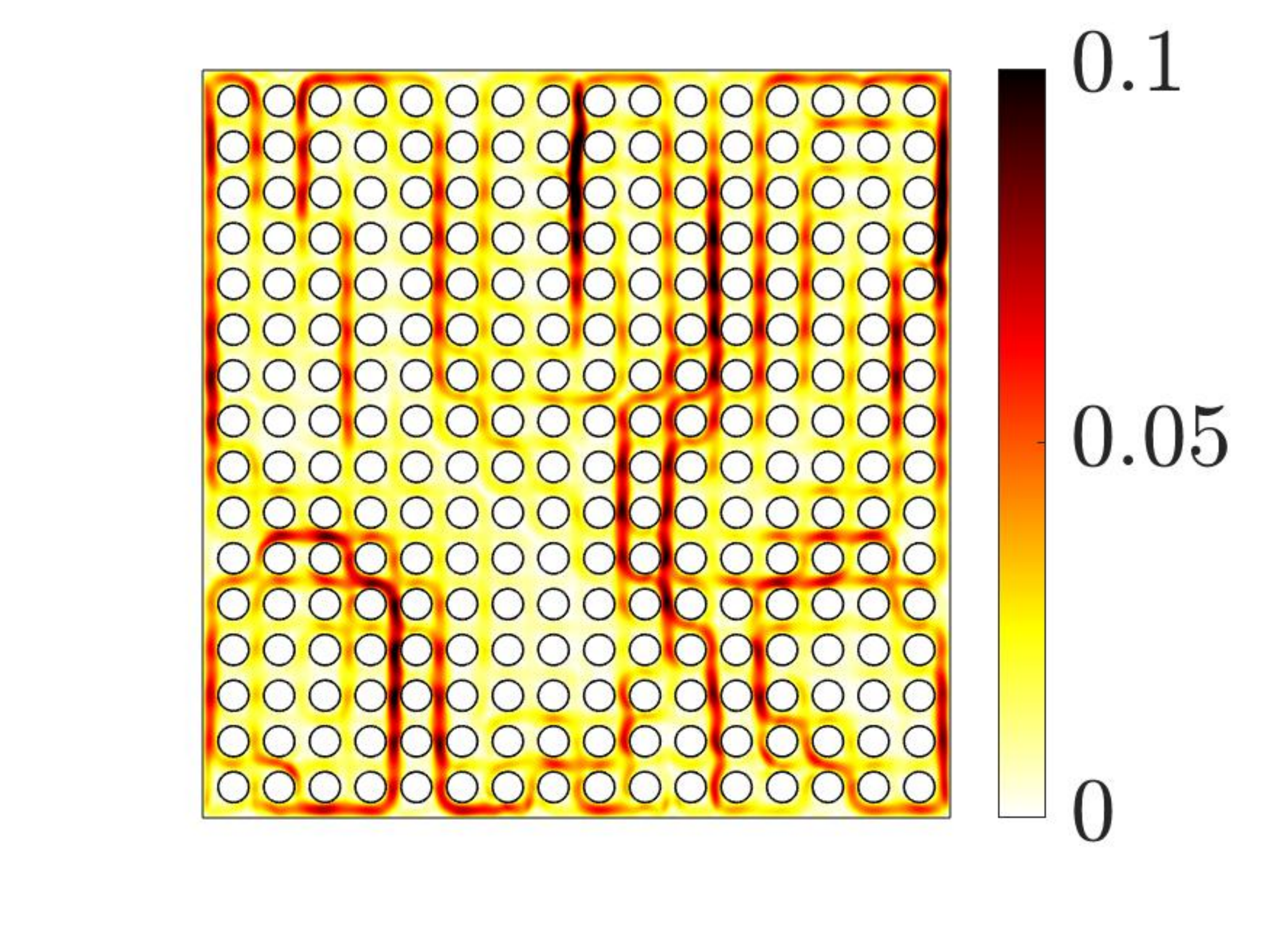}
	\put(-125,81){$(b)$}
	\put(-75.5,95){$|\vec{u}|$}
	\hspace{-2 mm}
	\includegraphics[width=0.33\linewidth]{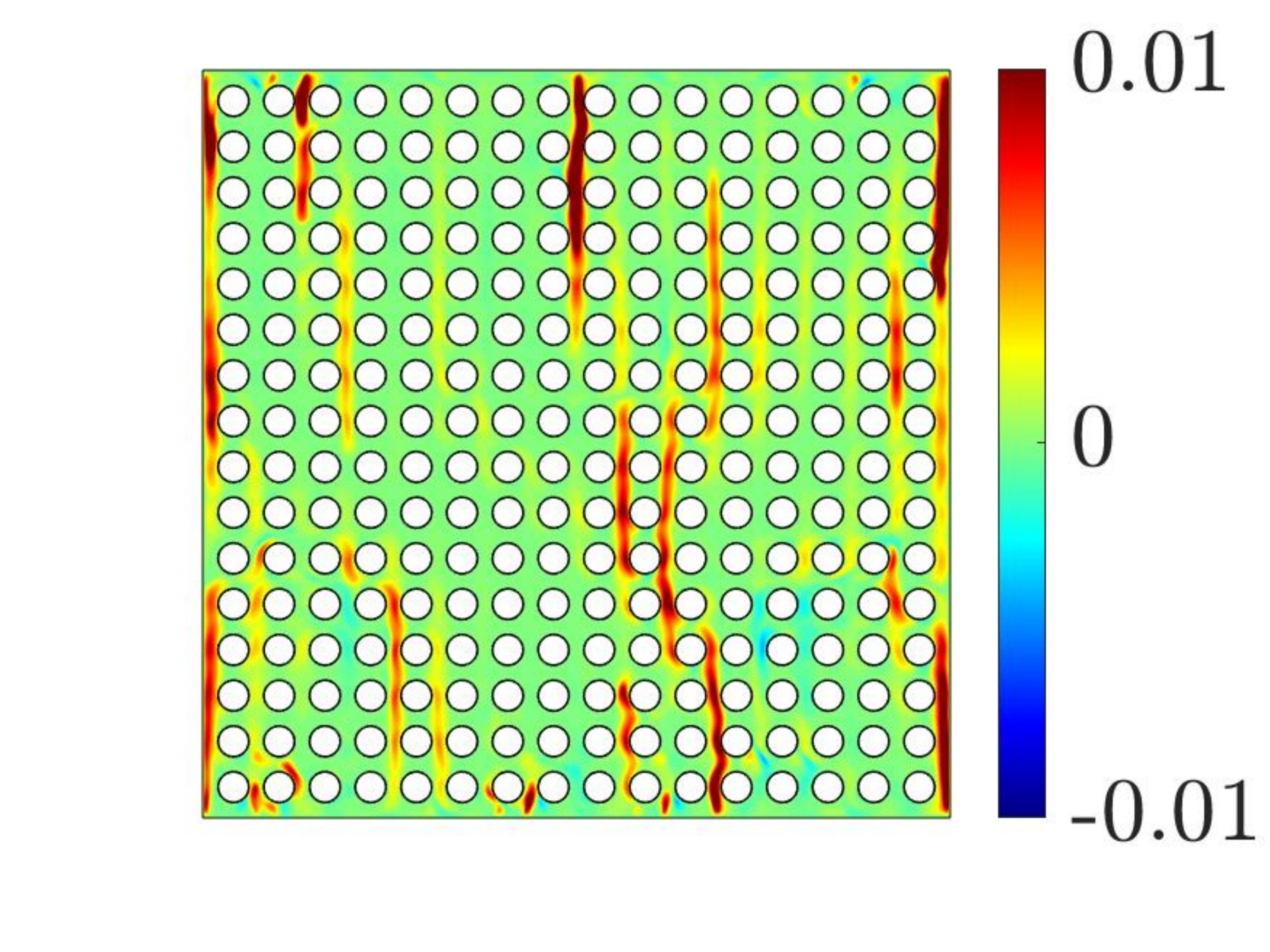}
	\put(-125,81){$(c)$}
	\put(-75.5,95){$v\cdot \delta T$}
	\\
	\includegraphics[width=0.33\linewidth]{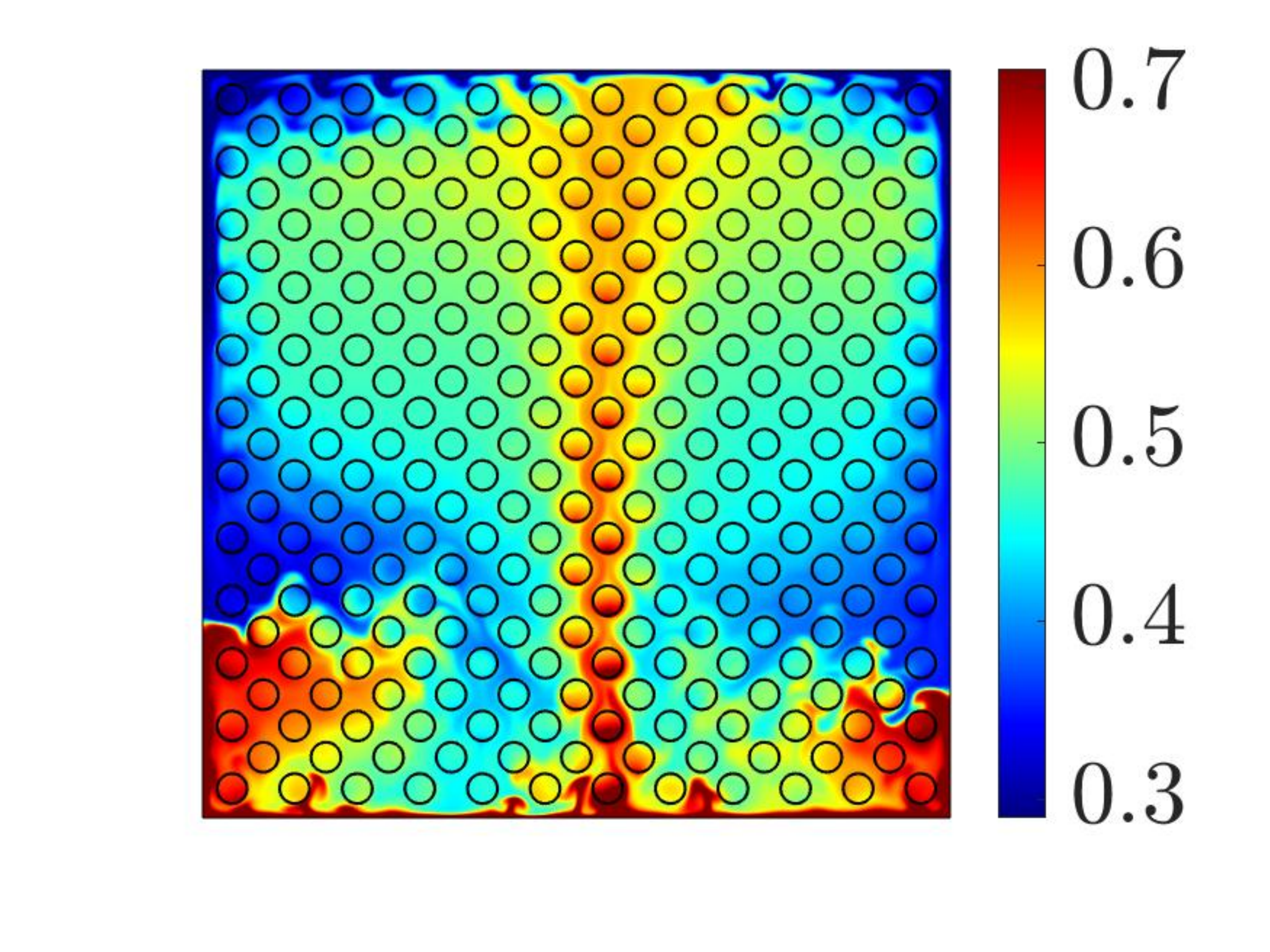}
	\put(-125,81){$(d)$}
	\put(-125,49){\rotatebox{90}{II}}
	\hspace{-1 mm}
	\includegraphics[width=0.33\linewidth]{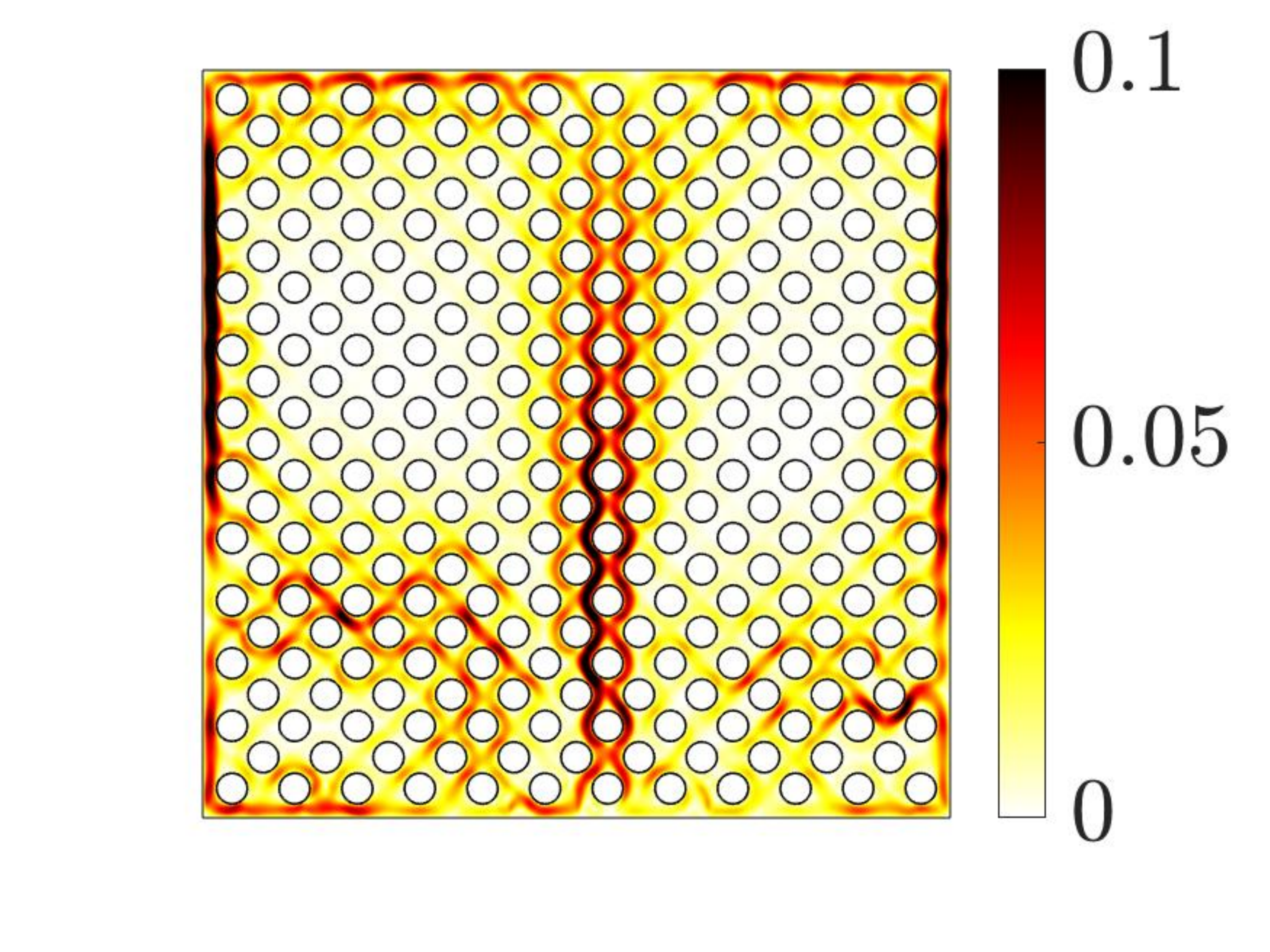}
	\put(-125,81){$(e)$}
	\hspace{-2 mm}
	\includegraphics[width=0.33\linewidth]{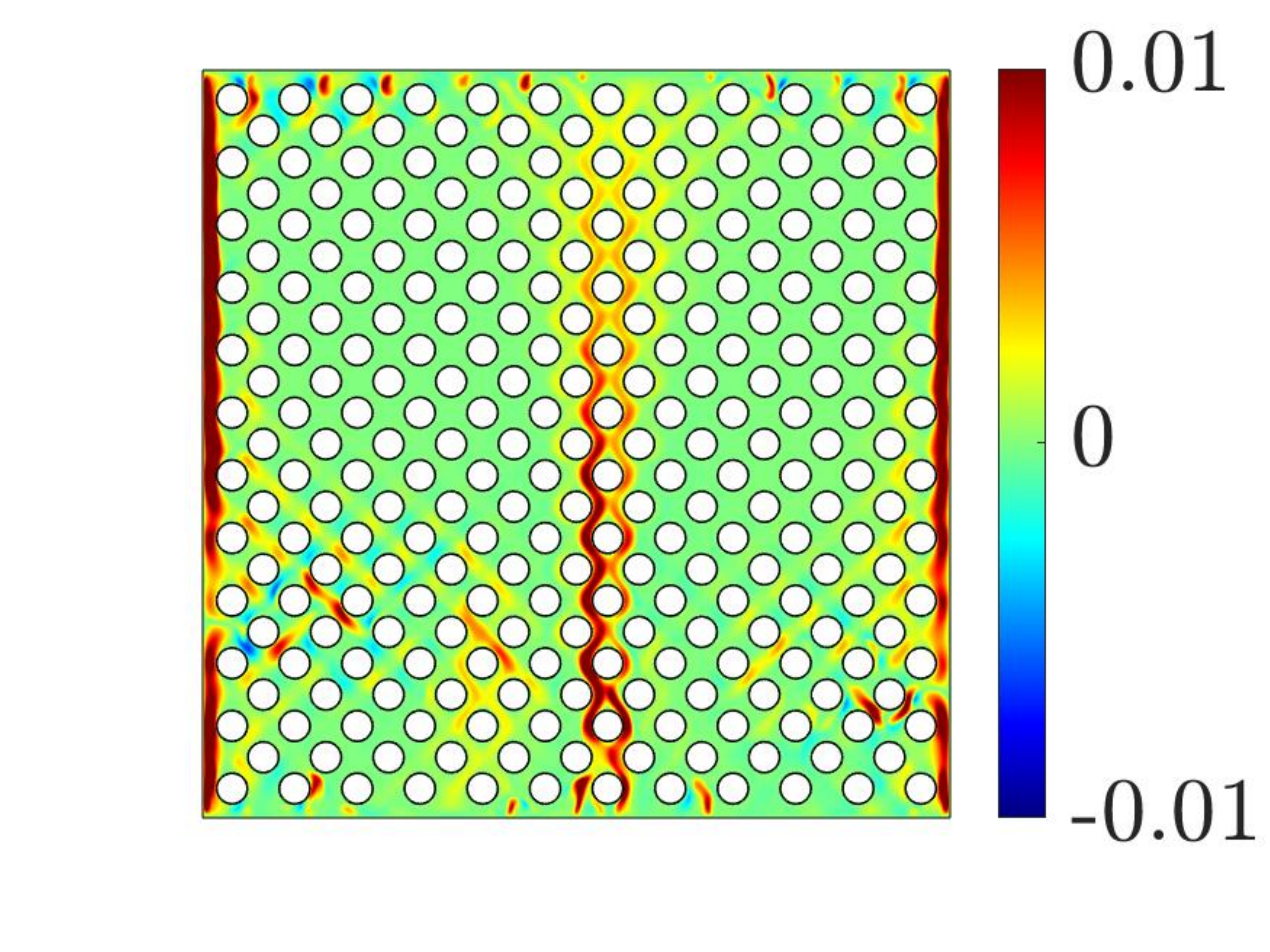}
	\put(-125,81){$(f)$}
	\\
	\includegraphics[width=0.33\linewidth]{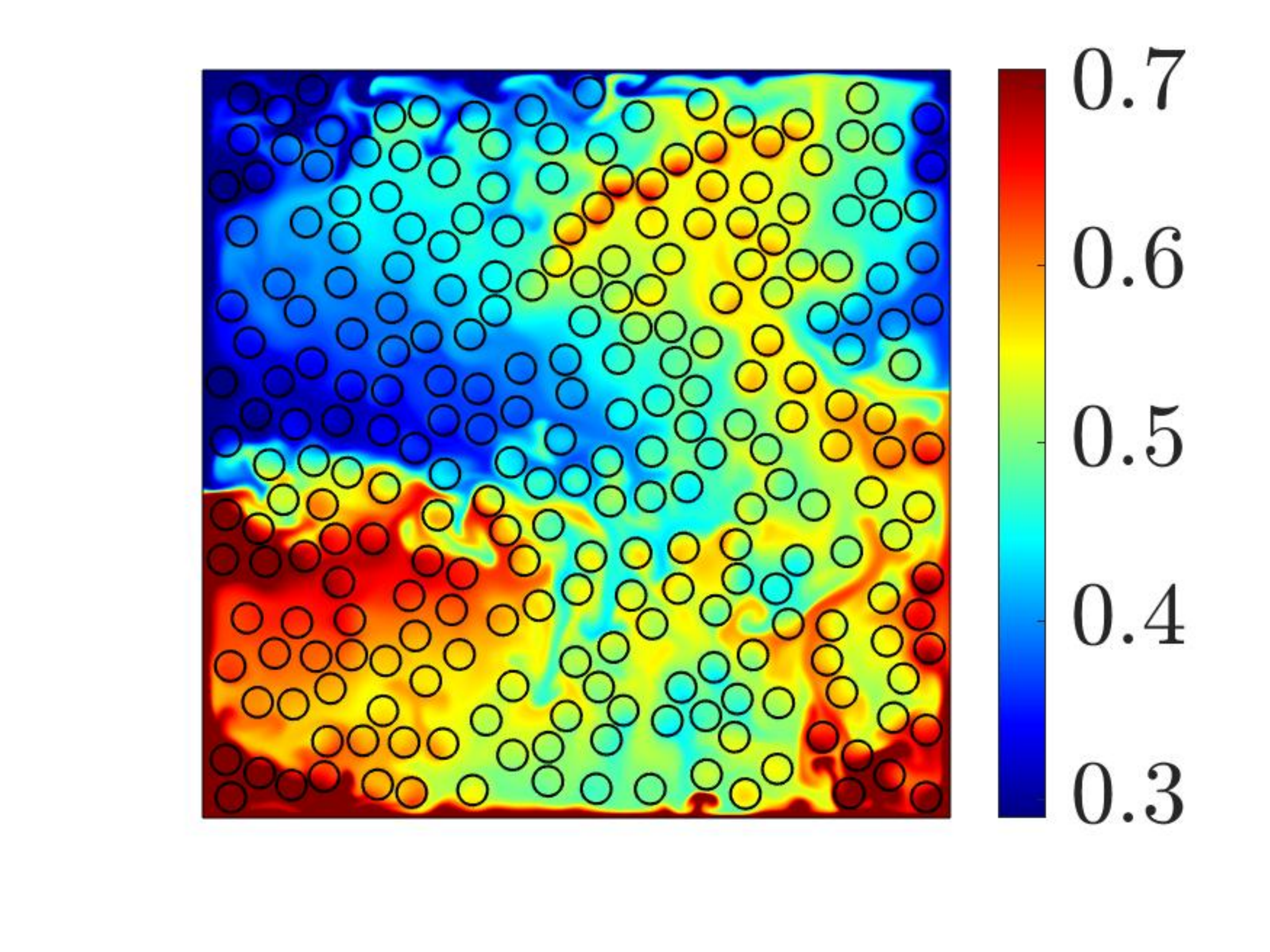}
	\put(-125,81){$(g)$}
	\put(-125,49){\rotatebox{90}{III}}
	\hspace{-1 mm}
	\includegraphics[width=0.33\linewidth]{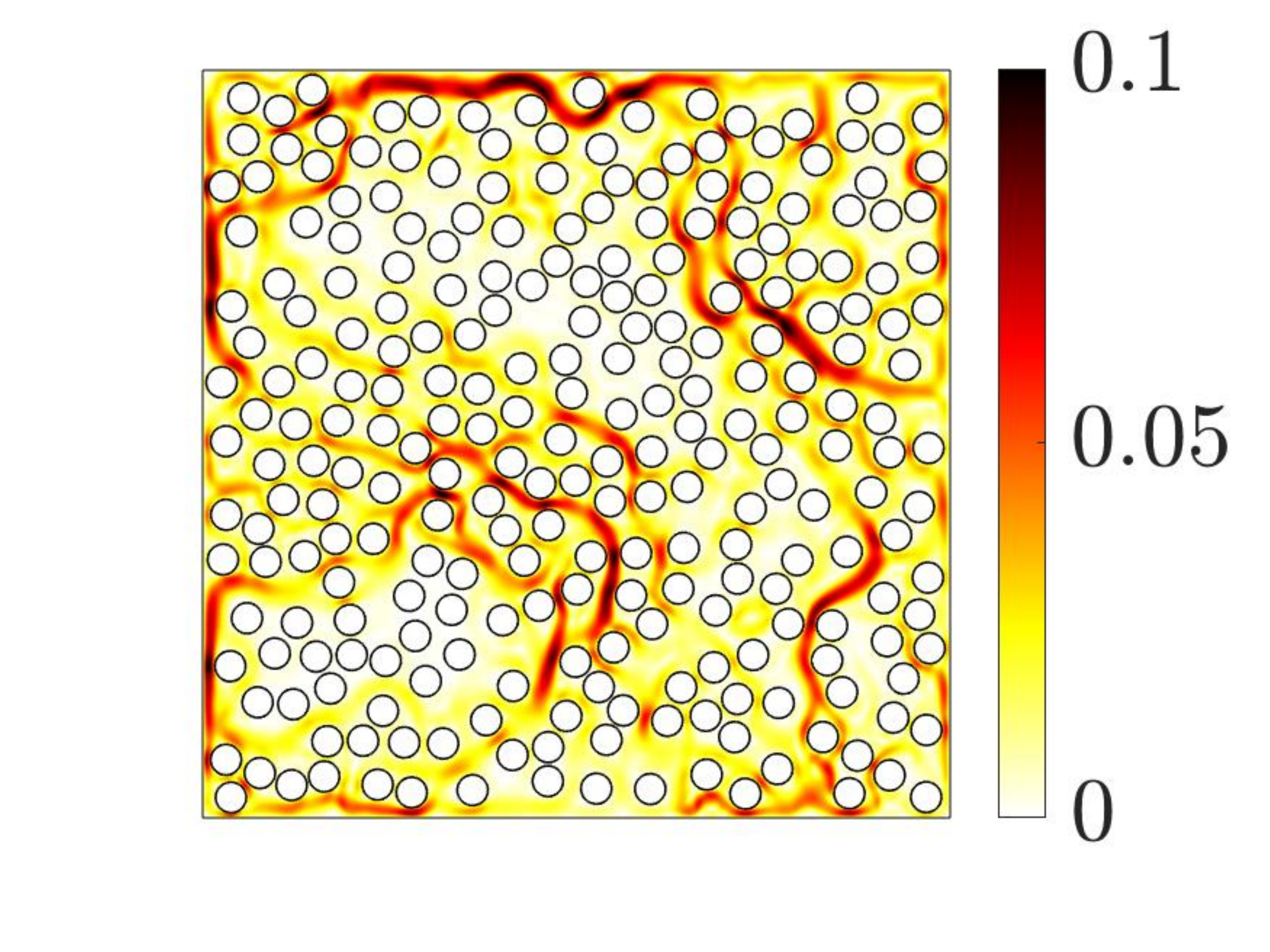}
	\put(-125,81){$(h)$}
	\hspace{-2 mm}
	\includegraphics[width=0.33\linewidth]{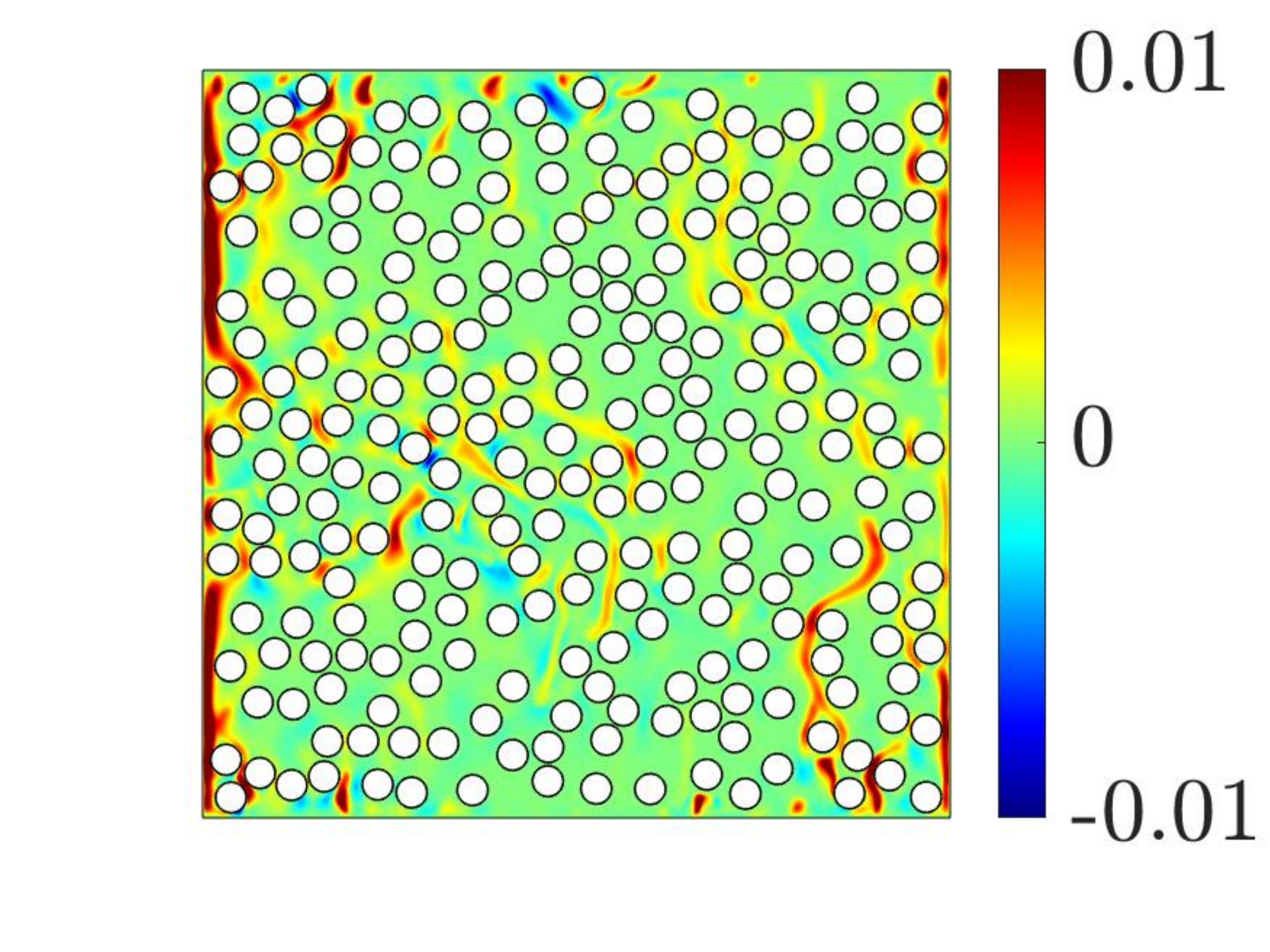}
	\put(-125,81){$(i)$}
	\caption{\label{obstacle_arrangement} {\color{black}Snapshots of $(a,d,g)$ temperature $T$, $(b,e,h)$ velocity magnitude $|\vec{u}|$, and $(c,f,i)$ the convective heat flux $v\cdot\delta T$ in the vertical direction for three different obstacle arrangements with similar porosities at $Ra=10^9$. $(a,b,c)$ Porous medium I consists of obstacles located on a square lattice, and the porosity $\phi=0.68$. $(d,e,f)$ Porous medium II is constructed by rotating the obstacle lattice of I by 45 degrees, with porosity $\phi=0.67$. $(g,h,i)$ Porous medium III is constructed by randomly placing the obstacles in the cell, with porosity $\phi=0.68$.}}
\end{figure}

How robust are our results with respect to the arrangement of the obstacles? To answer this question, we study how the obstacle arrangement influences the flow structure. We consider three different arrangements with similar porosities, i.e. two regular arrangements and one random arrangement. 
Simulations are performed at $Ra=10^9$.
The snapshots of temperature $T$, velocity magnitude $|\vec{u}|$, and the convective heat flux $v\cdot \delta T$ in the vertical direction are shown in figure \ref{obstacle_arrangement}. 
It is seen that the flow structure is significantly influenced by the obstacle arrangement. In figures \ref{obstacle_arrangement}$(a,b,c)$, the porous medium I is constructed by placing the circular obstacles on a square lattice, as done in the rest of this paper. As already discussed, thermal plumes can penetrate up and down without obstruction, forming vertical convection channels with strong fluid and heat transport.
In figures \ref{obstacle_arrangement}$(d,e,f)$, the regular porous medium II is constructed by rotating the lattice of porous medium I by 45 degrees. 
In such an obstacle arrangement, the vertical convection channels observed in porous medium I are obstructed by the obstacles, except the vertical channels close to the sidewalls. Despite that, we find that thermal plumes can penetrate deep into the bulk along zigzag channels aligning vertically.
Besides, strong fluid and heat transport occur close to the sidewalls, where the fluid can penetrate up and down without obstruction.
In figures \ref{obstacle_arrangement}$(g,h,i)$, a random porous medium is constructed (porous medium III), with the pore scale $l\ge l_{min}=0.005$, where $l_{min}$ is the minimum pore scale.
It is observed that the plume motion is less organized compared to those in the regular porous media I and II. Curved convection channels with strong flow emerge in the pores, with no preferred flowing directions. Except the localized vertical channels, the curved convection channels are less efficient at transporting heat in the vertical direction. 
The Nusselt numbers for the obstacle arrangements I, II and III are 51.96, 53.97 and 47.14, respectively. The lower $Nu$ in porous medium III is attributed to the less organized plume motion and less efficient vertical heat transfer.

\section{Energy dissipation rates}\label{sec:sec_energy_dissipation}

RB convection is a closed system with exact balance properties. The overall kinetic and thermal energy dissipation rates $\epsilon_{u,T}$ are related to the governing and response parameters by two exact relations \citep{shraiman1990heat,grossmann2000scaling,ahlers2009heat}:
\begin{equation}
\left\langle\epsilon_{u}\right\rangle_{V, t}=\frac{\nu^3}{L^4}(N u-1)Ra Pr^{-2},~~~ \left\langle\epsilon_{T}\right\rangle_{V, t}=\kappa\frac{\Delta^2}{L^2} N u,
\label{global_balance}
\end{equation}
where $\langle \cdot \rangle_{V,t}$ denotes the volume and time average, and
\begin{equation}
\epsilon_{u} = \frac{1}{2}\nu\sum_{ij}\left[ \frac{\partial u_j}{\partial x_i} + \frac{\partial u_i}{\partial x_j} \right]^{2}, ~~~
\epsilon_{T}=\kappa \sum_{i} \left[ \frac{\partial T}{\partial x_i} \right]^{2}.
\end{equation}
{\color{black}We note that the exact global relations (\ref{global_balance}) are not affected by the presence of the obstacles.}
{\color{black}Figures \ref{kinetic_energy_dissipation_rate} and \ref{thermal_energy_dissipation_rate} plot} typical snapshots of the non-dimensional kinetic and thermal energy dissipation rates $log_{10}\epsilon_{u,T}(\vec{x})$ {\color{black}on logarithmic scale} for $Ra=10^6,~10^8,~10^9$ and $\phi=1,~0.92$.
In the traditional RB convection with $\phi=1$, it is found that the characteristic length scales of the flow structures become smaller for larger $Ra$, as shown in figures \ref{kinetic_energy_dissipation_rate}($a-c$) and \ref{thermal_energy_dissipation_rate}($a-c$). A well-organized LSC develops at sufficiently large $Ra$, and velocity boundary layers appear close to both the horizontal plates and vertical sidewalls, which dominate the dissipation of kinetic energy.
There are no thermal boundary layers along the sidewalls due to the imposed adiabatic condition for temperature, and the dissipation of thermal energy is dominated by plumes in the bulk and the thermal boundary layers near the horizontal top and bottom plates.
The bulk contributions of both the kinetic and thermal energy dissipation rates are small \citep{grossmann2000scaling,grossmann2001thermal,grossmann2004fluctuations,zhang2017statistics}.
Results of convection in a regular porous medium are displayed in figures \ref{kinetic_energy_dissipation_rate}($d-f$) and \ref{thermal_energy_dissipation_rate}($d-f$).
When $Ra$ is small, convection is dominated by large-scale hot and cold plumes, penetrating to the top and bottom plates, respectively, along convection channels with fast velocities. Along these channels intense dissipation of kinetic energy occurs with length scale characterized by the pore scale.
As $Ra$ is increased, the length scales of the flow structures become smaller, and plumes wander in the bulk, leading to intense dissipation of kinetic energy in the bulk, in distinct contrast to the results of traditional RB convection with $\phi=1$. For sufficiently large $Ra$, the kinetic energy dissipation in the bulk is concentrated around the obstacle surfaces due to the interaction between plumes and the obstacle array, and the pore scale is no longer a proper length scale for the kinetic energy dissipation.
Due to the wandering motion of plumes in regular porous media, intense dissipation of thermal energy can occur locally in the bulk. However, {\color{black}the averaged bulk contribution} of thermal energy dissipation remains small as compared to the boundary-layer contribution.

\begin{figure}
	\centering
	\vspace{3 mm}
	\includegraphics[width=0.33\linewidth]{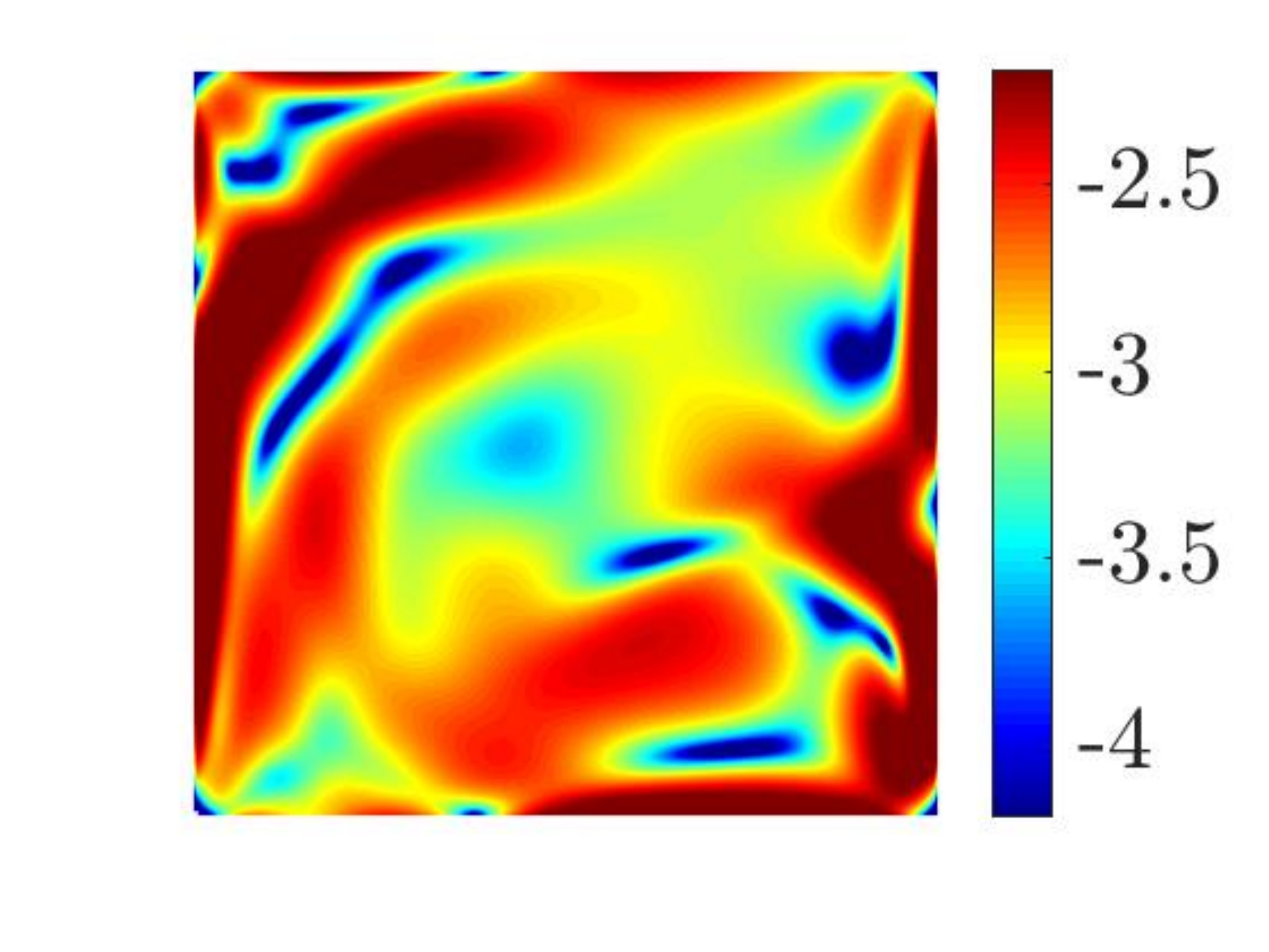}
	\put(-125,81){$(a)$}
	\put(-84,95){$Ra=10^6$}
	\put(-125,40){\rotatebox{90}{$\phi=1$}}
	\includegraphics[width=0.33\linewidth]{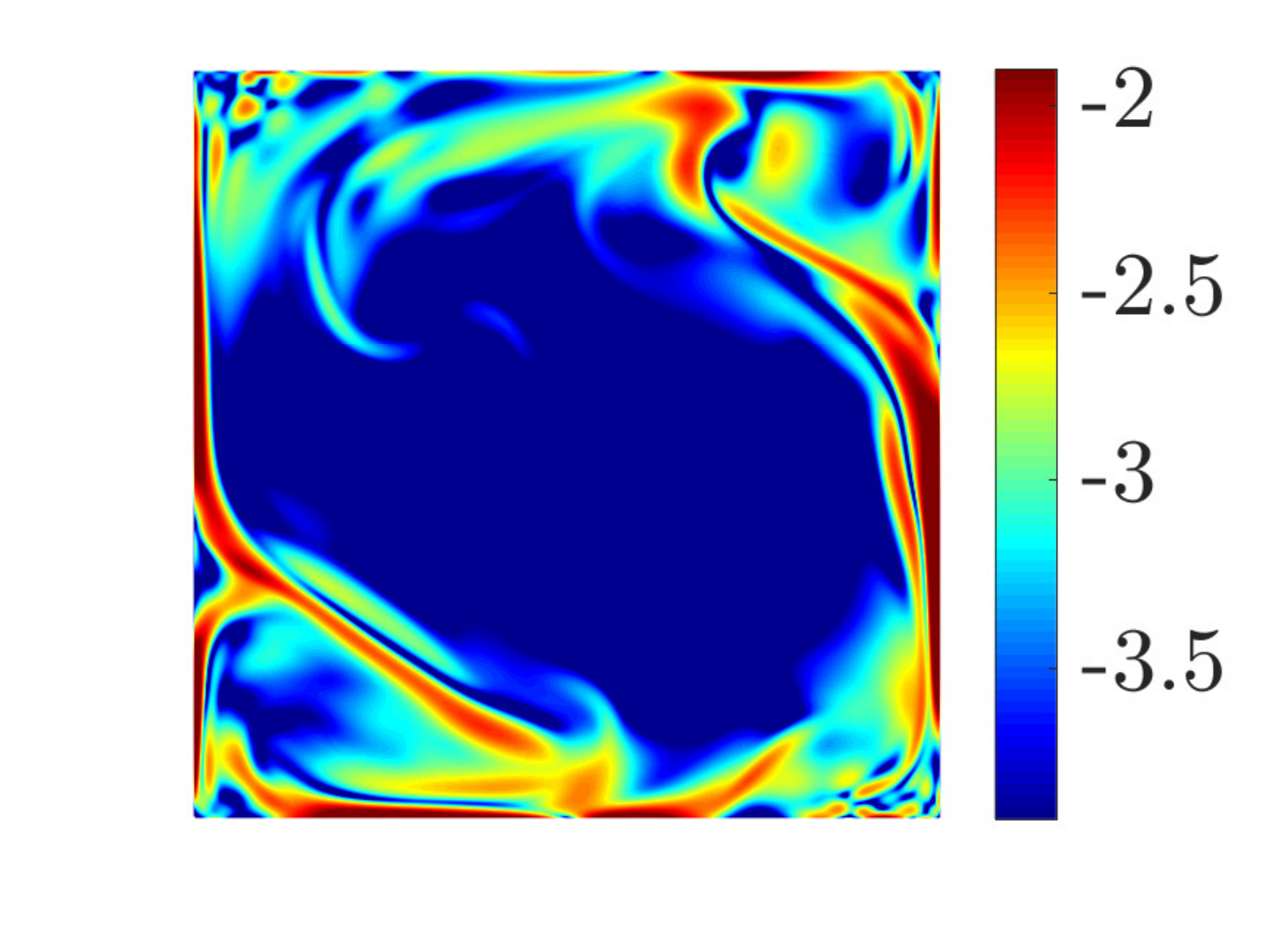}
	\put(-125,81){$(b)$}
	\put(-84,95){$Ra=10^8$}
	\includegraphics[width=0.33\linewidth]{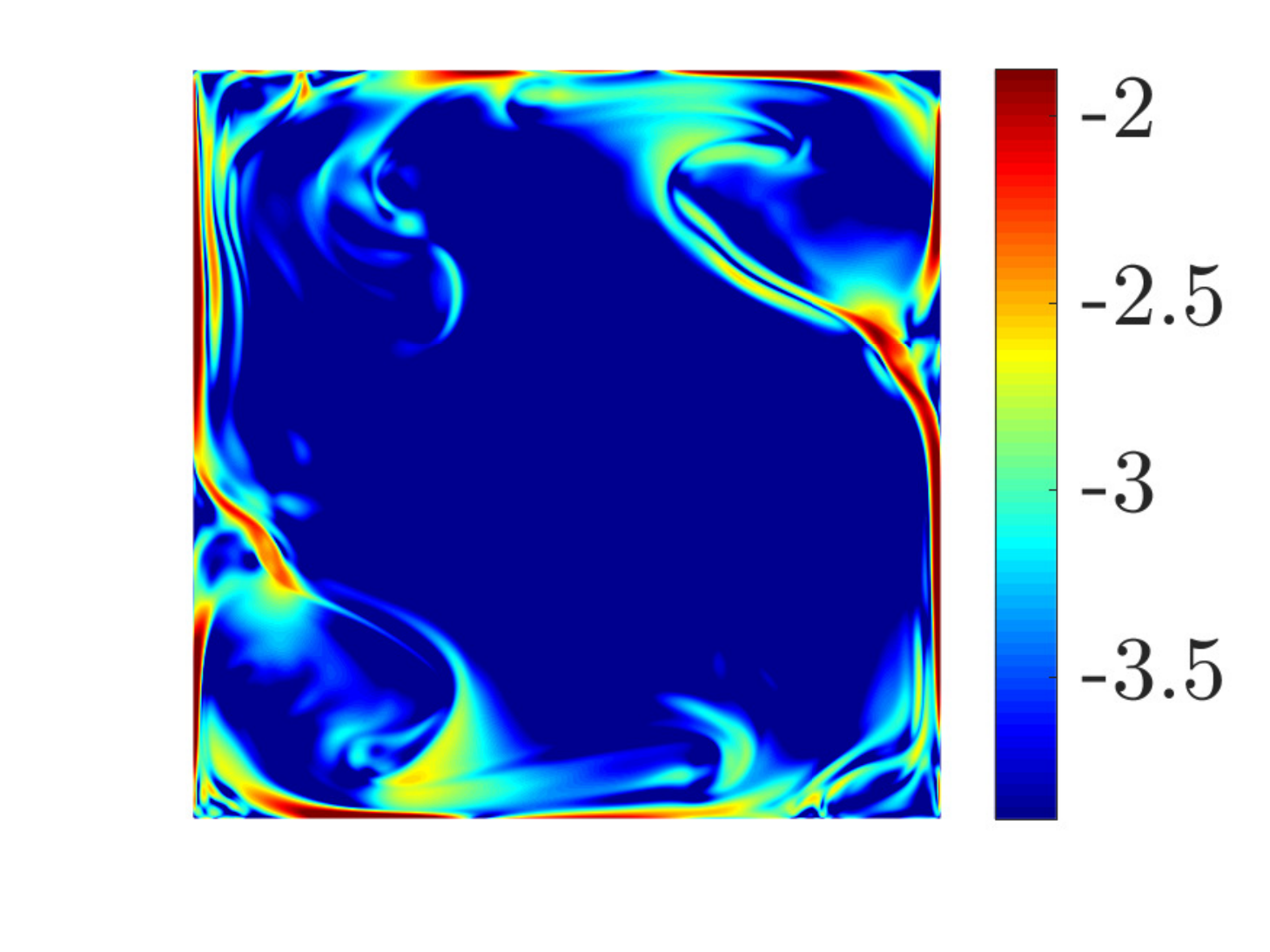}
	\put(-125,81){$(c)$}
	\put(-84,95){$Ra=10^9$}
	\\
	\includegraphics[width=0.33\linewidth]{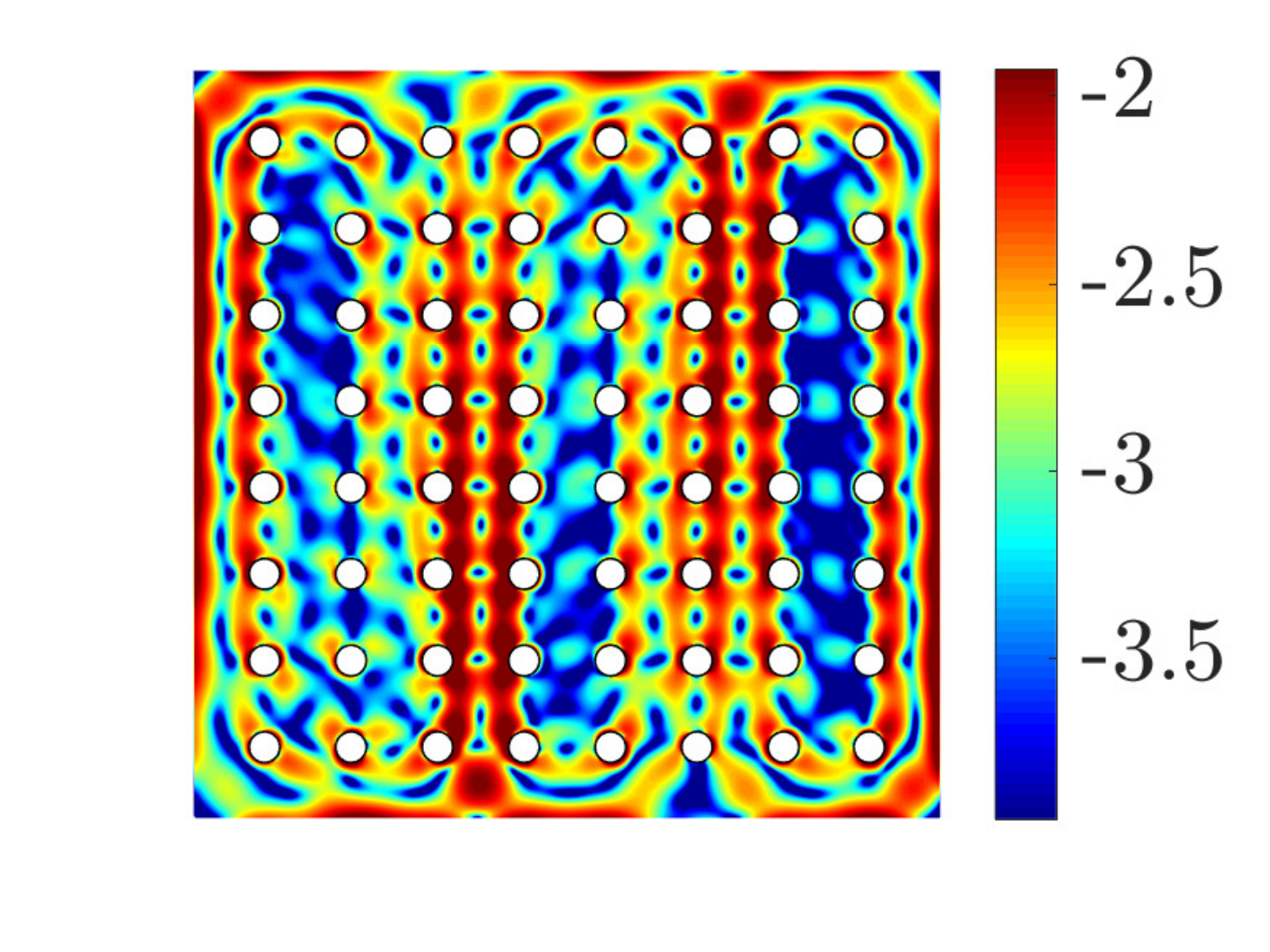}
	\put(-125,81){$(d)$}
	\put(-125,34){\rotatebox{90}{$\phi=0.92$}}
	\includegraphics[width=0.33\linewidth]{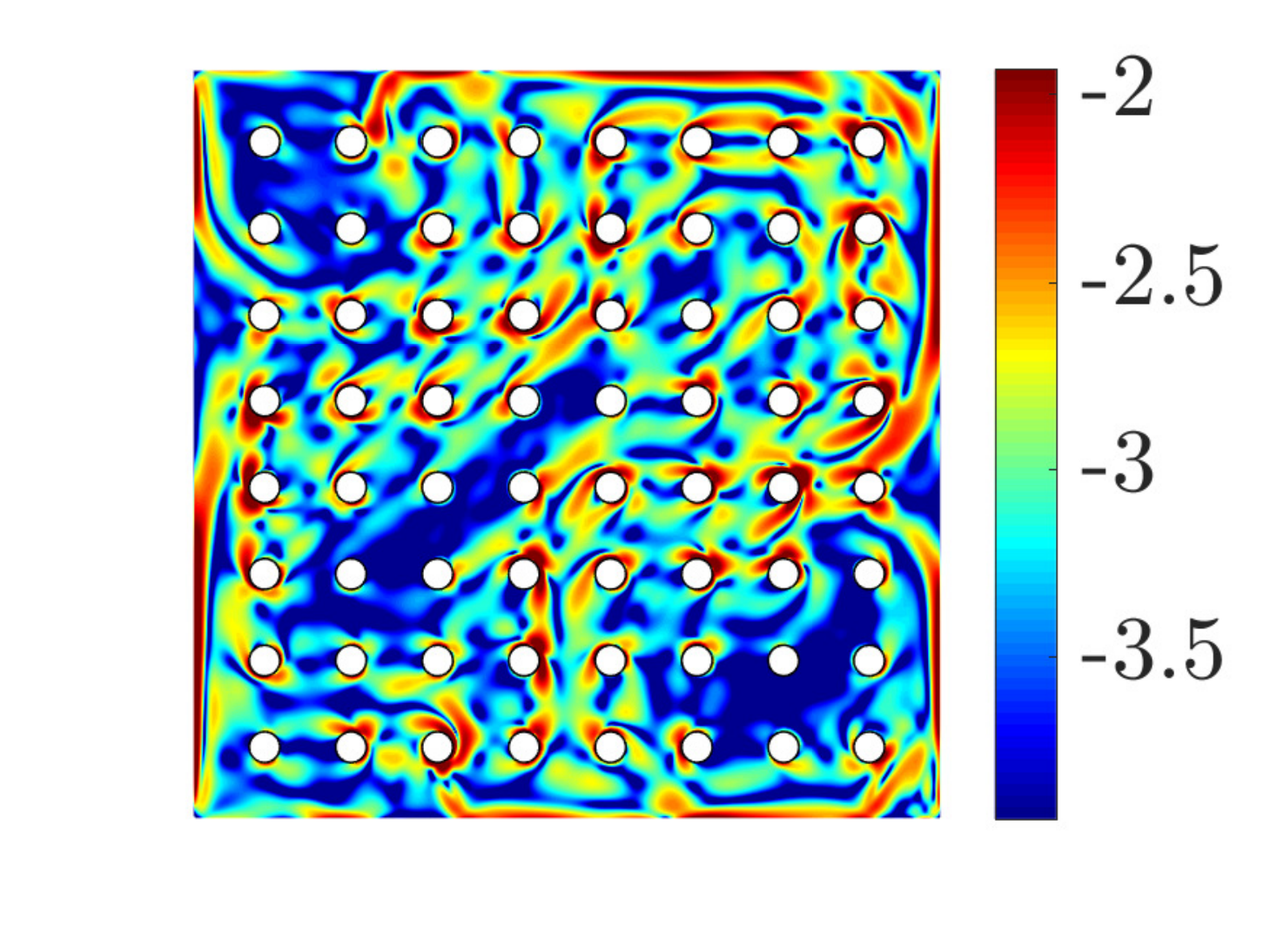}
	\put(-125,81){$(e)$}
	\includegraphics[width=0.33\linewidth]{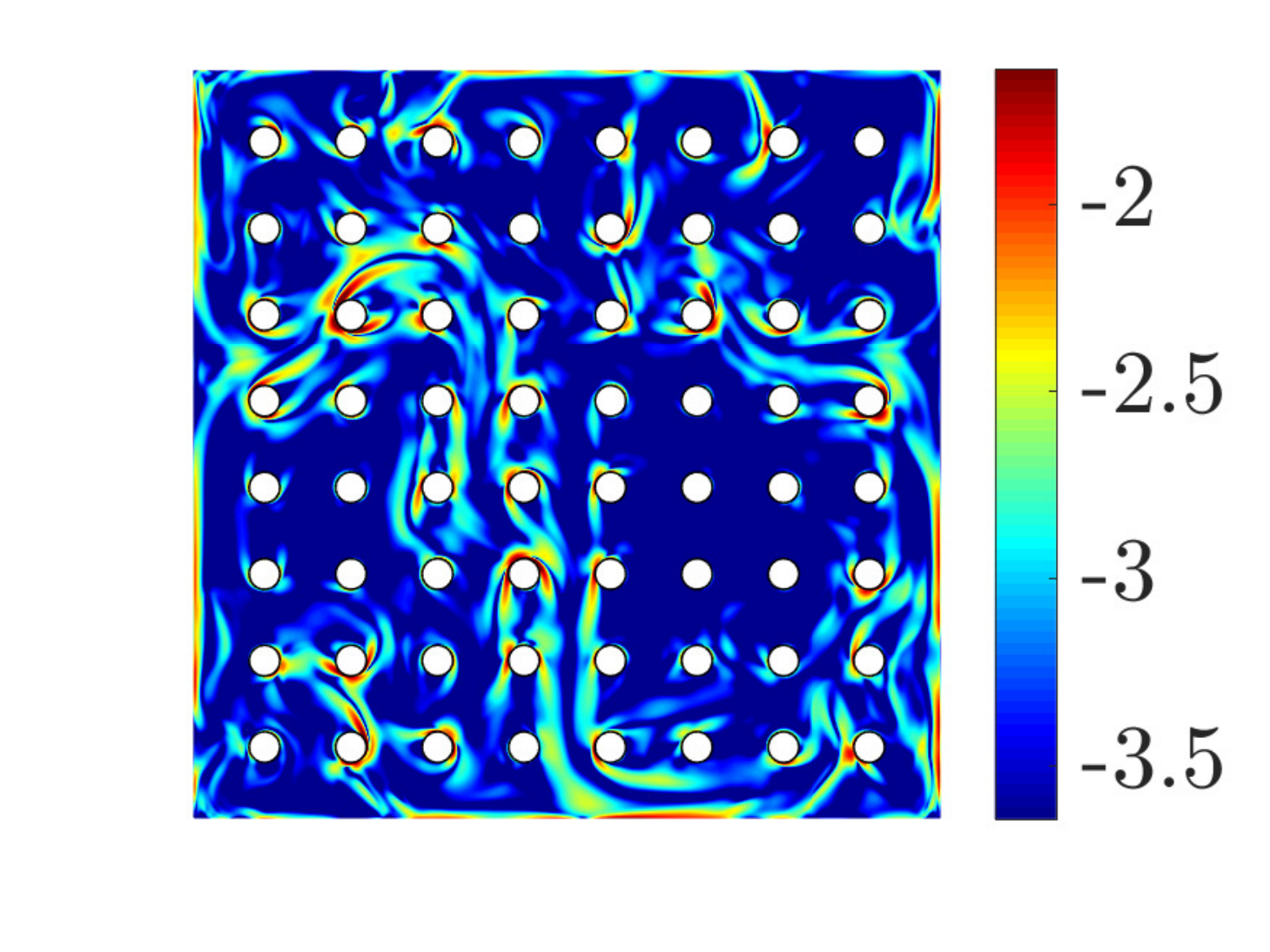}
	\put(-125,81){$(f)$}
	\caption{\label{kinetic_energy_dissipation_rate} Typical snapshots of the kinetic energy dissipation rates $log_{10}\epsilon_{u}(\vec{x})$ {\color{black}on logarithmic scale} for different $Ra$ and $\phi$. $Ra=10^6,~10^8,~10^9$ for the figures in the left, middle, and right columns, respectively. ($a-c$) $\phi=1$, ($d-f$) $\phi=0.92$.}
\end{figure}

\begin{figure}
	\centering
	\vspace{3 mm}
	\includegraphics[width=0.33\linewidth]{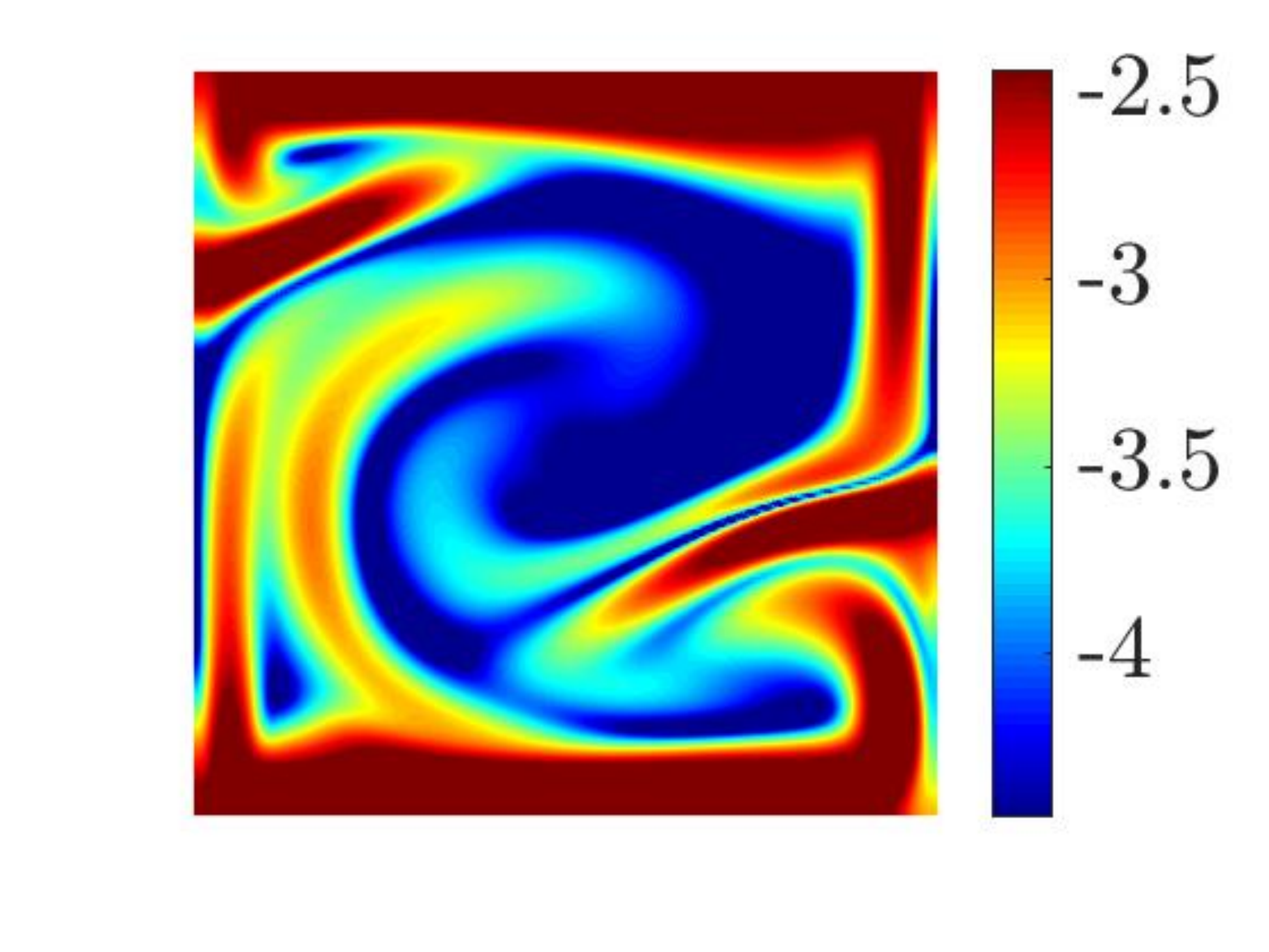}
	\put(-125,81){$(a)$}
	\put(-84,95){$Ra=10^6$}
	\put(-125,40){\rotatebox{90}{$\phi=1$}}
	\includegraphics[width=0.33\linewidth]{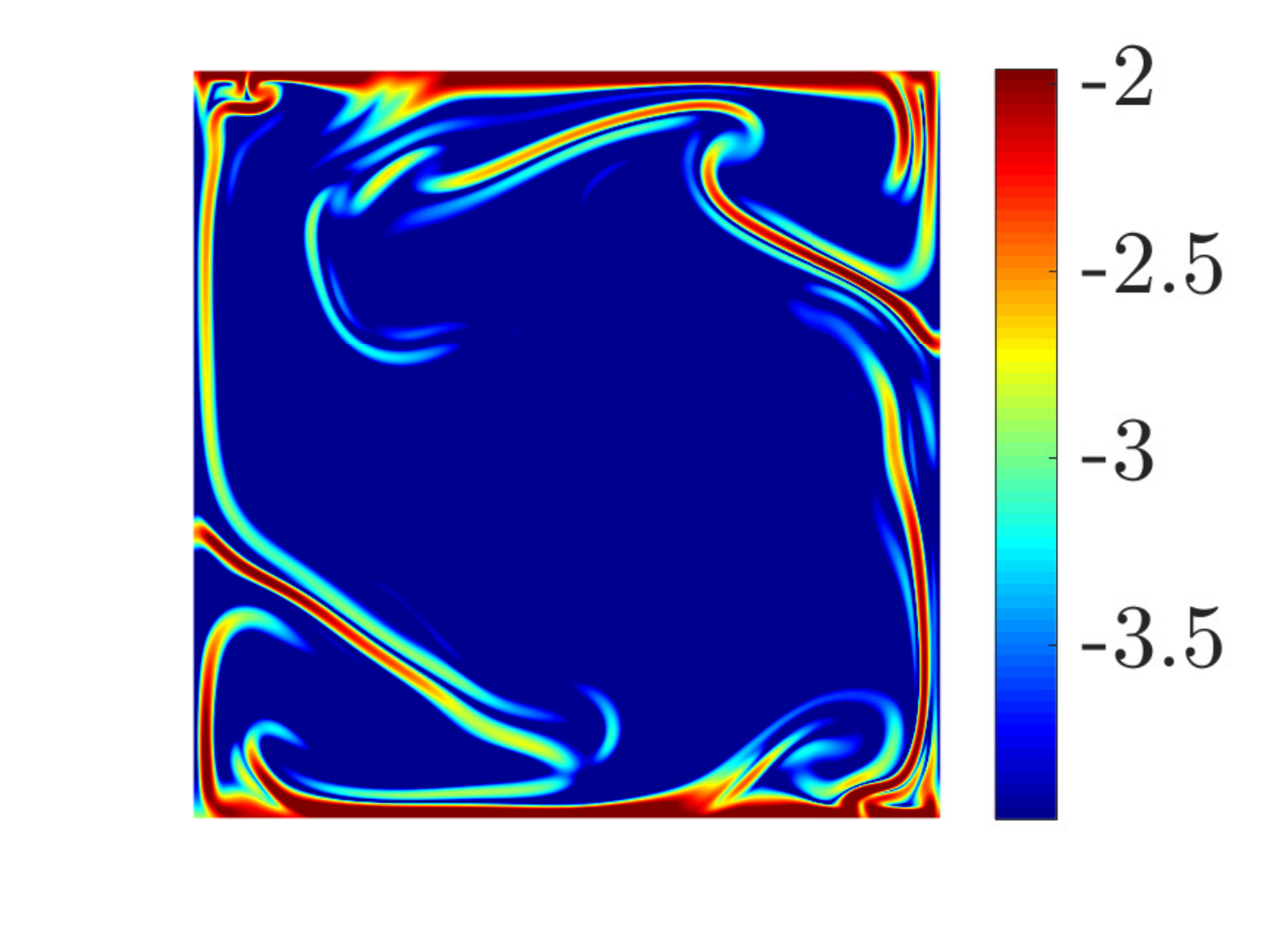}
	\put(-125,81){$(b)$}
	\put(-84,95){$Ra=10^8$}
	\includegraphics[width=0.33\linewidth]{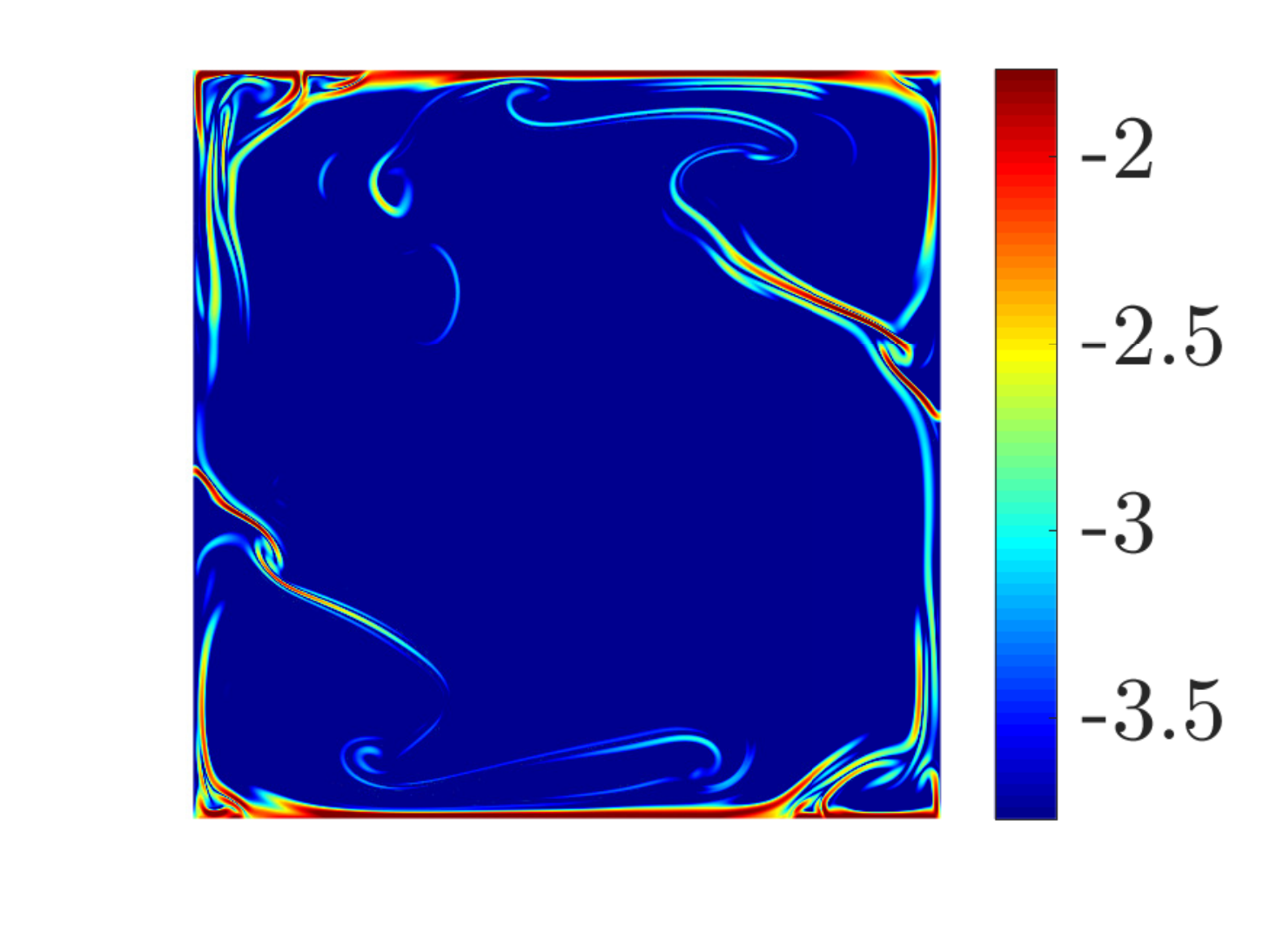}
	\put(-125,81){$(c)$}
	\put(-84,95){$Ra=10^9$}
	\\
	\includegraphics[width=0.33\linewidth]{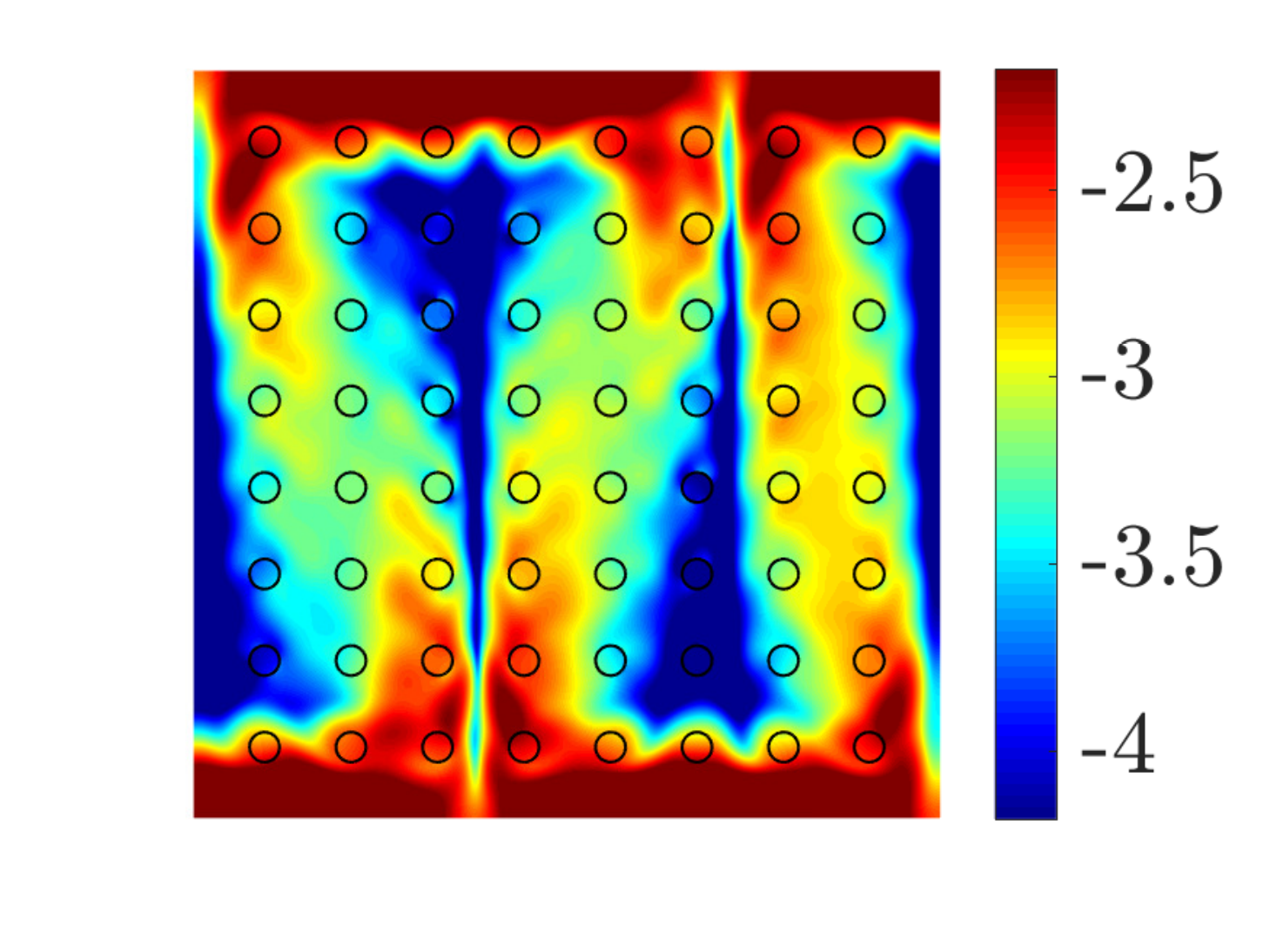}
	\put(-125,81){$(d)$}
	\put(-125,34){\rotatebox{90}{$\phi=0.92$}}
	\includegraphics[width=0.33\linewidth]{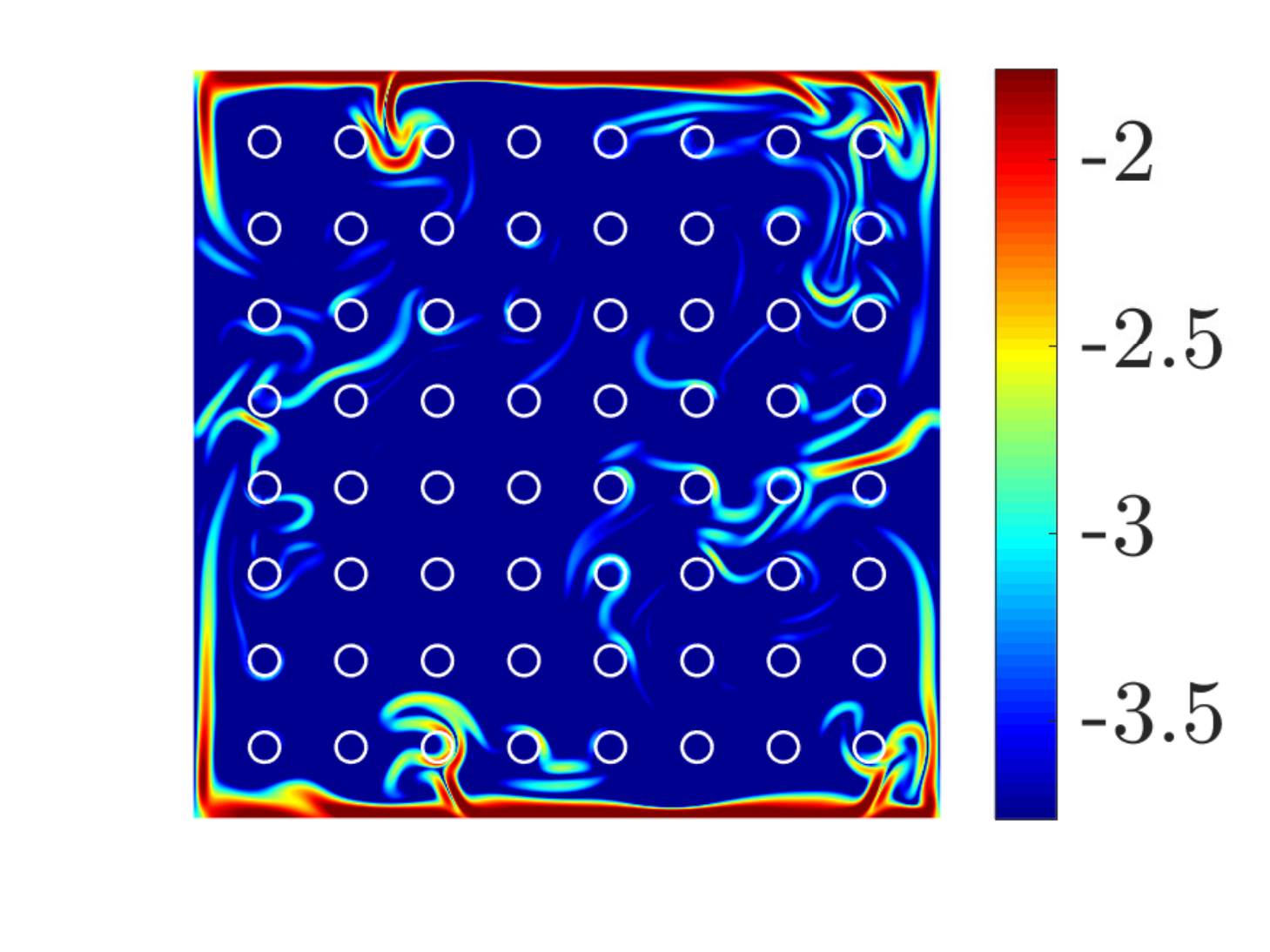}
	\put(-125,81){$(e)$}
	\includegraphics[width=0.33\linewidth]{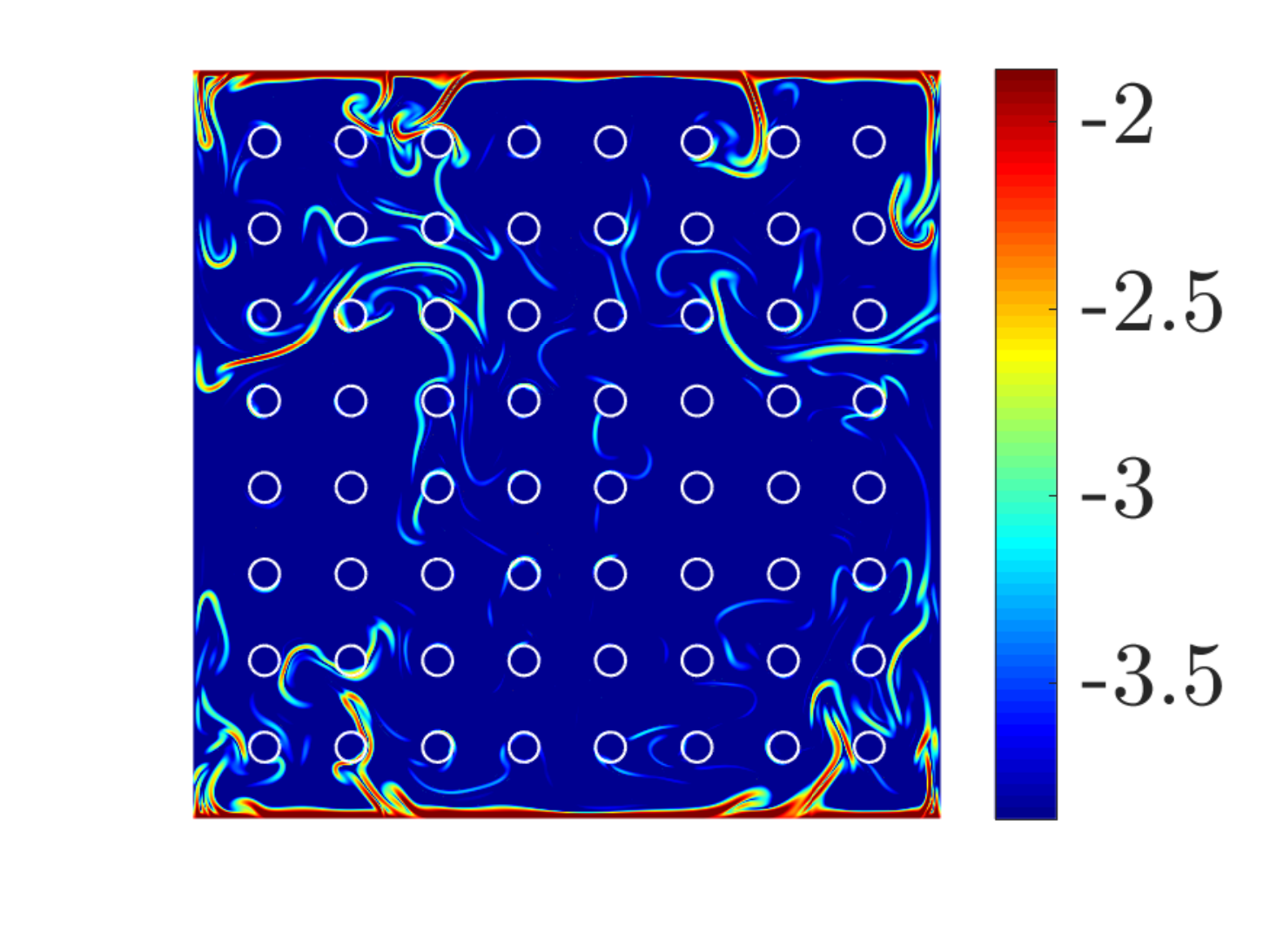}
	\put(-125,81){$(f)$}
	\caption{\label{thermal_energy_dissipation_rate} Typical snapshots of the thermal energy dissipation rates $log_{10}\epsilon_{T}(\vec{x})$ {\color{black}on logarithmic scale} for different $Ra$ and $\phi$. $Ra=10^6,~10^8,~10^9$ for the figures in the left, middle, and right columns, respectively. ($a-c$) $\phi=1$, ($d-f$) $\phi=0.92$. In $(e)$ and $(l)$, we show the obstacles as white thin circles. Note that the temperature field in them is well defined according to equation (\ref{temperature_eqn}), and correspondingly the thermal energy dissipation rate $\epsilon_T(\vec{x},t)$ is non-zero at these locations.}
\end{figure}

\begin{figure}
	\centering
	\includegraphics[width=0.45\linewidth]{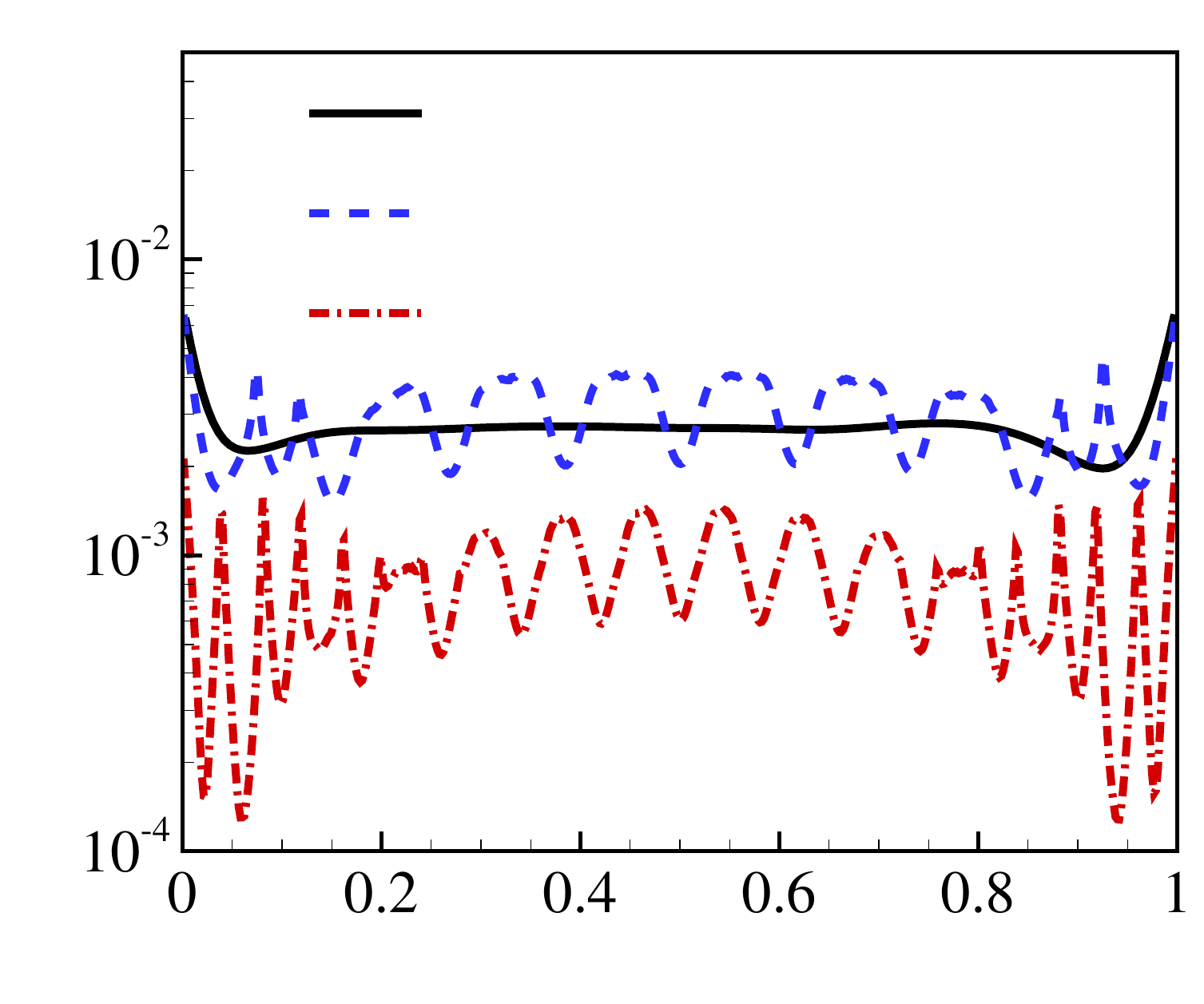}
	\put(-175,130){$(a)$}
	\put(-172,65){\rotatebox{90}{$\langle\epsilon_u\rangle_{x,t}$}}
	\put(-79,4){$z$}
	\put(-108,125){$\phi=1$}
	\put(-108,111){$\phi=0.92$}
	\put(-108,96.5){$\phi=0.82$}
	\hspace{2 mm}
	\includegraphics[width=0.45\linewidth]{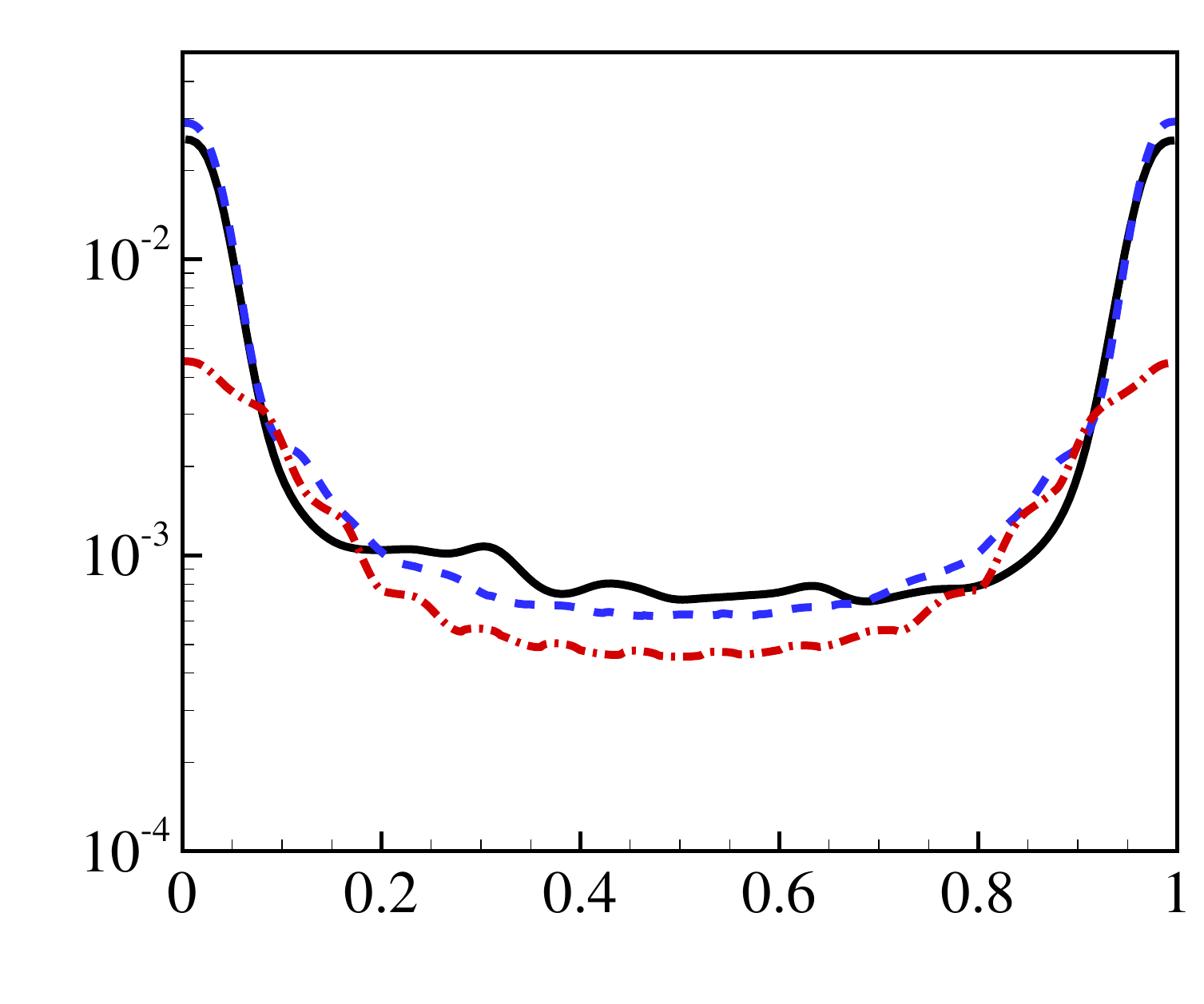}
	\put(-175,130){$(b)$}
	\put(-172,65){\rotatebox{90}{$\langle\epsilon_T\rangle_{x,t}$}}
	\put(-79,4){$z$}
	\caption{\label{dissipaiton_rate_porosity} Variations of the averaged ($a$) kinetic and ($b$) thermal energy dissipation rates $\langle\epsilon_{u,T}\rangle_{x,t}$ with $z$ for different $\phi$ at $Ra=10^6$.}
\end{figure}

\begin{figure}
	\centering
	\includegraphics[width=0.45\linewidth]{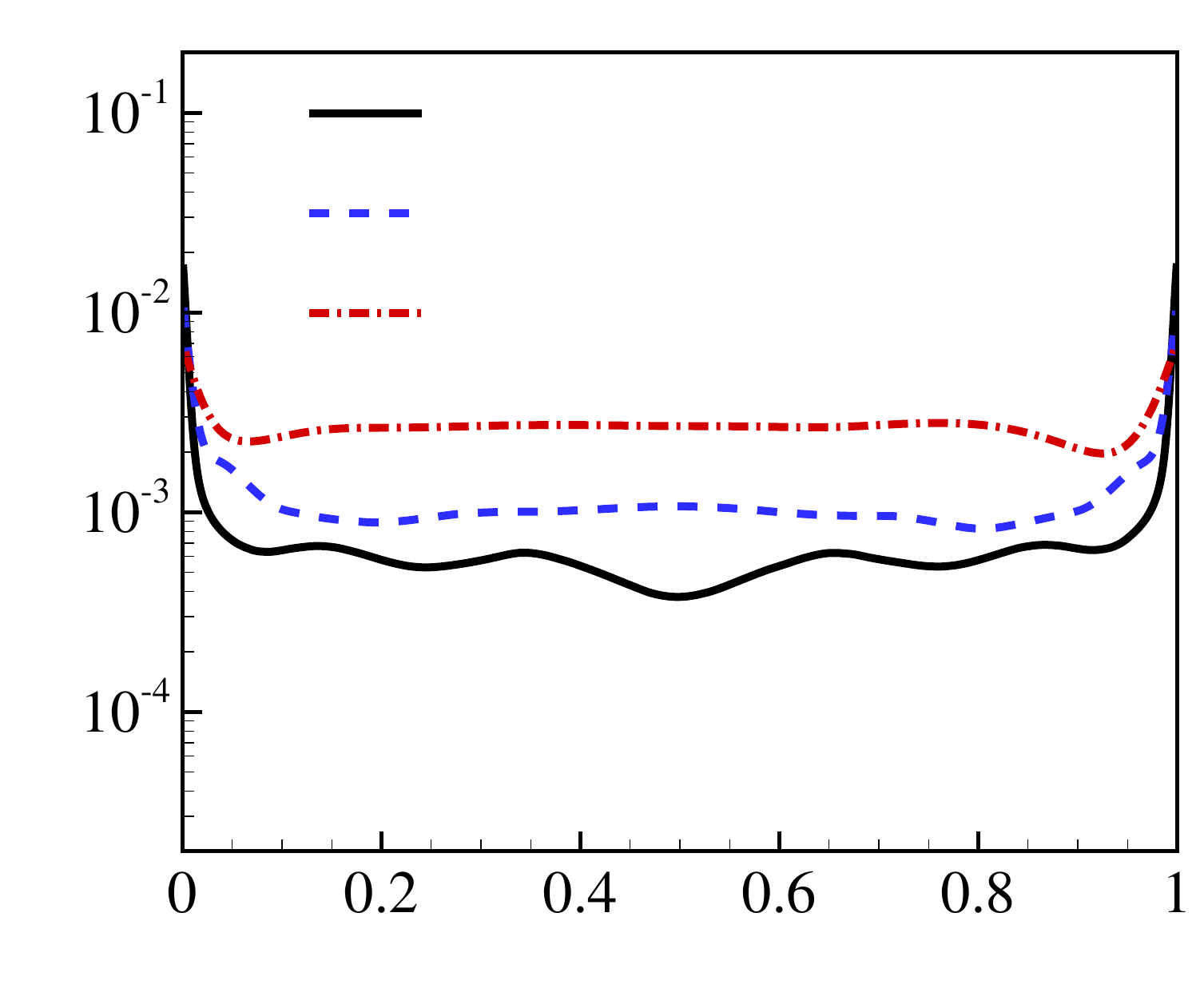}
	\put(-175,130){$(a)$}
	\put(-172,65){\rotatebox{90}{$\langle\epsilon_u\rangle_{x,t}$}}
	\put(-79,4){$z$}
	\put(-108,125){{\small $Ra=10^9$}}
	\put(-108,111){{\small $Ra=10^8$}}
	\put(-108,96){{\small $Ra=10^6$}}
	%	\put(-52,111){{\small $10^5$}}
	\hspace{2 mm}
	\includegraphics[width=0.45\linewidth]{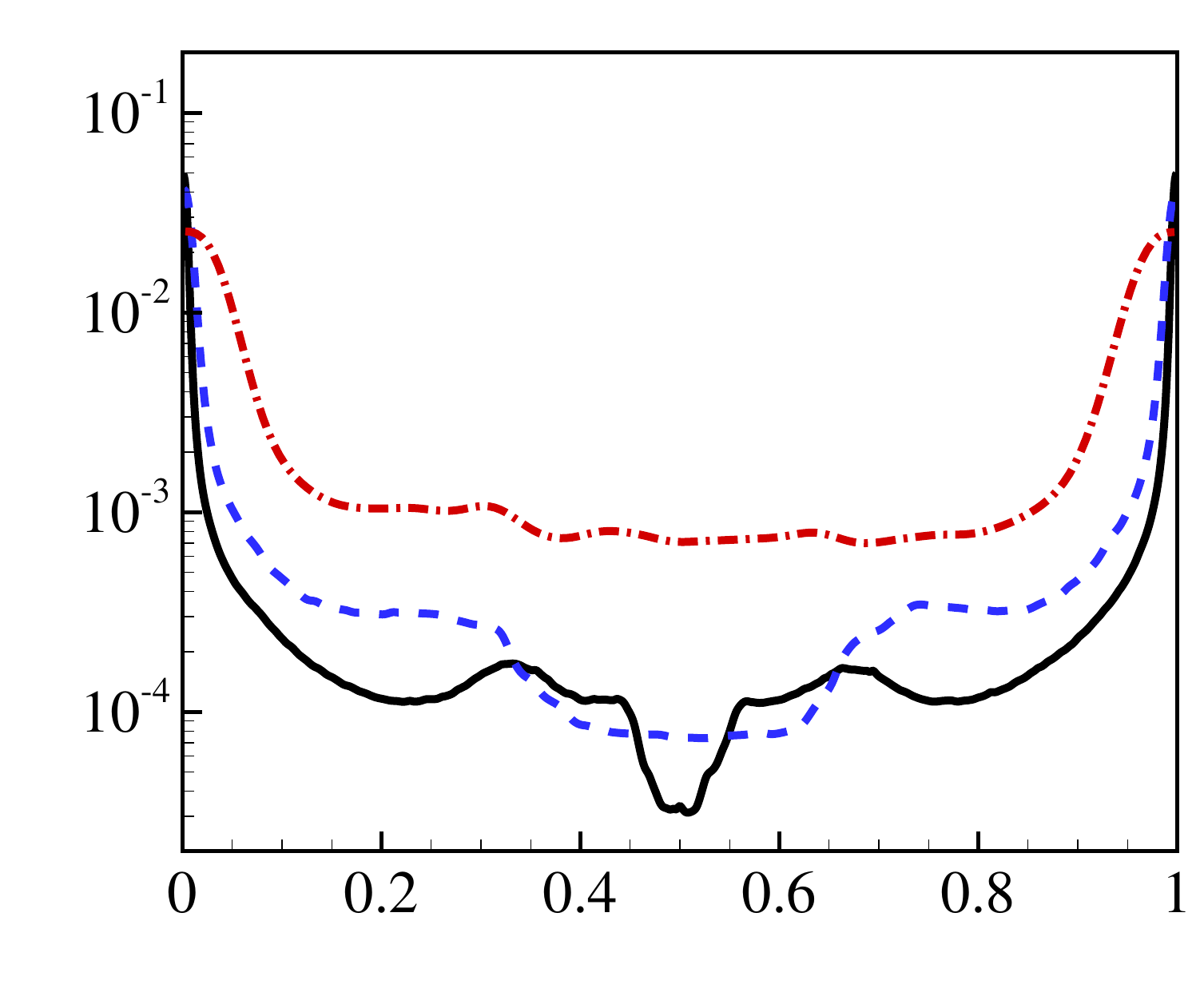}
	\put(-175,130){$(b)$}
	\put(-172,65){\rotatebox{90}{$\langle\epsilon_T\rangle_{x,t}$}}
	\put(-79,4){$z$}
	\\
	\includegraphics[width=0.45\linewidth]{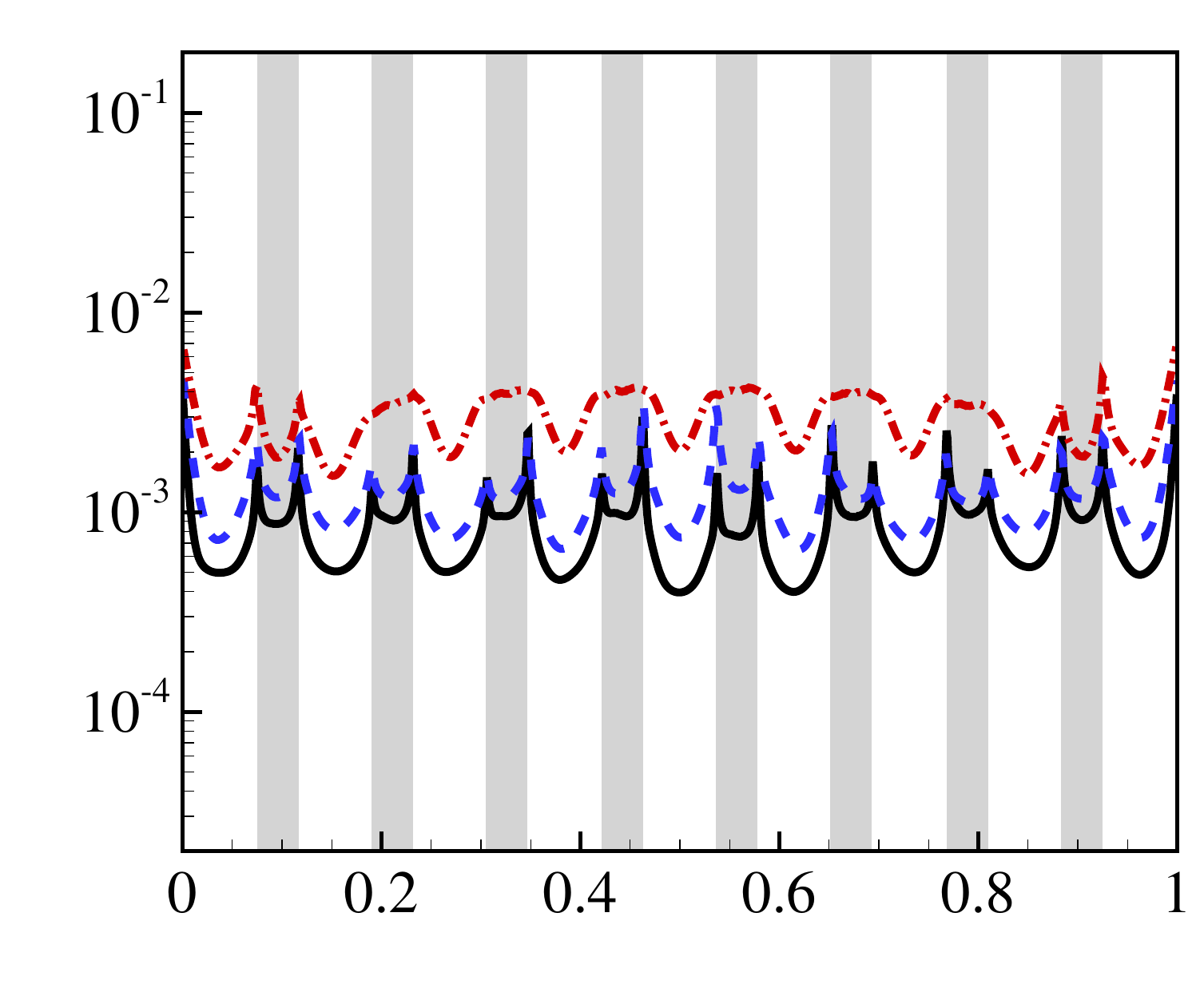}
	\put(-175,130){$(c)$}
	\put(-172,65){\rotatebox{90}{$\langle\epsilon_u\rangle_{x,t}$}}
	\put(-79,4){$z$}
	\hspace{2 mm}
	\includegraphics[width=0.45\linewidth]{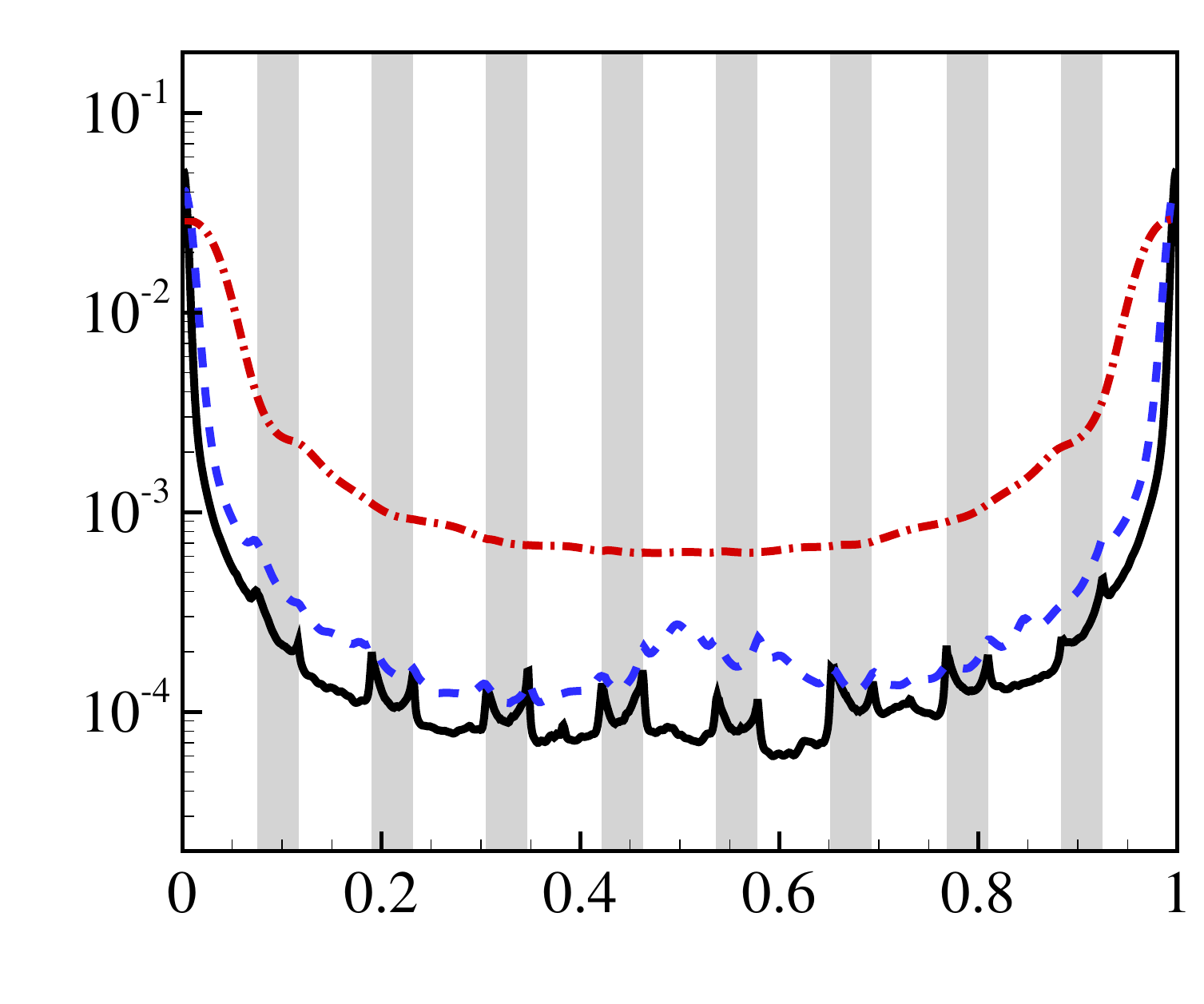}
	\put(-175,130){$(d)$}
	\put(-172,65){\rotatebox{90}{$\langle\epsilon_T\rangle_{x,t}$}}
	\put(-79,4){$z$}
	\caption{\label{dissipation_rate_Ra} Variations of the averaged ($a,c$) kinetic and ($b,d$) thermal energy dissipation rates $\langle\epsilon_{u,T}\rangle_{x,t}$ with $z$ for different $Ra$ at $(a,b)$ $\phi=1$ and $(c,d)$ $\phi=0.92$. In ($c,d$) the grey areas indicate obstacle positions.}
\end{figure}

To further quantify the spatial distributions of the kinetic and thermal energy dissipation rates, we plot the profiles of averaged energy dissipation rates $\langle\epsilon_{u,T}\rangle_{x,t}$ along the vertical direction $z$ in figure \ref{dissipation_rate_Ra}, showing the contributions of the bulk and boundary-layer regions.
 {\color{black} Using equations (\ref{global_balance}) $Nu$ can be obtained by integrating these profiles of $\langle\epsilon_{u,T}\rangle_{x,t}$.}
 Figure \ref{dissipation_rate_Ra}($a,b$) plots the results of traditional RB convection with $\phi=1$. The contributions of the boundary-layer region dominate for both the kinetic and thermal energy dissipation rates, consistent with the observations in figures \ref{kinetic_energy_dissipation_rate} and \ref{thermal_energy_dissipation_rate}.
 {\color{black}From figure \ref{dissipation_rate_Ra}($a$) it seems to be less obvious that the kinetic energy dissipation is dominated by regions close to the walls. We note that $\langle\epsilon_{u}\rangle_{x,t}$ contains the kinetic energy dissipation from the velocity boundary layers near the sidewalls. Contrary to the velocity, the temperature field has no boundary layers close to the sidewalls, which are adiabatic.}
The results of convection in regular porous media are shown in figures \ref{dissipation_rate_Ra}($c,d$). For these cases the obstacle array has a significant influence on the distribution of $\langle\epsilon_{u}\rangle_{x,t}$ {\color{black}due to the no-slip condition at the pores}, while $\langle\epsilon_{T}\rangle_{x,t}$ is only mildly affected.
The relative contribution of the bulk region of $\epsilon_{u}$ is significantly enhanced, while $\epsilon_{T}$ remains boundary-layer dominated.
For relatively large $Ra$, intense dissipation of kinetic energy occurs around the obstacle surfaces.
These observations are also consistent with those seen in figures \ref{kinetic_energy_dissipation_rate} and \ref{thermal_energy_dissipation_rate}.

In figure \ref{optimal_spacing} it is shown that the heat transfer is enhanced when $\phi$ is slightly decreased from 1 at fixed $Ra$, and further decrease of $\phi$ reduces the heat transfer compared with the $\phi=1$ case.
To show the influence of the obstacle array on the heat transfer from a local perspective, we plot the profiles of $\langle\epsilon_{u,T}\rangle_{x,t}$ for various $\phi$ at $Ra=10^6$ in figure \ref{dissipaiton_rate_porosity}.
As $\phi$ is slightly decreased from 1, the kinetic energy dissipation is enhanced in the bulk, which is the local manifestation of heat transfer enhancement due to the obstacle array.
When $\phi$ is further decreased, the pore scale decreases accordingly and convection is significantly slowed down due to the drag of the obstacle array. Correspondingly, $\langle\epsilon_{u}\rangle_{x,t}$ is decreased in both the boundary-layer and the bulk regions, and $\langle\epsilon_{T}\rangle_{x,t}$ is also decreased, particularly in the boundary-layer region. %These modifications of $\langle\epsilon_{u,T}\rangle_{x,t}$ are the local manifestation of heat transfer reduction at small enough $\phi$. %The critical $\phi$ for the crossover of two heat transfer regimes decreases as $Ra$ is increased.
Due to the smallness of the spatial coherence length and strong convection at large $Ra$, smaller porosity (or pore scale) is needed to significantly slow down the convection and reduce the overall energy dissipation rates.

For fixed $\phi$, $Nu$ increases with $Ra$ with an effective power law $Nu\sim Ra^{0.65}$ when $Ra$ is relatively small, and it changes to $Nu\sim Ra^{0.30}$ for large enough $Ra$.
For small $Ra$ with the spatial coherence length larger than the pore scale, turbulence is suppressed in the pores.
{\color{black}In the presence of the obstacle array one basically gets additional laminar-type boundary layers in the bulk region, namely, the laminar shear flows in the pores}, {\color{black} and viscosity dominates such that the flow becomes of Prandtl-Blasius-Pohlhausen type.} Intense dissipation of kinetic energy occurs along the convection channels.
Thus, the overall kinetic energy dissipation rate can be estimated as $\langle\epsilon_u\rangle_{V,t}\sim \phi\nu U_{rms}^2/l^2$, {\color{black}corresponding to the so-called $\infty$-regime in \cite{grossmann2001thermal} for large $Pr$ and small $Ra$, but with $l$ replacing the box size $L$}. Based on the exact relation between $\langle\epsilon_u\rangle_{V,t}$ and $Nu$, we can further estimate $Nu$ as 
\begin{equation}
\centering
N u \approx c \cdot \phi\left(\frac{L}{l}\right)^{4} Pr^{2} Re^{2} R a^{-1}+1,
\label{Nu_estimate}
\end{equation}
where $Re=U_{rms} l/\nu$ and $c$ is an undetermined constant \citep{grossmann2000scaling,grossmann2001thermal,grossmann2004fluctuations}.
Plots of $Re(Ra)$ in the small-$Ra$ regime are shown in figure \ref{nu_based_on_velo}$(a)$ for four fixed $\phi$, and then $Nu$ can be estimated using equation (\ref{Nu_estimate}). With $c=8.0$, it is found that $Nu$ obtained using equation (\ref{Nu_estimate}) is consistent with the results based on the definition of $Nu$ for different $Ra$ and $\phi$, as shown in figure \ref{nu_based_on_velo}$(b)$, {\color{black}which provides additional confirmation of our observations regarding the modifications of the heat transfer and flow structure by the porous media}. %In the high-$Ra$ regime the scaling exponent of traditional RB convection ($\approx 0.30$) is recovered. 
As $Ra$ is increased, the characteristic scales of the flow structures become smaller and intense dissipation of kinetic energy tends to occur around the obstacle surfaces. For large enough $Ra$, convection in the pores becomes chaotic or even turbulent, the pore scale is no longer a proper length scale for the kinetic energy dissipation rate, {\color{black}and the estimate of $Nu$ through equation (\ref{Nu_estimate}) is no longer applicable.}

\begin{figure}
	\centering
	\includegraphics[width=0.48\linewidth]{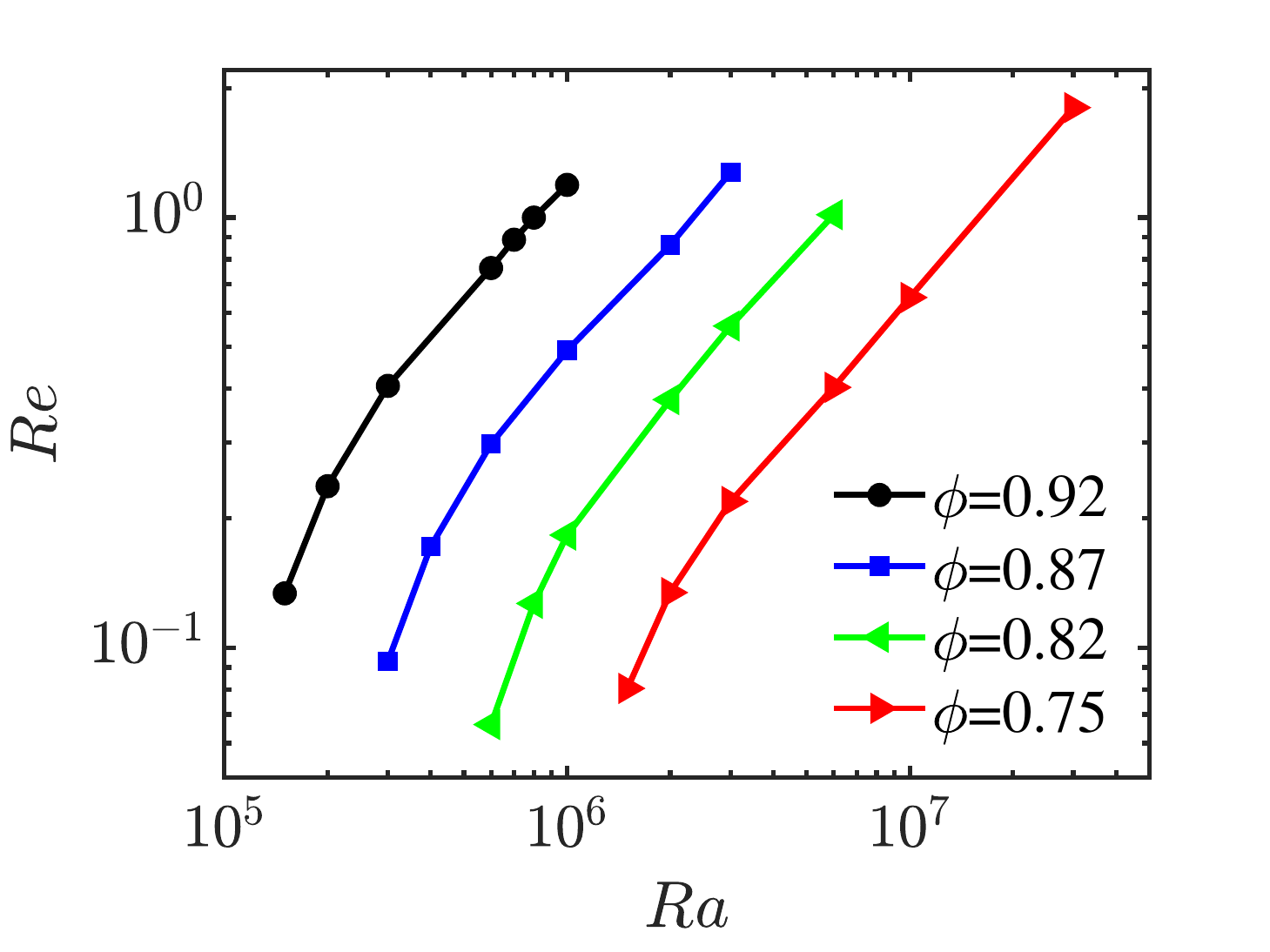}
	\put(-180,121){$(a)$}
	\includegraphics[width=0.48\linewidth]{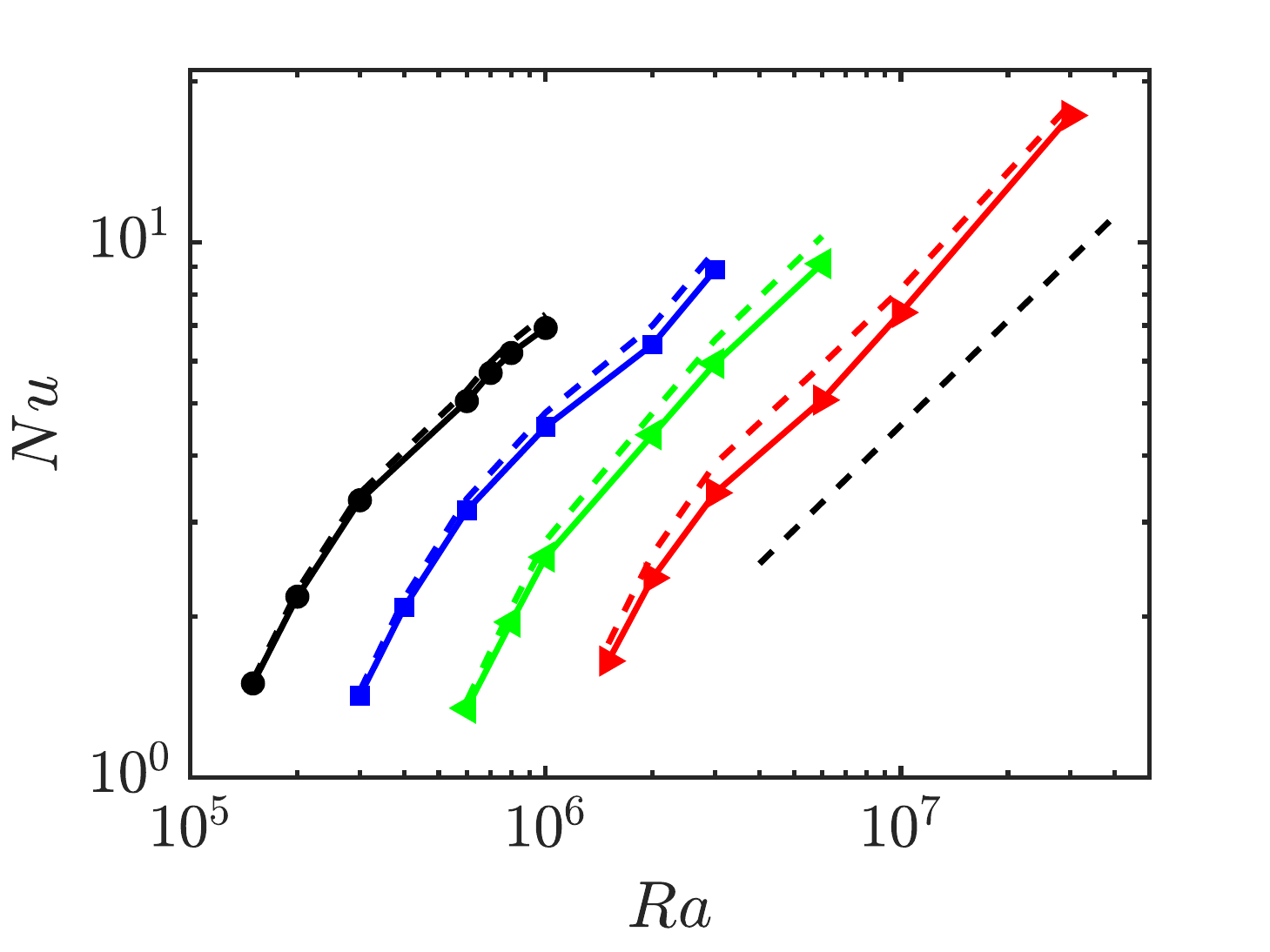}
	\put(-180,121){$(b)$}
	\put(-50,66){$Ra^{0.65}$}
	\caption{\label{nu_based_on_velo} $(a)$ $Re(Ra)$ in the small-$Ra$ regime. $(b)$ Comparison of $Nu$ obtained based on the {\color{black}heat flux averaging $Nu=-\langle\partial_z T\rangle_{x,t}$ over the horizontal plates} (dashed lines) and equation (\ref{Nu_estimate}) with $c=8.0$ (solid lines with symbols) for the same four cases for the porosity $\phi$ shown in $(a)$.}
\end{figure}

\section{Summary}\label{sec:summary}
We have studied the heat transfer and flow structure of 2D RB convection in regular porous media using pore-scale modeling. The porous medium is comprised of circular, solid obstacles {\color{black}located} on a square lattice. The heat transfer between the fluid and solid phases is considered.

The obstacle array has two competing effects on the heat transfer. On the one hand, the flow becomes more coherent with the correlation between temperature fluctuation and vertical velocity enhanced and the counter-gradient convective heat transfer suppressed, leading to heat transfer enhancement. On the other hand, the convection strength is reduced due the impedance of the obstacle array, leading to heat transfer reduction.
{\color{black}The coexistence of the two competing effects leads to the non-monotonic {\color{black}behaviour} of $Nu(\phi)$ as $\phi$ is decreased from 1, as shown in figure \ref{optimal_spacing}.} The heat transfer enhancement is consistent with the counterintuitive observation that an appropriate strength of a stabilizing force can enhance heat transfer by increasing flow coherence \citep{chong2017confined}. 
 Significant enhancement of the heat transfer due to the increased flow coherence was also observed in the confined inclined convection in low-$Pr$ fluids \citep{zwirner2018confined}. Due to the emergence of system-sized plume columns and the interaction of these plume columns with the opposed boundary layers in inclined convection, an increase of the heat transfer by a factor of approximately 2.3 can be realized.
The observations of heat transfer enhancement in these distinct systems are manifestations of a universal mechanism to enhance turbulent transfer by increasing flow coherence.

{\color{black}The influence of porosity on flow properties is dependent on $Ra$, and two different heat transfer regimes are observed at fixed $\phi$. In the small-$Ra$ regime where viscosity dominates, $Nu$ is decreased as compared to the $\phi=1$ case with a steep effective power law $Nu\sim Ra^{0.65}$, while in the large-$Ra$ regime $Nu$ is increased as compared to $\phi=1$ with the classical power law $Nu\sim Ra^{0.30}$. The scaling crossover occurs when the thickness of the thermal boundary layer $\delta_{th}$ is comparable to the pore scale $l$.}

The influence of the obstacle array on the heat transfer is also analyzed from the local perspective, namely, the energy dissipation rates. It is found that the bulk contribution of kinetic energy dissipation rate is enhanced as $\phi$ is slightly decreased from 1; while when $\phi$ is small enough, convection is significantly slowed down by the obstacle array, and the kinetic energy dissipation rate is decreased in the whole cell. 
For small $Ra$, due to the large coherence length of the flow structures and the impedance of the obstacle array, turbulence is suppressed in the pores and additional laminar-type boundary layers appear in the bulk. {\color{black}It is shown that also for porous-media convection $Nu$ can be estimated from $Re$ through equation (\ref{Nu_estimate}) similarly to that suggested by \cite{grossmann2001thermal,grossmann2004fluctuations} for the viscosity-dominated so-called $\infty$-regime.}
{\color{black}Regarding the turbulent modulations, the results of this study suggest that it is possible to modulate the heat transfer with porous structure and that the transitional $Ra$ between different heat transfer regimes can be changed by manipulating the pore scale.}

{\color{black}The influence of the obstacle arrangement on the flow structure is also studied. Three different arrangements with similar porosities are considered. The flow structure is found to be significantly influenced by the obstacle arrangement. In regular porous media, thermal plumes can penetrate deep into the bulk along convection channels aligning vertically, resulting in strong fluid and heat transport. When the obstacles are randomly located in the cell, the plume motion is less organized, and curved convection channels with strong flow emerge in the pores, with reduced efficiency of vertical heat transfer.}

In this study we {\color{black}have mainly focused} on the heat transfer and flow structure of 2D RB convection in regular porous media.
In the future, it will be of interest to extend the study to the 3D case, to allow for a one-to-one comparison with experiment.
Following the comparative study of \cite{chong2017confined} and \cite{lim2019quasistatic}, {\color{black}it is also desirable to investigate further turbulent flows with different stabilizing forces, to get an even more complete understanding of various stabilizing-destabilizing flow systems.}

\section*{Acknowledgements}
This work was supported by the Natural Science Foundation of China (Grant Nos. 11988102, 91852202, 11861131005 and 11672156), D.L.'s ERC-Advanced Grant (Grant No. 740479), and {\color{black}Tsinghua University Initiative Scientific Research Program (Grant No. 20193080058)}. Z.W. acknowledges the support of {\color{black}the Natural Science Foundation of China (Grant No. 11621202)}. 
{\color{black}S.L. acknowledges the project funded by the China Postdoctoral Science Foundation.}
We are grateful to the Dutch Supercomputing Consortium SURFsara and Chinese supercomputer Tianhe-2 for the allocation of computing time. 

\section*{Declaration of interests}
The authors report no conflict of interest.

%\bibliographystyle{jfm}
%\bibliography{ref_porous}

\end{document}